\documentclass[12pt]{article}
\pdfoutput=1
\usepackage{latexsym}
\usepackage{amssymb, amsfonts, amsmath}
\usepackage{indentfirst}
\usepackage{bbm}
\usepackage{amssymb}
\usepackage{verbatim}
\usepackage{amsmath, amsthm, amssymb}
\usepackage{mathrsfs}
\usepackage{amsfonts}
\usepackage{dsfont}
\usepackage{csquotes}
\usepackage{hyperref}
\usepackage{url}
\usepackage{xcolor}
\usepackage{xpatch}
\usepackage[english]{babel}

\usepackage{graphicx} 
\usepackage[export]{adjustbox}
\usepackage{float}
\graphicspath{ {images/} }

\usepackage[style=phys, articletitle=true, biblabel=brackets, backend=bibtex,
chaptertitle=false, pageranges=false, eprint=true,%
natbib=true, maxbibnames=10, giveninits=true, sorting=none]{biblatex} 

\DeclareFieldFormat[article]{journaltitle}{\mkbibitalic{#1\isdot}} 
\makeatletter
\newrobustcmd{\mkbibfixedbrackets}[1]{%
	\begingroup
	\blx@blxinit
	\blx@setsfcodes
	\bibleftbracket#1\bibrightbracket
	\endgroup}

\renewbibmacro*{doi+eprint+url}{%
	\iftoggle{bbx:doi}
	{\printfield{doi}}
	{}%
	\ifboolexpr{test {\iffieldequalstr{eprinttype}{arxiv}} or test {\iffieldequalstr{eprinttype}{arXiv}}}
	{\setunit{\addspace}\newblock}
	{\newunit\newblock}%
	\iftoggle{bbx:eprint}
	{\usebibmacro{eprint}}
	{}%
	\newunit\newblock
	\iftoggle{bbx:url}
	{\usebibmacro{url+urldate}}
	{}}

\xpatchbibdriver{online}
{\newunit\newblock
	\iftoggle{bbx:eprint}
	{\usebibmacro{eprint}}
	{}}
{\ifboolexpr{test {\iffieldequalstr{eprinttype}{arxiv}} or test {\iffieldequalstr{eprinttype}{arXiv}}}
	{\setunit{\addspace}\newblock}
	{\newunit\newblock}%
	\iftoggle{bbx:eprint}
	{\usebibmacro{eprint}}
	{}}
{}{}

\DeclareFieldFormat{eprint:arxiv}{%
	\mkbibbrackets{%
		\ifhyperref
		{\href{http://arxiv.org/\abx@arxivpath/#1}{%
				arXiv\addcolon
				\nolinkurl{#1}%
				\iffieldundef{eprintclass}
				{}
				{\addspace\UrlFont{\mkbibfixedbrackets{\thefield{eprintclass}}}}}}
		{arXiv\addcolon
			\nolinkurl{#1}%
			\iffieldundef{eprintclass}
			{}
			{\addspace\UrlFont{\mkbibfixedbrackets{\thefield{eprintclass}}}}}}}
\makeatother

\parskip=\medskipamount

\newcommand{\cA}{{\cal A}}
\newcommand{\cB}{{\cal B}}

\newcommand{\cF}{{\cal F}}

\newcommand{\cH}{{\cal H}}
\newcommand{\cI}{{\cal I}}

\newcommand{\cN}{{\cal N}}

\newcommand{\cT}{{\cal T}}

\def\a{\alpha}

\def\b{\beta}

\def\d{\delta}
\def\e{\epsilon}
\def\f{\phi}
\def\g{\gamma}
\def\G{\Gamma}

\def\l{\lambda}

\def\q{\theta}
\def\r{\rho}
\def\s{\sigma}

\def\x{\xi}

\def\D{\Delta}
\def\F{\Phi}
\def\J{\Psi}

\def\P{\Pi}

\newcommand{\ve}{\varepsilon}                            

\newcommand{\pa}{\partial}                           

\newcommand{\be}{\begin{equation}}
\newcommand{\ee}{\end{equation}}
\newcommand{\bea}{\begin{eqnarray}}
\newcommand{\eea}{\end{eqnarray}}

\newcommand{\ba}{\begin{array}}
\newcommand{\ea}{\end{array}}

\def\double #1{#1{\hbox{\kern-2pt $#1$}}}

\newcommand{\bsubeq}{\begin{subequations}}
\newcommand{\esubeq}{\end{subequations}}

\numberwithin{equation}{section}

\begin{document}

\begin{titlepage}
\begin{flushright}
Oct, 2022
\end{flushright}
\vspace{2mm}

\begin{center}
\Large \bf Three-point functions of conserved currents in 3D CFT: general formalism for arbitrary spins
\end{center}

\begin{center}
{\bf
Evgeny I. Buchbinder and Benjamin J. Stone}

{\footnotesize{
{\it Department of Physics M013, The University of Western Australia\\
35 Stirling Highway, Crawley W.A. 6009, Australia}} ~\\
}
\end{center}
\begin{center}
\texttt{Email: evgeny.buchbinder@uwa.edu.au, \\ benjamin.stone@research.uwa.edu.au}
\end{center}

\vspace{4mm}
\begin{abstract}
\baselineskip=14pt
\noindent We analyse the general structure of the three-point functions involving conserved bosonic and fermionic higher-spin currents in three-dimensional conformal field theory. Using the constraints of conformal symmetry and conservation equations, we use a computational formalism to analyse the general structure of $\langle J^{}_{s_{1}} J'_{s_{2}} J''_{s_{3}} \rangle$, where $J^{}_{s_{1}}$, $J'_{s_{2}}$ and $J''_{s_{3}}$ are conserved currents with spins $s_{1}$, $s_{2}$ and $s_{3}$ respectively (integer or half-integer). The calculations are completely automated for any chosen spins and are limited only by computer power. We find that the correlation function is in general fixed up to two independent ``even" structures, and one ``odd" structure, subject to a set of triangle inequalities. We also analyse the structure of three-point functions involving higher-spin currents and fundamental scalars and spinors. 


\end{abstract}
\end{titlepage}

\newpage
\renewcommand{\thefootnote}{\arabic{footnote}}
\setcounter{footnote}{0}

\tableofcontents
\vspace{1cm}
\bigskip\hrule


\section{Introduction}\label{section1}

It is widely understood that in any conformal field theory, the general structure of three-point correlation functions is determined up to finitely many parameters by conformal symmetry. However, it remains a non-trivial problem to construct explicit solutions for three-point functions for various classes of primary operators. Among the most important primary operators are conserved currents, whose scale dimension saturates the unitarity bound. The fundamental examples of conserved currents in any conformal field theory are the energy-momentum tensor and vector currents; the three-point functions of these currents were analysed in \cite{Osborn:1993cr, Erdmenger:1996yc}, where a systematic approach to study correlation functions of primary operators was introduced (see also refs.~\cite{Polyakov:1970xd, Schreier:1971um, Migdal:1971xh, Migdal:1971fof, Ferrara:1972cq, Ferrara:1973yt, Koller:1974ut, Mack:1976pa, Fradkin:1978pp, Stanev:1988ft} for earlier works). The analysis was performed in general dimensions, however, it did not consider higher-spin conserved currents, which can exist in more general conformal field theories. It also did not account for the possibility of parity-violating structures, which appear in the three-point functions of the energy-momentum tensor and vector currents in three-dimensions. These structures were found in~\cite{Giombi:2011rz}, where correlation functions of higher-spin conserved currents were considered, and were also found to contain parity-violating structures. Soon after, it was proven in~\cite{Maldacena:2011jn} that under certain assumptions (which are, however, violated in the presence of fermionic higher-spin currents) that all correlation functions involving the energy-momentum tensor and higher-spin currents are equal to those of free theories. This is an extension of the Coleman-Mandula theorem \cite{Coleman:1967ad} to conformal field theories; it was originally proven in three dimensions and was later generalised to four- and higher-dimensional cases in~\cite{Zhiboedov:2012bm, Stanev:2012nq, Stanev:2013qra, Alba:2013yda, Alba:2015upa}. There are also approaches 
to the construction of correlation functions of conserved currents which make use of embedding formalisms~\cite{Weinberg:2010fx, Costa:2011dw, Costa:2011mg, Weinberg:2012mz, Costa:2014rya, Fortin:2020des} (see also \cite{Goldberger:2011yp, Goldberger:2012xb} for supersymmetric extensions), while others carry out the calculations in momentum space~\cite{Bzowski:2013sza,Coriano:2018bbe,Bzowski:2017poo,Isono:2019ihz, Bautista:2019qxj, Jain:2020puw, Jain:2020rmw, Jain:2021gwa, Jain:2021vrv, Jain:2021wyn}. Results have also been obtained within the framework of the AdS/CFT correspondence (see e.g. \cite{Freedman:1998tz, Giombi:2009wh, Fitzpatrick:2011ia, Didenko:2012tv, Skvortsov:2022wzo}). The study of correlation functions of conserved currents has also been extended to superconformal field theories in diverse dimensions \cite{Park:1997bq, Osborn:1998qu, Park:1998nra, Park:1999pd, Park:1999cw, Kuzenko:1999pi, Nizami:2013tpa, Buchbinder:2015qsa, Buchbinder:2015wia, Kuzenko:2016cmf, Buchbinder:2021gwu, Buchbinder:2021izb, Buchbinder:2021kjk, Buchbinder:2021qlb, Jain:2022izp, Buchbinder:2022kmj}.

The general structure of the three-point functions of conserved higher-spin, bosonic, vector currents was proposed by Giombi, Prakash and Yin \cite{Giombi:2011rz} in three dimensions, and further analysis was undertaken by Stanev~\cite{Stanev:2012nq, Stanev:2013eha, Stanev:2013qra} (see also~\cite{Elkhidir:2014woa,Buchbinder:2022cqp}) in the four dimensional case, and by Zhiboedov \cite{Zhiboedov:2012bm} in general dimensions. 
Despite the obvious success, the analysis in \cite{Giombi:2011rz, Stanev:2012nq,Zhiboedov:2012bm} appears to have some limitations. First, the results only apply to conserved currents of integer spin. Second, it is unclear how the results comprise all linearly independent structures for a given choice of spins. In particular, in \cite{Stanev:2012nq, Zhiboedov:2012bm}, the conserved three-point functions are presented in the form of generating functions which are proposed (to best of our understanding) without proof of the latter.

In this paper, we develop a formalism to study the general structure of the three-point correlation function
\begin{equation}
	\langle J^{}_{s_{1}}(x_{1}) \, J'_{s_{2}}(x_{2}) \, J''_{s_{3}}(x_{3}) \rangle \, ,
\end{equation}
in three-dimensional conformal field theory, 
assuming only the constraints imposed by conformal symmetry and conservation equations. Here by $J^{}_{s}$ we denote a conserved current of spin $s$. 
Our formalism is suitable for both integer and half-integer spin.
Within our approach we reproduce all known results concerning the structure of three-point functions of bosonic conserved currents and also extend the results to three-point functions involving currents of an arbitrary 
half-integer spin. We also apply it to correlation functions of scalar/spinor operators thus covering essentially all possible three-point function in three-dimensional conformal field 
theory. Our method is exhaustive; first we construct all possible structures for the correlation function for a given set of spins $s_1, s_2$ and $s_3$, consistent with its conformal properties. We then systematically extract the linearly independent structures and then, finally, impose the conservation equations and symmetries under permutations of spacetime points. As a result we obtain the three-point function in a very explicit form which can be explicitly presented even for relatively high spins.\footnote{A similar analysis can also be done in the four-dimensional case and will appear elsewhere.}
Our method can be applied for arbitrary $s_1, s_2$ and $s_3$ and is limited only by computer power. Due to these limitations we were able to 
carry out computations up to $s_{i} = 20$, however, with a sufficiently powerful computer one could probably extend our results up to $s_{i} \sim 50$ as in~\cite{Stanev:2012nq}.
We demonstrate that in all cases with $s_{i} \leq 20$, including examples involving conserved {\it half-integer} spin currents, that the correlation function is fixed up to the following form:
\begin{equation}
	\langle J^{}_{s_{1}} J'_{s_{2}} J''_{s_{3}} \rangle = a_{1} \, \langle J^{}_{s_{1}} J'_{s_{2}} J''_{s_{3}} \rangle_{E_1} + a_{2} \, \langle J^{}_{s_{1}} J'_{s_{2}} J''_{s_{3}} \rangle_{E_2} + b \, \langle J^{}_{s_{1}} J'_{s_{2}} J''_{s_{3}} \rangle_{O} \, .
\end{equation}
where $\langle J^{}_{s_{1}} J'_{s_{2}} J''_{s_{3}} \rangle_{E_1}$ and $\langle J^{}_{s_{1}} J'_{s_{2}} J''_{s_{3}} \rangle_{E_2}$ are parity-even solutions (in the bosonic case 
corresponding to free bosonic and fermionic theories respectively), while $\langle J^{}_{s_{1}} J'_{s_{2}} J''_{s_{3}} \rangle_{O}$ is a parity-violating (or parity-odd) solution. Parity-odd solutions are unique to three dimensions, and have been shown to correspond to Chern-Simons theories interacting with parity-violating matter \cite{Aharony:2011jz, Giombi:2011kc, Maldacena:2012sf, Jain:2012qi, GurAri:2012is, Aharony:2012nh, Giombi:2016zwa, Chowdhury:2017vel, Sezgin:2017jgm, Skvortsov:2018uru, Inbasekar:2019wdw}.\footnote{The parity-odd terms in correlation functions involving scalars and spinors can also arise in theories without a Chern-Simons term, for example in theories with fermions in three-dimensions, because $\bar{\psi}\psi$ is a parity-odd pseudoscalar, see e.g. \cite{Prakash:2017hwq,Prakash:2022gvb}. We are grateful to S. Prakash for pointing this out.}
Further, the existence of the odd solution depends on a set of triangle inequalities:
\begin{align}
	s_{1} \leq s_{2} + s_{3} \, , && s_{2} \leq s_{1} + s_{3} \, , && s_{3} \leq s_{1} + s_{2} \, .
\end{align}
When the triangle inequalities are simultaneously satisfied there are two even solutions, and one odd solution. However, when any one of the above relations 
is not satisfied there are only two even solutions; the odd solution is incompatible with conservation equations. 

The analysis quickly becomes cumbersome due to the proliferation of tensor indices; to streamline the calculations we develop a hybrid, 
index-free formalism which combines the approach of Osborn and Petkou~\cite{Osborn:1993cr} and a method based on contraction of tensor indices with auxiliary spinors. 
This method is widely used throughout the literature to construct correlation functions involving more complicated tensor operators. Our particular approach, 
however, describes the correlation function completely in terms of a polynomial which is a function of a single conformally covariant three-point building block, $X$, and the auxiliary spinor 
variables $u$, $v$, and $w$. Hence, one does not have to work with the spacetime points explicitly when imposing conservation equations. To find all solutions for the polynomial, we construct a generating function which produces an exhaustive list of all possible linearly dependent structures for a given set of spins using \textit{Mathematica}. With the use of pattern-matching functions, we then systematically apply linear dependence relations to this set of structures to form a linearly-independent ansatz for the correlation function. Once this ansatz is obtained, we impose conservation equations and any symmetries due to permutation of spacetime points. The tensor structures (related to the leading singular OPE coefficient, as in \cite{Osborn:1993cr}) may then be read off by acting on the polynomials with appropriate partial derivatives in the auxiliary spinors. The computational approach we have developed is essentially automatic and limited only by computer power; one simply chooses the spins of the fields and the solution for the three-point function consistent with conservation and point-switch symmetries is generated. 

The results of this paper are organised as follows. In section \ref{section2} we review the essentials of the group theoretic formalism used to construct correlation functions of 
primary operators in three dimensions. In section \ref{section3} we develop the formalism necessary to impose all constraints arising from conservation equations and point switch symmetries on three-point functions. 
In particular, we introduce an index-free, auxiliary spinor formalism which allows us to construct a generating function for the three-point functions, and we outline the important aspects of our computational approach. Section \ref{section4} is then devoted to the analysis of three-point functions involving bosonic conserved currents. We show that we reproduce the known results 
previously found and proposed in~\cite{Maldacena:2011jn,Giombi:2011rz,Zhiboedov:2012bm}.
In section \ref{section5} we analyse the structure of correlation functions involving fermionic currents. We present an explicit analysis for three-point correlation functions involving 
combinations of a ``supersymmetry-like" spin-3/2 current, the energy-momentum tensor and the conserved vector current. 
The results are then expanded to include higher-spin conserved currents. In section \ref{section6}, for completeness, we perform the analysis of correlation functions involving 
combinations of scalars, spinors and conserved higher-spin currents. Finally, in section \ref{section7}, we comment on the general results in the context of superconformal field theories. 
The appendices are devoted to mathematical conventions, various useful identities and extra results for higher-spin conserved currents. In particular, in appendix \ref{AppB} we
present some extra results for higher-spin three-point functions to illustrate that our method produces very explicit results even for relatively high spins.





\section{Conformal building blocks}\label{section2}

In this section we will review the pertinent aspects of the group theoretic formalism used to compute correlation functions of primary operators in three dimensional conformal field theories. For a more detailed review of the formalism as applied to correlation functions of bosonic primary fields, the reader may consult \cite{Osborn:1993cr}. 

\subsection{Two-point building blocks}\label{subsection2.1}

Consider 3D Minkowski space $\mathbb{M}^{1,2}$, parameterised by coordinates $ x^{m} $, where $m = 0, 1, 2$ are Lorentz indices. Given two points, $x_{1}$ and $x_{2}$, we can define the covariant two-point function
\begin{equation} \label{Two-point building blocks 1}
	x_{12}^{m} = (x_{1} - x_{2})^{m} \, , \hspace{10mm} x_{21}^{m} = - x_{12}^{m} \, . 
\end{equation}
Next, following Osborn and Petkou \cite{Osborn:1993cr}, we introduce the conformal inversion tensor, $I_{mn}$, which is defined as follows:
\begin{align} \label{Inversion tensor}
I_{mn}(x) = \eta_{mn} - 2 \, \frac{ x_{m} x_{n}}{x^{2}} \, , \hspace{10mm} I_{m a}(x) \, I^{a n}(x) = \d_{m}^{n} \, .
\end{align}
This object played a pivotal role in the construction of correlation functions in \cite{Osborn:1993cr}, as the full conformal group may be generated by 
considering Poincar\'e transformations supplemented by inversions. However, in the context of this work, we require an analogous operator for the spinor representation. 
Hence, we convert the vector two-point functions \eqref{Two-point building blocks 1} into spinor notation using the conventions outlined in appendix \ref{AppA}:
\begin{align}
	x_{12 \, \a \b} &= (\g^{m})_{\a \b} x_{12 \, m} \, , & x_{12}^{\a \b} &= (\g^{m})^{\a \b} x_{12 \, m} \, , & x_{12}^{2} &= - \frac{1}{2} x_{12}^{\a \b} x^{}_{12 \, \a \b} \, .
\end{align}
In this form the two-point functions possess the following useful properties:
\begin{align}  \label{Two-point building blocks - properties 1} 
 	x_{12 \, \a \b} = x_{12 \, \b \a} \, , \hspace{10mm} x_{12}^{\a \s} x^{}_{12 \, \s \b} = - x_{12}^{2} \d_{\b}^{\a} \, . 
\end{align}
Hence, we find
\begin{equation} \label{Two-point building blocks 4}
	(x_{12}^{-1})^{\a \b} = - \frac{x_{12}^{\a \b}}{x_{12}^{2}} \, .
\end{equation}
We also introduce the normalised two-point functions, denoted by $\hat{x}_{12}$,
\begin{align} \label{Two-point building blocks 3}
		\hat{x}_{12 \, \a \b} = \frac{x_{12 \, \a \b}}{( x_{12}^{2})^{1/2}} \, , \hspace{10mm} \hat{x}_{12}^{\a \s} \hat{x}^{}_{12 \, \s \b} = - \d_{\b}^{\a} \, . 
\end{align}
From here we can now construct an operator analogous to the conformal inversion tensor acting on the space of symmetric traceless spin-tensors of arbitrary rank. Given a two-point function $x$, we define the operator
\begin{equation} \label{Higher-spin inversion operators a}
\cI_{\a(k) \b(k)}(x) = \hat{x}_{(\a_{1} (\b_{1}} \dots \hat{x}_{ \a_{k}) \b_{k})}  \, ,
\end{equation}
along with its inverse
\begin{equation} \label{Higher-spin inversion operators b}
\cI^{\a(k) \b(k)}(x) = \hat{x}^{(\a_{1} (\b_{1}} \dots \hat{x}^{ \a_{k}) \b_{k})} \, .
\end{equation}
The spinor indices may be raised and lowered using the standard conventions as follows:
\begin{subequations}
	\begin{align}
		\cI_{\a(k)}{}^{\b(k)}(x) &= \ve^{\b_{1} \g_{1}} \dots \ve^{\b_{k} \g_{k}} \, \cI_{\a(k) \g(k)}(x) \, .
	\end{align}
\end{subequations}
Now due to the property
\begin{equation}
	\cI_{\a(k) \b(k)}(-x) = (-1)^{k} \cI_{\a(k) \b(k)}(x) \, ,
\end{equation}
the following identity holds for products of inversion tensors:
\begin{subequations} \label{Higher-spin inversion operators - properties}
	\begin{align}
		\cI_{\a(k) \s(k)}(x_{12}) \, \cI^{\s(k) \b(k)}(x_{21}) &= \d_{(\a_{1}}^{(\b_{1}} \dots \d_{\a_{k})}^{\b_{k})} \, .
	\end{align}
\end{subequations}
The objects \eqref{Higher-spin inversion operators a}, \eqref{Higher-spin inversion operators b} prove to be essential in the construction of correlation functions of primary operators with arbitrary spin. Indeed, the vector representation of the inversion tensor may be recovered in terms of the spinor two-point functions as follows:
\begin{equation}
	I_{m n}(x) = - \frac{1}{2} \, \text{Tr}( \g_{m} \, \hat{x} \, \g_{n} \, \hat{x} ) \, .
\end{equation}
%
%


\subsection{Three-point building blocks}\label{subsection2.2}

Given three distinct points in Minkowski space, $x_{i}$, with $i = 1,2,3$, we define conformally covariant three-point functions in terms of the two-point functions as in \cite{Osborn:1993cr}
\begin{align}
	X_{ij} &= \frac{x_{ik}}{x_{ik}^{2}} - \frac{x_{jk}}{x_{jk}^{2}} \, , & X_{ji} &= - X_{ij} \, ,  & X_{ij}^{2} &= \frac{x_{ij}^{2}}{x_{ik}^{2} x_{jk}^{2} } \, , 
\end{align}
where $(i,j,k)$ is a cyclic permutation of $(1,2,3)$. For example we can have
\begin{equation}
	X_{12}^{m} = \frac{x_{13}^{m}}{x_{13}^{2}} - \frac{x_{23}^{m}}{x_{23}^{2}} \, , \hspace{10mm} X_{12}^{2} = \frac{x_{12}^{2}}{x_{13}^{2} x_{23}^{2} } \, .
\end{equation}
There are several useful identities involving the two- and three-point functions along with the conformal inversion tensor. For example one can prove the algebraic identities
\begin{subequations}
	\begin{align} \label{Inversion tensor identities - vector case 1}
		I_{m}{}^{a}(x_{13}) \, I_{a n}(x_{23}) &= I_{m}{}^{a}(x_{12}) \, I_{a n}(X_{13}) \, ,  & I_{m n}(x_{23}) \, X_{12}^{n} &= \frac{x_{12}^{2}}{x_{13}^{2}} \, X_{13 \, m} \, ,
	\end{align} \vspace{-5mm}
	\begin{align} \label{Inversion tensor identities - vector case 2}
		I_{m}{}^{a}(x_{23}) \, I_{a n}(x_{13}) &= I_{m}{}^{a}(x_{21}) \, I_{a n}(X_{32}) \, ,  & I_{m n}(x_{13}) \, X_{12}^{n} &= \frac{x_{12}^{2}}{x_{23}^{2}} \, X_{32 \, m} \, .
	\end{align}
\end{subequations}
The three-point functions also possess the following differential properties:
\begin{align}
	\pa_{(1) \, m} X_{12 \, n} = \frac{1}{x_{13}^{2}} I_{m n}(x_{13}) \, , \hspace{10mm} \pa_{(2) \, m} X_{12 \, n} = - \frac{1}{x_{23}^{2}} I_{m n}(x_{23}) \, . \label{Inversion tensor identities - vector case 3}
\end{align}
Converting to spinor notation, the three-point functions may be represented as follows:
\begin{equation}
	X_{ij , \, \a \b} = (\g_{m})_{\a \b} X_{ij}^{m} \, , \hspace{10mm} X_{ij , \, \a \b} = - (x^{-1}_{ik})_{\a \s} x_{ij}^{\s \g} (x^{-1}_{jk})_{\g \b} \, .
\end{equation}
These objects satisfy properties similar to the two-point functions, as in \eqref{Two-point building blocks - properties 1}. It is also convenient to define the normalised three-point functions, $\hat{X}_{ij}$, and the inverses, $(X_{ij}^{-1})$,
\begin{equation}
	\hat{X}_{ij , \, \a \b} = \frac{X_{ij , \, \a \b}}{( X_{ij}^{2})^{1/2}} \, , \hspace{10mm}	(X_{ij}^{-1})^{\a \b} = - \frac{X_{ij}^{\a \b}}{X_{ij}^{2}} \, .
\end{equation}  
Now given an arbitrary three-point building block, $X$, it is useful to construct the following higher-spin inversion operator:
\begin{equation}
	\cI_{\a(k) \b(k)}(X) = \hat{X}_{ (\a_{1} (\b_{1}} \dots \hat{X}_{\a_{k}) \b_{k})}  \, , \label{Inversion tensor identities - three point functions a}
\end{equation}
along with its inverse
\begin{equation}
	\cI^{\a(k) \b(k)}(X) = \hat{X}^{(\a_{1} (\b_{1}} \dots \hat{X}^{ \a_{k}) \b_{k})} \, . \label{Inversion tensor identities - three point functions b}
\end{equation}
These operators possess properties similar to the two-point higher-spin inversion operators \eqref{Higher-spin inversion operators a}, \eqref{Higher-spin inversion operators b}. There are also some useful algebraic identities relating the two- and three-point functions at various points, such as
\begin{equation}
	 \cI_{\a \s}(x_{13}) \, \cI^{\s \g}(x_{12}) \, \cI_{\g \b}(x_{23}) = \cI_{\a \b}(X_{12}) \, , \hspace{5mm}  \cI^{\a \s}(x_{13}) \, \cI_{\s \g}(X_{12})  \, \cI^{\g \b}(x_{13}) = \cI^{\a \b}(X_{32})  \, . \label{Inversion tensor identities - spinor case}
\end{equation}
These identities (and cyclic permutations of them) are analogous to \eqref{Inversion tensor identities - vector case 1}, \eqref{Inversion tensor identities - vector case 2}, and also admit higher-spin generalisations, for example
\begin{equation}
	\cI^{\a(k) \s(k)}(x_{13}) \, \cI_{\s(k) \g(k)}(X_{12}) \, \cI^{\g(k) \b(k)}(x_{13}) = \cI^{\a(k) \b(k)}(X_{32})  \, . \label{Inversion tensor identities - higher spin case}
\end{equation}
In addition, similar to \eqref{Inversion tensor identities - vector case 3}, there are also the following identities:
\begin{equation}
	\pa_{(1) \, \a \b} X_{12}^{ \g \d} = - \frac{2}{x_{13}^{2}} \, \cI_{(\a}{}^{\g}(x_{13}) \,  \cI_{\b)}{}^{\d}(x_{13}) \, , \hspace{5mm} \pa_{(2) \, \a \b} X_{12}^{ \g \d} = \frac{2}{x_{23}^{2}} \, \cI_{(\a}{}^{\g}(x_{23}) \,  \cI_{\b)}{}^{\d}(x_{23}) \, . \label{Three-point building blocks - differential identities}
\end{equation}
These identities allow us to account for the fact that correlation functions of primary fields obey differential constraints which can arise due to conservation equations. Indeed, given a tensor field $\cT_{\cA}(X)$, there are the following differential identities which arise as a consequence of \eqref{Three-point building blocks - differential identities}:
\begin{subequations}
	\begin{align}
		\pa_{(1) \, \a \b} \cT_{\cA}(X_{12}) &= \frac{1}{x_{13}^{2}} \, \cI_{\a}{}^{\g}(x_{13}) \,  \cI_{\b}{}^{\d}(x_{13}) \, \frac{ \pa}{ \pa X_{12}^{ \g \d}} \, \cT_{\cA}(X_{12}) \, ,  \label{Three-point building blocks - differential identities 2} \\[2mm]
		\pa_{(2) \, \a \b} \cT_{\cA}(X_{12}) &= - \frac{1}{x_{23}^{2}} \, \cI_{\a}{}^{\g}(x_{23}) \,  \cI_{\b}{}^{\d}(x_{23}) \, \frac{ \pa}{ \pa X_{12}^{ \g \d}} \, \cT_{\cA}(X_{12}) \, . \label{Three-point building blocks - differential identities 3}
	\end{align}
\end{subequations}

\section{General formalism for correlation functions of primary operators}\label{section3}

In this section we develop a formalism to construct correlation functions of higher-spin primary operators in 3D conformal field theories. We utilise a hybrid method which combines auxiliary spinors with the approach of Osborn \& Petkou \cite{Osborn:1993cr}.

\subsection{Two-point functions}\label{subsection3.1}

Let $\F_{\cA}$ be a primary field with dimension $\D$, where $\cA$ denotes a collection of Lorentz spinor indices. The two-point correlation function of $\F_{\cA}$ is fixed by conformal symmetry to the form
\begin{equation} \label{Two-point correlation function}
	\langle \F_{\cA}(x_{1}) \, \F^{\cB}(x_{2}) \rangle = c \, \frac{\cI_{\cA}{}^{\cB}(x_{12})}{(x_{12}^{2})^{\D}} \, , 
\end{equation} 
where $\cI$ is an appropriate representation of the inversion tensor and $c$ is a constant real parameter. The denominator of the two-point function is determined by the conformal dimension of $\F_{\cA}$, which guarantees that the correlation function transforms with the appropriate weight under scale transformations.

\subsection{Three-point functions}\label{subsection3.2}

Now concerning three-point correlation functions, let $\F$, $\J$, $\P$ be primary fields with scale dimensions $\D_{1}$, $\D_{2}$ and $\D_{3}$ respectively. The three-point function may be constructed using the general ansatz
\begin{align}
	\langle \F_{\cA_{1}}(x_{1}) \, \J_{\cA_{2}}(x_{2}) \, \P_{\cA_{3}}(x_{3}) \rangle = \frac{ \cI^{(1)}{}_{\cA_{1}}{}^{\cA'_{1}}(x_{13}) \,  \cI^{(2)}{}_{\cA_{2}}{}^{\cA'_{2}}(x_{23}) }{(x_{13}^{2})^{\D_{1}} (x_{23}^{2})^{\D_{2}}}
	\; \cH_{\cA'_{1} \cA'_{2} \cA_{3}}(X_{12}) \, , \label{Three-point function - general ansatz}
\end{align} 
where the tensor $\cH_{\cA_{1} \cA_{2} \cA_{3}}$ encodes all information about the correlation function, and is related to the leading singular OPE coefficient \cite{Osborn:1993cr}. It is highly constrained by conformal symmetry as follows:
\begin{enumerate}
	\item[\textbf{(i)}] Under scale transformations of Minkowski space $x^{m} \mapsto x'^{m} = \l^{-2} x^{m}$, the three-point building blocks transform as $X^{m} \mapsto X'^{m} = \l^{2} X^{m}$. As a consequence, the correlation function transforms as 
	\begin{equation}
		\langle \F_{\cA_{1}}(x_{1}') \, \J_{\cA_{2}}(x_{2}') \, \P_{\cA_{3}}(x_{3}') \rangle = (\l^{2})^{\D_{1} + \D_{2} + \D_{3}} \langle \F_{\cA_{1}}(x_{1}) \, \J_{\cA_{2}}(x_{2}) \,  \P_{\cA_{3}}(x_{3}) \rangle \, ,
	\end{equation}
	which implies that $\cH$ obeys the scaling property
	\begin{equation}
		\cH_{\cA_{1} \cA_{2} \cA_{3}}(\l^{2} X) = (\l^{2})^{\D_{3} - \D_{2} - \D_{1}} \, \cH_{\cA_{1} \cA_{2} \cA_{3}}(X) \, , \hspace{5mm} \forall \l \in \mathbb{R} \, \backslash \, \{ 0 \} \, .
	\end{equation}
	This guarantees that the correlation function transforms correctly under scale transformations.
	
	\item[\textbf{(ii)}] If any of the fields $\F$, $\J$, $\P$ obey differential equations, such as conservation laws in the case of conserved currents, then the tensor $\cH$ is also constrained by differential equations which may be derived with the aid of identities \eqref{Three-point building blocks - differential identities 2}, \eqref{Three-point building blocks - differential identities 3}.
	
	\item[\textbf{(iii)}] If any (or all) of the operators $\F$, $\J$, $\P$ coincide, the correlation function possesses symmetries under permutations of spacetime points, e.g.
	\begin{equation}
		\langle \F_{\cA_{1}}(x_{1}) \, \F_{\cA_{2}}(x_{2}) \, \P_{\cA_{3}}(x_{3}) \rangle = (-1)^{\e(\F)} \langle \F_{\cA_{2}}(x_{2}) \, \F_{\cA_{1}}(x_{1}) \, \P_{\cA_{3}}(x_{3}) \rangle \, ,
	\end{equation}
	where $\e(\F)$ is the Grassmann parity of $\F$. As a consequence, the tensor $\cH$ obeys constraints which will be referred to as ``point-switch identities".
	
\end{enumerate}
The constraints above fix the functional form of $\cH$ (and therefore the correlation function) up to finitely many independent parameters. Hence, using the general formula \eqref{H ansatz}, the problem of computing three-point correlation functions is reduced to deriving the general structure of the tensor $\cH$ subject to the above constraints.

\subsubsection{Conserved currents}\label{subsection2.3}

In this paper we are primarily interesting in the structure of three-point correlation functions of conserved currents. In three-dimensional conformal field theory, a conserved current with spin $s$ (integer or half-integer), is defined as a totally symmetric spin-tensor, $J_{\a_{1} \dots \a_{2s} }(x) = J_{(\a_{1} \dots \a_{2s}) }(x)$, satisfying a conservation equation of the form:
\begin{equation} \label{Conserved current}
	(\gamma^{m})^{\a_{1} \a_{2}} \pa_{m} J_{\a_{1} \a_{2} \dots \a_{2s}} = 0 \, .
\end{equation}
Conserved currents are primary fields, as they possesses the following infinitesimal conformal transformation properties \cite{Buchbinder:1998qv}:
\begin{equation}
	\delta J_{\a_{1} \dots \a_{2s}}(x) = - \xi J_{\a_{1} \dots \a_{2s}}(x) - \Delta_{J} \, \s(x) J_{\a_{1} \dots \a_{2s}}(x) + 2s \, \omega_{( \a_{1} }{}^{\delta}(x) \, J_{\a_{2} \dots \a_{2s}) \delta}(x) \, , 
\end{equation}
where $\xi$ is a conformal Killing vector field, and $\s(x)$, $\omega_{\a \b}(x)$ are local parameters defined in terms of $\xi$, which are associated with local scale and Lorentz transformations. The dimension $\Delta_{J}$ is uniquely fixed by the conservation condition \eqref{Conserved current}, as it may be shown that this condition is primary provided that $\Delta_{J} = s + 1$. This is also the dimension of the two-point correlation function \eqref{Two-point correlation function}, in the case of conserved currents.

\subsubsection{Comments on differential constraints}\label{subsubsection3.2.1}

An important aspect of this construction which requires further elaboration is that it is sensitive to the configuration of the fields in the correlation function. Indeed, depending on the exact way in which one constructs the general ansatz \eqref{H ansatz}, it can be difficult to impose conservation equations on one of the three fields due to a lack of useful identities such as \eqref{Three-point building blocks - differential identities 2}, \eqref{Three-point building blocks - differential identities 3}. To illustrate this more clearly, consider the following example; suppose we want to determine the solution for the correlation function $\langle \F_{\cA_{1}}(x_{1}) \, \J_{\cA_{2}}(x_{2}) \, \P_{\cA_{3}}(x_{3}) \rangle$, with the ansatz
\begin{equation} \label{H ansatz}
	\langle \F_{\cA_{1}}(x_{1}) \, \J_{\cA_{2}}(x_{2}) \, \P_{\cA_{3}}(x_{3}) \rangle = \frac{ \cI^{(1)}{}_{\cA_{1}}{}^{\cA'_{1}}(x_{13}) \,  \cI^{(2)}{}_{\cA_{2}}{}^{\cA'_{2}}(x_{23}) }{(x_{13}^{2})^{\D_{1}} (x_{23}^{2})^{\D_{2}}}
	\; \cH_{\cA'_{1} \cA'_{2} \cA_{3}}(X_{12}) \, . 
\end{equation} 
All information about this correlation function is encoded in the tensor $\cH$, however, this particular formulation of the three-point function prevents us from imposing conservation on the field $\P$ in a straightforward way. To rectify this issue we reformulate the ansatz with $\P$ at the front as follows:
\begin{equation} \label{Htilde ansatz}
	\langle \P_{\cA_{3}}(x_{3}) \, \J_{\cA_{2}}(x_{2}) \, \F_{\cA_{1}}(x_{1}) \rangle = \frac{ \cI^{(3)}{}_{\cA_{3}}{}^{\cA'_{3}}(x_{31}) \,  \cI^{(2)}{}_{\cA_{2}}{}^{\cA'_{2}}(x_{21}) }{(x_{31}^{2})^{\D_{3}} (x_{21}^{2})^{\D_{2}}}
	\; \tilde{\cH}_{\cA_{1} \cA'_{2} \cA'_{3} }(X_{32}) \, . 
\end{equation} 
In this case, all information about this correlation function is now encoded in the tensor $\tilde{\cH}$, which is a completely different solution compared to $\cH$. Conservation on $\P$ can now be imposed by treating $x_{3}$ as the first point with the aid of identities analogous to \eqref{Three-point building blocks - differential identities}, \eqref{Three-point building blocks - differential identities 2}, \eqref{Three-point building blocks - differential identities 3}. We now require an equation relating the tensors $\cH$ and $\tilde{\cH}$, which correspond to different representations of the same correlation function. Since the two ansatz above must be equal, we obtain the following:
\begin{align} \label{Htilde and H relation}
	\tilde{\cH}_{\cA_{1} \cA_{2}  \cA_{3} }(X_{32}) &= (x_{13}^{2})^{\D_{3} - \D_{1}} \bigg(\frac{x_{21}^{2}}{x_{23}^{2}} \bigg)^{\hspace{-1mm} \D_{2}} \, \cI^{(1)}{}_{\cA_{1}}{}^{\cA'_{1}}(x_{13}) \, \cI^{(2)}{}_{\cA_{2}}{}^{\cB_{2}}(x_{12}) \,  \cI^{(2)}{}_{\cB_{2}}{}^{\cA'_{2}}(x_{23}) \nonumber \\[-2mm]
	& \hspace{50mm} \times \cI^{(3)}{}_{\cA_{3}}{}^{\cA'_{3}}(x_{13}) \, \cH_{\cA'_{1} \cA'_{2} \cA'_{3}}(X_{12}) \, ,
\end{align}
where we have absorbed any signs due to Grassmann parity into the overall normalisation of $\tilde{\cH}$. In general, this equation is impractical to work with due to the presence of both two- and three-point functions, hence, further simplification is required. At this point it is convenient to partition our solution into ``even" and ``odd" sectors as follows:
\begin{equation}
		\cH_{\cA_{1} \cA_{2}  \cA_{3} }(X) = \cH^{(+)}_{\cA_{1} \cA_{2}  \cA_{3} }(X) + \cH^{(-)}_{\cA_{1} \cA_{2}  \cA_{3} }(X) \, , 
\end{equation}
where $\cH^{(+)}$ contains all structures involving an even number of spinor metrics, $\ve_{\a \b}$, and $\cH^{(-)}$ contains structures involving an odd number of spinor metrics. With this choice of convention, as a consequence of \eqref{Inversion tensor identities - spinor case}, \eqref{Inversion tensor identities - higher spin case}, the following relation holds:
\begin{align} \label{Hc and H relation}
	\cH^{(\pm)}_{\cA_{1} \cA_{2} \cA_{3}}(X_{32}) &= \pm \, (x_{13}^{2} X_{32}^{2})^{\D_{3} - \D_{2} - \D_{1}} \cI^{(1)}{}_{\cA_{1}}{}^{\cA'_{1}}(x_{13}) \, \cI^{(2)}{}_{\cA_{2}}{}^{\cA'_{2}}(x_{13}) \nonumber \\
	& \hspace{45mm} \times \cI^{(3)}{}_{\cA_{3}}{}^{\cA'_{3}}(x_{13}) \, \cH^{(\pm)}_{\cA'_{1} \cA'_{2} \cA'_{3}}(X_{12}) \, .
\end{align}
This equation is an extension of (2.14) in \cite{Osborn:1993cr} to spin-tensor representations, and it allows us to construct an equation relating the different representations of the correlation function. After substituting \eqref{Hc and H relation} directly into \eqref{Htilde and H relation}, we apply identities such as \eqref{Inversion tensor identities - spinor case} to obtain the following relation between $\cH$ and $\tilde{\cH}$:
\begin{equation} \label{Htilde and Hc relation}
	\tilde{\cH}^{(\pm)}_{\cA_{1} \cA_{2} \cA_{3} }(X) = \pm \, (X^{2})^{\D_{1} - \D_{3}} \, \cI^{(2)}{}_{\cA_{2}}{}^{\cA'_{2}}(X) \, \cH^{(\pm)}_{\cA_{1} \cA'_{2} \cA_{3}}(X) \, . 
\end{equation}
We see here that $\cI$ acts as an intertwining operator between the different representations of the correlation function. Once $\tilde{\cH}$ is obtained we can then impose conservation on $\Pi$ as if it were located at the ``first point'', using identities analogous to \eqref{Three-point building blocks - differential identities}. It is also important to note that the even and odd sectors of the correlation function are linearly independent, and therefore may be considered separately in the constraint analysis. Another result that follows from the properties \eqref{Inversion tensor identities - three point functions a}, \eqref{Inversion tensor identities - three point functions b} and \eqref{Inversion tensor identities - higher spin case} of the inversion tensor is
\begin{align} \label{H inversion}
	\cH^{(\pm)}_{\cA_{1} \cA_{2} \cA_{3}}(X) = \pm \, \cI^{(1)}{}_{\cA_{1}}{}^{\cA'_{1}}(X) \, \cI^{(2)}{}_{\cA_{2}}{}^{\cA'_{2}}(X) \, \cI^{(3)}{}_{\cA_{3}}{}^{\cA'_{3}}(X) \, \cH^{(\pm)}_{\cA'_{1} \cA'_{2} \cA'_{3}}(X) \, .
\end{align}
That is, ``even" structures are invariant under the action of $\cI$, while ``odd" structures are pseudo-invariant under the action of $\cI$.

If we now consider the correlation function of three conserved primaries $J^{}_{\a(I)}$, $J'_{\b(J)}$, $J''_{\g(K)}$, where $I=2s_{1}$, $J=2s_{2}$, $K=2s_{3}$, then the general ansatz is
\begin{align} \label{Conserved correlator ansatz}
	\langle J^{}_{\a(I)}(x_{1}) \, J'_{\b(J)}(x_{2}) \, J''_{\g(K)}(x_{3}) \rangle = \frac{ \cI_{\a(I)}{}^{\a'(I)}(x_{13}) \,  \cI_{\b(J)}{}^{\b'(J)}(x_{23}) }{(x_{13}^{2})^{\D_{1}} (x_{23}^{2})^{\D_{2}}}
	\; \cH_{\a'(I) \b'(J) \g(K)}(X_{12}) \, ,
\end{align} 
where $\D_{i} = s_{i} + 1$. The constraints on $\cH$ are then as follows:
\begin{enumerate}
	\item[\textbf{(i)}] {\bf Homogeneity:}
	\begin{equation}
		\cH_{\a(I) \b(J) \g(K)}(\l^{2} X) = (\l^{2})^{\D_{3} - \D_{2} - \D_{1}} \, \cH_{\a(I) \b(J) \g(K)}(X) \, , \hspace{5mm} \forall \l \in \mathbb{R} \, \backslash \, \{ 0 \} \, .
	\end{equation}

	\item[\textbf{(ii)}] {\bf Differential constraints:} \\
	After application of the identities \eqref{Three-point building blocks - differential identities 2}, \eqref{Three-point building blocks - differential identities 3} we obtain the following constraints:
	\begin{subequations}
		\begin{align}
			\text{Conservation at $x_{1}$:} && \pa^{\a_{1} \a_{2}}_{X} \cH_{\a_{1} \a_{2} \a(I - 2) \b(J) \g(K)}(X) &= 0 \, , \\
			\text{Conservation at $x_{2}$:} && \pa^{\b_{1} \b_{2}}_{X} \cH_{\a(I) \b_{1} \b_{2} \b(J-2) \g(K)}(X) &= 0 \, , \\
			\text{Conservation at $x_{3}$:} && \pa^{\g_{1} \g_{2}}_{X} \tilde{\cH}_{\a(I) \b(J) \g_{1} \g_{2} \g(K-2)  }(X) &= 0 \, ,
		\end{align}
	\end{subequations}
	where 
	\begin{equation}
		\tilde{\cH}^{(\pm)}_{\a(I) \b(J) \g(K) }(X) = \pm \, (X^{2})^{\D_{1} - \D_{3}} \, \cI_{\b(J)}{}^{\b'(J)}(X) \, \cH^{(\pm)}_{\a(I) \b'(J) \g(K)}(X) \, . 
	\end{equation}

	\item[\textbf{(iii)}] {\bf Point-switch symmetries:} \\
	If the fields $J$ and $J'$ coincide, then we obtain the following point-switch identity
	\begin{equation}
		\cH_{\a(I) \b(I) \g(K)}(X) = (-1)^{\e(J)} \cH_{\b(I) \a(I) \g(K)}(-X) \, ,
	\end{equation}
	where $\e(J)$ is the Grassmann parity of $J$. Likewise, if the fields $J$ and $J''$ coincide, then we obtain the constraint
	\begin{equation}
		\tilde{\cH}_{\a(I) \b(J) \g(I) }(X) = (-1)^{\e(J)} \cH_{\g(I) \b(J) \a(I)}(-X) \, .
	\end{equation}
\end{enumerate}
In practice, imposing the constraints above on correlation functions involving higher-spin currents quickly becomes unwieldy using the tensor formalism, particularly due to the sheer number of possible tensor structures for a given set of spins. Hence, in the next subsections we will develop an index-free formalism to handle the computations efficiently.

\subsubsection{Auxiliary spinor formalism}\label{subsubsection3.2.2}

To study and impose constraints on correlation functions of primary fields with general spins it is often advantageous to use the formalism of auxiliary spinors to streamline the calculations. Suppose we must analyse the constraints on a general spin-tensor $\cH_{\cA_{1} \cA_{2} \cA_{3}}(X)$, where $\cA_{1} = \{ \a_{1}, ... , \a_{I} \}, \cA_{2} = \{ \b_{1}, ... , \b_{J} \}, \cA_{3} = \{ \g_{1}, ... , \g_{K} \}$ represent sets of totally symmetric spinor indices associated with the fields at points $x_{1}$, $x_{2}$ and $x_{3}$ respectively. We introduce sets of commuting auxiliary spinors for each point; $u$ at $x_{1}$, $v$ at $x_{2}$, and $w$ at $x_{3}$, where the spinors satisfy 
\begin{align}
u^2 &= \varepsilon_{\a \b} \, u^{\a} u^{\b}=0\,,  &
v^{2} &= \varepsilon_{\a \b} \, v^{\a} v^{\b}=0\,, & w^{2} &= \varepsilon_{\a \b} \, w^{\a} w^{\b}=0\,. 
\label{extra1}
\end{align}
Now if we define the objects
\begin{subequations}
	\begin{align}
		\mathbf{U}^{\cA_{1}} &\equiv \mathbf{U}^{\a(I)} = u^{\a_{1}} \dots u^{\a_{I}} \, , \\
		\mathbf{V}^{\cA_{2}} &\equiv \mathbf{V}^{\b(J)} = v^{\b_{1}} \dots v^{\b_{J}} \, , \\
		\mathbf{W}^{\cA_{3}} &\equiv \mathbf{W}^{\g(K)} = w^{\g_{1}} \dots w^{\g_{K}} \, ,
	\end{align}
\end{subequations}
then the generating polynomial for $\cH$ is constructed as follows:
\begin{equation} \label{H - generating polynomial}
	\cH(X; u,v,w) = \cH_{ \cA_{1} \cA_{2} \cA_{3} }(X) \, \mathbf{U}^{\cA_{1}} \mathbf{V}^{\cA_{2}} \mathbf{W}^{\cA_{3}} \, . \\
\end{equation}
%
There is in fact a one-to-one mapping between the space of symmetric traceless spin tensors and the polynomials constructed using the above method. Indeed, the tensor $\cH$ is extracted from the polynomial by acting on it with the following partial derivative operators:
\begin{subequations}
	\begin{align}
		\frac{\pa}{\pa \mathbf{U}^{\cA_{1}} } &\equiv \frac{\pa}{\pa \mathbf{U}^{\a(I)}} = \frac{1}{I!} \frac{\pa}{\pa u^{\a_{1}} } \dots \frac{\pa}{\pa u^{\a_{I}}}  \, , \\
		\frac{\pa}{\pa \mathbf{V}^{\cA_{2}} } &\equiv \frac{\pa}{\pa \mathbf{V}^{\b(J)}} = \frac{1}{J!} \frac{\pa}{\pa v^{\b_{1}} } \dots \frac{\pa}{\pa v^{\b_{J}}} \, , \\
		\frac{\pa}{\pa \mathbf{W}^{\cA_{3}} } &\equiv \frac{\pa}{\pa \mathbf{W}^{\g(K)}} = \frac{1}{K!} \frac{\pa}{\pa w^{\g_{1}} } \dots \frac{\pa}{\pa w^{\g_{K}}} \, . 
	\end{align}
\end{subequations}
The tensor $\cH$ is then extracted from the polynomial as follows:
\begin{equation}
	\cH_{\cA_{1} \cA_{2} \cA_{3}}(X) = \frac{\pa}{ \pa \mathbf{U}^{\cA_{1}} } \frac{\pa}{ \pa \mathbf{V}^{\cA_{2}}} \frac{\pa}{ \pa \mathbf{W}^{\cA_{3}} } \, \cH(X; u, v, w) \, .
\end{equation}
Auxiliary vectors/spinors are widely used 
in the construction of correlation functions throughout the literature (see e.g.~\cite{Giombi:2011rz, Costa:2011mg, Stanev:2012nq, Zhiboedov:2012bm, Nizami:2013tpa, Elkhidir:2014woa}), however, usually the entire correlator is contracted with auxiliary variables and as a result one produces a polynomial 
depending on all three spacetime points and the auxiliary spinors. In contrast, our approach contracts the auxiliary spinors with the tensor $\cH_{ \cA_{1} \cA_{2} \cA_{3} }(X)$, which depends on only a single variable. This is advantageous as it becomes quite straightforward to impose constraints on the correlation function (particularly conservation), since $\cH$ does not depend on any of the spacetime points explicitly. After converting the constraints summarised in the previous subsection into the auxiliary spinor formalism, we obtain:
\begin{enumerate}
	\item[\textbf{(i)}] {\bf Homogeneity:}
	\begin{equation}
		\cH(\l^{2} X ; u(I), v(J), w(K)) = (\l^{2})^{\D_{3} - \D_{2} - \D_{1}} \, \cH(X; u(I), v(J), w(K)) \, ,
	\end{equation}
	where we have used the notation $u(I)$, $v(J)$, $w(K)$ to keep track of the homogeneity of the auxiliary spinors $u$, $v$ and $w$.
	\item[\textbf{(ii)}] {\bf Differential constraints:}
	\begin{subequations} \label{Conservation equations}
		\begin{align}
			\text{Conservation at $x_{1}$:} && \frac{\pa}{\pa X_{\a \b}} \frac{\pa}{\pa u^{\a}} \frac{\pa}{\pa u^{\b}} \, \cH(X; u(I), v(J), w(K)) &= 0 \, , \\
			\text{Conservation at $x_{2}$:} && \frac{\pa}{\pa X_{\a \b}} \frac{\pa}{\pa v^{\a}} \frac{\pa}{\pa v^{\b}} \, \cH(X; u(I), v(J), w(K)) &= 0 \, , \\
			\text{Conservation at $x_{3}$:} && \frac{\pa}{\pa X_{\a \b}} \frac{\pa}{\pa w^{\a}} \frac{\pa}{\pa w^{\b}} \, \tilde{\cH}(X; u(I), v(J), w(K)) &= 0 \, .
		\end{align}
	\end{subequations}
	In the auxiliary spinor formalism, $\tilde{\cH} = \tilde{\cH}^{(+)} + \tilde{\cH}^{(-)}$ is computed as follows:
	\begin{equation}
		\tilde{\cH}^{(\pm)}(X; u(I), v(J), w(K) ) = \pm \, \frac{1}{J!} (X^{2})^{\D_{1} - \D_{3}} ( v \hat{X} \pa_{t})^{J} \cH^{(\pm)}(X; u(I), t(J), w(K)) \, , 
	\end{equation}
	where $(v \hat{X} \pa_{t}) = v^{\a} \hat{X}_{\a}{}^{\b} \frac{\pa}{\pa t^{\b}}$.
	\item[\textbf{(iii)}] {\bf Point switch symmetries:} \\
	If the fields $\F$ and $\J$ coincide (hence $I = J$), then we obtain the following point-switch constraint
	\begin{equation} \label{Point switch A}
		\cH(X; u(I), v(I), w(K)) = (-1)^{\e(\F)} \cH(-X; v(I), u(I), w(K)) \, ,
	\end{equation}
	where, again, $\e(\F)$ is the Grassmann parity of $\F$. Similarly, if the fields $\F$ and $\P$ coincide (hence $I = K$) then we obtain the constraint
	\begin{equation} \label{Point switch B}
		\tilde{\cH}(X; u(I), v(J), w(I)) = (-1)^{\e(\F)} \cH(- X; w(I), v(J), u(I)) \, .
	\end{equation}
\end{enumerate}

\subsubsection{Generating function method}\label{subsubsection3.2.3}

The approach outlined above proves to be quite tractable, computationally speaking, as the polynomial, \eqref{H - generating polynomial}, is now constructed out of scalar combinations of $X$, and the auxiliary spinors $u$, $v$ and $w$ with the appropriate homogeneity. At this point it is convenient to introduce the following ``primitive" structures:
\begin{subequations} \label{Basis scalar structures}
	\begin{align}
		P_{1} &= \ve_{\a \b} v^{\a} w^{\b} \, , & P_{2} &= \ve_{\a \b} w^{\a} u^{\b} \, , & P_{3} &= \ve_{\a \b} u^{\a} v^{\b} \, , \\
		Q_{1} &= \hat{X}_{\a \b} v^{\a} w^{\b} \, , & Q_{2} &= \hat{X}_{\a \b} w^{\a} u^{\b} \, , & Q_{3} &= \hat{X}_{\a \b} u^{\a} v^{\b} \, , \\
		Z_{1} &= \hat{X}_{\a \b} u^{\a} u^{\b} \, , & Z_{2} &= \hat{X}_{\a \b} v^{\a} v^{\b} \, , & Z_{3} &= \hat{X}_{\a \b} w^{\a} w^{\b} \, .
	\end{align}
\end{subequations}
The most general ansatz for the polynomial $\cH$ is comprised of all possible combinations of the above structures which possess the correct homogeneity in $u$, $v$ and $w$. In general, it is a non-trivial technical problem to come up with an exhaustive list of possible solutions for the polynomial $\cH$ for a given set of spins. However, this problem can be simplified by introducing a generating function for the polynomial $\cH(X; u, v, w)$:
\begin{align} \label{Generating function}
	\cF(X; \G) &= X^{\d} P_{1}^{k_{1}} P_{2}^{k_{2}} P_{3}^{k_{3}} Q_{1}^{l_{1}} Q_{2}^{l_{2}} Q_{3}^{l_{3}} Z_{1}^{m_{1}} Z_{2}^{m_{2}} Z_{3}^{m_{3}} \, ,
\end{align}
where $\d = \D_{3} - \D_{2} - \D_{1}$, and the non-negative integers, $ \G = \{ k_{i}, l_{i}, m_{i}\}$, $i=1,2,3$, are solutions to the following linear system:
\begin{subequations} \label{Diophantine equations}
	\begin{align}
		k_{2} + k_{3} + l_{2} + l_{3} + 2m_{1} &= I \, , \\
		k_{1} + k_{3} + l_{1} + l_{3} + 2m_{2} &= J \, , \\
		k_{1} + k_{2} + l_{1} + l_{2} + 2m_{3} &= K \, ,
	\end{align}
\end{subequations}
and $I = 2s_{1}$, $J = 2s_{2}$, $K = 2s_{3}$ specify the spin-structure of the correlation function. These equations are obtained by comparing the homogeneity of the auxiliary spinors $u$, $v$, $w$ in the generating function \eqref{Generating function}, against the index structure of the tensor $\cH$. The solutions correspond to a linearly dependent basis of possible structures in which the polynomial $\cH$ can be decomposed. Using \textit{Mathematica}, it is straightforward to generate all possible solutions to \eqref{Diophantine equations} for fixed (and in some cases arbitrary) values of the spins. 

Now let us assume there exists a finite number of solutions $\G_{i}$, $i = 1, ..., N$ to \eqref{Diophantine equations} for a given choice of $I,J,K$. The set of solutions $\G = \{ \G_{i} \}$ may be partitioned into ``even" and ``odd" sets $\G^{+}$ and $\G^{-}$ respectively by counting the number of spinor metrics, $\ve_{\a \b}$, present in a particular solution. Since only the $P_{i}$ contain $\ve_{\a \b}$, we define
\begin{align}
	\G^{+} = \G|_{ \, k_{1} + k_{2} + k_{3} \, ( \hspace{-1mm}\bmod 2 ) = 0} \, , && \G^{-} = \G|_{ \, k_{1} + k_{2} + k_{3} \, ( \hspace{-1mm} \bmod 2 ) = 1} \, .
\end{align}
Hence, the even solutions are those such that $k_{1} + k_{2} + k_{3} = \text{even}$ (i.e contains an even number of spinor metrics), while the odd solutions are those such that $k_{1} + k_{2} + k_{3} = \text{odd}$ (contains an odd number of spinor metrics).\footnote{This convention agrees with the known result that ``odd" solutions typically contain the Levi-Civita tensor, while the ``even" solutions do not.} Let $|\G^{+}| = N^{+}$ and $|\G^{-}| = N^{-}$, with $N = N^{+} + N^{-}$, then the most general ansatz for the polynomial $\cH$ in \eqref{H - generating polynomial} is as follows:
\begin{subequations} \label{H decomposition}
	\begin{equation}
		\cH(X; u, v, w) = \cH^{(+)}(X; u, v, w) + \cH^{(-)}(X; u, v, w) \, ,
	\end{equation}
	\vspace{-7mm}
	\begin{align}
		\cH^{(+)}(X; u, v, w) = \sum_{i=1}^{N^{+}} A_{i} \, \cF(X; \G^{+}_{i}) \, , && \cH^{(-)}(X; u, v, w) = \sum_{i=1}^{N^{-}} B_{i} \, \cF(X; \G^{-}_{i}) \, ,
	\end{align}
\end{subequations}
where $A_{i}$ and $B_{i}$ are a set of real constants. Since the even and odd sectors of the correlation function do not mix with eachother, they may be considered independently.

Using the above method it is quite simple to generate all the possible structures for a given set of spins $\{s_{1}, s_{2}, s_{3} \}$, however, at this stage we must recall that the solutions generated using this approach are linearly dependent. To form a linearly independent set of solutions we must systematically take into account the following non-linear relations between the primitive structures: 
\begin{subequations}
	\begin{align} \label{Linear dependence 1}
		P_{1} Z_{1} + P_{2} Q_{3} + P_{3} Q_{2} &= 0 \, , \\
		P_{2} Z_{2} + P_{1} Q_{3} + P_{3} Q_{1} &= 0 \, , \\
		P_{3} Z_{3} + P_{1} Q_{2} + P_{2} Q_{1} &= 0 \, ,
	\end{align}
\end{subequations}
\vspace{-10mm}
\begin{subequations}
	\begin{align} \label{Linear dependence 2}
		Q_{1} Z_{1} - Q_{2} Q_{3} - P_{2} P_{3} &= 0 \, , \\
		Q_{2} Z_{2} - Q_{1} Q_{3} - P_{1} P_{3} &= 0 \, , \\
		Q_{3} Z_{3} - Q_{1} Q_{2} - P_{1} P_{2} &= 0 \, ,
	\end{align}
\end{subequations}
\vspace{-10mm}
\begin{subequations}
	\begin{align} \label{Linear dependence 3}
		Z_{2} Z_{3} + P_{1}^{2} - Q_{1}^{2} &= 0 \, , \\
		Z_{1} Z_{3} + P_{2}^{2} - Q_{2}^{2} &= 0 \, , \\
		Z_{1} Z_{2} + P_{3}^{2} - Q_{3}^{2} &= 0 \, ,
	\end{align}
\end{subequations}
\vspace{-10mm}
\begin{align} \label{Linear dependence 4}
	P_{1} P_{2} P_{3} + P_{1} Q_{2} Q_{3} + P_{2} Q_{1} Q_{3} + P_{3} Q_{1} Q_{2} &= 0 \, .
\end{align}
This appears to be an exhaustive list of relations, and similar results have been obtained in other approaches which make use of auxiliary spinors \cite{Giombi:2011rz}. Applying the relations above to a set of linearly dependent polynomial structures is relatively straightforward to implement using Mathematica's built-in pattern matching capabilities.

Now that we have taken care of linear-dependence, it now remains to impose conservation on all three points in addition to the various point-switch symmetries. Introducing the $P$, $Q$ and $Z$ objects proves to streamline this analysis significantly. First let us consider conservation; we define the following three differential operators:
\begin{align}
	D_{1} = \frac{\pa}{\pa X_{\a \b}} \frac{\pa}{\pa u^{\a}} \frac{\pa}{\pa u^{\b}} \, , && D_{2} = \frac{\pa}{\pa X_{\a \b}} \frac{\pa}{\pa v^{\a}} \frac{\pa}{\pa v^{\b}} \, , && D_{3} = \frac{\pa}{\pa X_{\a \b}} \frac{\pa}{\pa w^{\a}} \frac{\pa}{\pa w^{\b}} \, .
\end{align}
To impose conservation on $x_{1}$, (for either sector) we compute
\begin{align}
	D_{1} \cH(X; u,v,w) &= D_{1} \Bigg\{ \sum_{i=1}^{N} c_{i} \, \cF(X; \G_{i}) \Bigg\} \nonumber \\
	&= \sum_{i=1}^{N} c_{i} \, D_{1} \cF(X; \G_{i}) \, .
\end{align}
We then solve for the $c_{i}$ such that the result above vanishes. It's apparent that it would be extremely useful to obtain an explicit expression for $D_{1} \cF(X; \G)$, as this would allow us to impose conservation in a simple manner; this proves to be very cumbersome to carry out by hand, however it is possible to obtain an exact result computationally (which we will not present here as it is $\sim 200$ terms long). Hence, given a particular solution $\cF(X;\G_{i})$, we can compute $D_{1} \cF(X; \G_{i})$. The fact that $D_{1} \cF(X; \G)$ can also be expressed using the primitive structures \eqref{Basis scalar structures} is due to the following reasoning: let $\mathbf{P}[ X(\d); u(I), v(J), w(K) ]$ represent the space of polynomials which are homogeneous degree $\d$ in $X$, $I$ in $u$, $J$ in $v$ and $K$ in $w$; any polynomial in this space can naturally be constructed in terms of the primitives \eqref{Basis scalar structures}. The operator $D_{1}$ may then be interpreted as follows:
\begin{align}
	D_{1} : \mathbf{P}[ X(\d); u(I), v(J), w(K) ] \longmapsto \mathbf{P}[ X(\d-1); u(I-2), v(J), w(K) ] \, .
\end{align}
Hence, $D_{1}$ is a map from $\mathbf{P}[ X(\d); u(I), v(J), w(K) ]$ to $\mathbf{P}[ X(\d-1); u(I-2), v(J), w(K) ]$, that is, the space of polynomials homogeneous degree $\d-1$ in $X$, $I-2$ in $u$, $J$ in $v$ and $K$ in $w$. Any polynomial in this space can naturally be constructed using the same primitives defined in \eqref{Basis scalar structures}. Analogous results also apply for $D_{2} \cF(X; \G)$.

However, to impose conservation on $x_{3}$ we must first obtain an explicit expression for $\tilde{\cH}$ in terms of $\cH$, that is, we must compute (e.g. for the even sector)
\begin{equation}
	\tilde{\cH}(X; u(I), v(J), w(K) ) = \frac{1}{J!} (X^{2})^{\D_{1} - \D_{3}} ( v \hat{X} \pa_{t})^{J} \cH(X; u(I), t(J), w(K)) \, .
\end{equation}
Recalling the fact that any solution for $\cH$ can be written in the form of the generating function $\cF(X; \G)$, we compute
\begin{align}
	\tilde{\cF}(X; \G) &= \frac{1}{J!} (X^{2})^{\D_{1} - \D_{3}} ( v \hat{X} \pa_{t})^{J} \cF(X; \G) \nonumber \\
	&= \frac{1}{J!} (X^{2})^{\D_{1} - \D_{3}} ( v \hat{X} \pa_{t})^{J} \big\{ X^{\d} P_{1}^{k_{1}} P_{2}^{k_{2}} P_{3}^{k_{3}} Q_{1}^{l_{1}} Q_{2}^{l_{2}} Q_{3}^{l_{3}} Z_{1}^{m_{1}} Z_{2}^{m_{2}} Z_{3}^{m_{3}} \big\} \nonumber \\
	&= \frac{1}{J!} X^{\D_{1} - \D_{2} - \D_{3}} P_{2}^{k_{2}} Q_{2}^{l_{2}} Z_{1}^{m_{1}}  Z_{3}^{m_{3}} ( v \hat{X} \pa_{t})^{J} \big\{ P_{1}^{k_{1}} P_{3}^{k_{3}} Q_{1}^{l_{1}} Q_{3}^{l_{3}} Z_{2}^{m_{2}} \big\} \, .
\end{align}
Since $P_{1}$, $P_{3}$, $Q_{1}$, $Q_{3}$ and $Z_{2}$ are the only objects with $t$ dependence, if we make use of the fact that $k_{1} + k_{3} + l_{1} + l_{3} + 2m_{2} = J$, in addition to the identities
\begin{subequations}
	\begin{align}
		( v \hat{X} \pa_{t}) P_{1} &= - Q_{1} \, , & ( v \hat{X} \pa_{t}) P_{3} &= Q_{3} \, , \\
		( v \hat{X} \pa_{t}) Q_{1} &= - P_{1} \, , & ( v \hat{X} \pa_{t}) Q_{3} &= P_{3} \, , \\
		( v \hat{X} \pa_{t})^{2} Z_{2} &= 2 Z_{2} \, , 
	\end{align}
\end{subequations}
then it may be shown that
\begin{align}
	\tilde{\cF}(X; \G) &= X^{\tilde{\d}} (-Q_{1})^{k_{1}} P_{2}^{k_{2}} Q_{3}^{k_{3}} (-P_{1})^{l_{1}} Q_{2}^{l_{2}} P_{3}^{l_{3}} Z_{1}^{m_{1}} Z_{2}^{m_{2}} Z_{3}^{m_{3}} \, , \nonumber \\
	&= (-1)^{ k_{1} + l_{1}} X^{\tilde{\d}} P_{1}^{l_{1}} P_{2}^{k_{2}} P_{3}^{l_{3}} Q_{1}^{k_{1}} Q_{2}^{l_{2}} Q_{3}^{k_{3}} Z_{1}^{m_{1}} Z_{2}^{m_{2}} Z_{3}^{m_{3}} \, ,
\end{align}
where $\tilde{\d} = \D_{1} - \D_{2} - \D_{3}$. Hence we arrive at the following result
\begin{equation}
	\tilde{\cF}(X; \G) = (-1)^{ k_{1} + l_{1}} \cF(X; \G)|_{\d \rightarrow \tilde{\d}, \, k_{1} \leftrightarrow l_{1}, \, k_{3} \leftrightarrow l_{3} } \, .
\end{equation}
Therefore the computation of $\tilde{\cH}$ is actually quite straightforward: we take each term in the ansatz for $\cH$ and make appropriate swaps of the primitive structures. This also simplifies imposing conservation at $x_{3}$, as we can now use the same generating function that we used for conservation at $x_{1}$ and $x_{2}$ as follows:
\begin{equation}
	D_{3} \tilde{\cF}(X; \G) = (-1)^{ k_{1} + l_{1}} D_{3} \cF(X; \G)|_{\d \rightarrow \tilde{\d}, \, k_{1} \leftrightarrow l_{1}, \, k_{3} \leftrightarrow l_{3} } \, .
\end{equation}
Now that we have exact expressions for $D_{i} \cF(X; \G)$, it remains to find out how point-switch symmetries act on the primitive structures. For permutation of spacetime points $x_{1}$ and $x_{2}$, we have $X \rightarrow - X$, $u \leftrightarrow v$. This results in the following replacement rules for the basis objects \eqref{Basis scalar structures}:
\begin{subequations} \label{Point switch A - basis}
	\begin{align} 
		P_{1} &\rightarrow - P_{2} \, , & P_{2} &\rightarrow -P_{1} \, , & P_{3} &\rightarrow -P_{3} \, , \\
		Q_{1} &\rightarrow - Q_{2} \, , & Q_{2} &\rightarrow - Q_{1} \, , & Q_{3} &\rightarrow - Q_{3} \, , \\
		Z_{1} &\rightarrow - Z_{2} \, , & Z_{2} &\rightarrow - Z_{1} \, , & Z_{3} &\rightarrow - Z_{3} \, .
	\end{align}
\end{subequations}
Likewise, for permutation of spacetime points $x_{1}$ and $x_{3}$ we have $X \rightarrow - X$, $u \leftrightarrow w$, resulting in the following replacements:
\begin{subequations} \label{Point switch B - basis}
	\begin{align} 
		P_{1} &\rightarrow - P_{3} \, , & P_{2} &\rightarrow -P_{2} \, , & P_{3} &\rightarrow -P_{1} \, , \\
		Q_{1} &\rightarrow - Q_{3} \, , & Q_{2} &\rightarrow - Q_{2} \, , & Q_{3} &\rightarrow - Q_{1} \, , \\
		Z_{1} &\rightarrow - Z_{3} \, , & Z_{2} &\rightarrow - Z_{2} \, , & Z_{3} &\rightarrow - Z_{1} \, .
	\end{align}
\end{subequations}
We have now developed all the formalism necessary to analyse the structure of three-point correlation functions in 3D CFT. In the remaining sections of this paper we will analyse the three-point functions of conserved higher-spin currents (for both integer and half-integer spin) using the following method:
\begin{enumerate}
	\item We construct all possible (linearly dependent) structures for $\cH(X; u,v,w)$ for a given set of spins, which is governed by the solutions to \eqref{Diophantine equations}. The solutions are sorted into even and odd sectors and analysed separately.
	\item In each sector, we apply an algorithm to the set of dependent structures which systematically reduces it to a linearly independent set through repeated application of the identities \eqref{Linear dependence 1}, \eqref{Linear dependence 2}, \eqref{Linear dependence 3}, \eqref{Linear dependence 4}. This is sufficient to form the most general linearly independent ansatz.
	\item Using the method outlined in subsection \ref{subsubsection3.2.3}, we impose the conservation equations \eqref{Conservation equations} on each sector.
	 
	\item Once the general form of the polynomial $\cH(X; u,v,w)$ (associated with the conserved three-point function $\langle J^{}_{s_{1}} J'_{s_{2}} J''_{s_{3}} \rangle$) is obtained for a given set of spins $\{s_{1},s_{2}, s_{3}\}$, we then impose any symmetries under permutation of spacetime points, that is, \eqref{Point switch A} and \eqref{Point switch B} (if applicable). In certain cases, imposing these constraints can eliminate the remaining structures.
\end{enumerate}
Due to computational limitations such as CPU clock speed and available RAM, we could carry out this explicit analysis up to $s_{i} = 20$, 
however, with more optimisation of the code and sufficient computational resources this approach should hold for arbitrary spins. 
Since there are an enormous number of possible three-point functions with $s_{i} \leq 20$, we present the final results for $\cH(X; u,v,w)$ for some particularly interesting examples, as the solutions and coefficient constraints become cumbersome to present beyond low spin cases. We are primarily interested in counting the number of independent tensor structures after imposing all the constraints. 

The results in the next sections are organised as follows: in section \ref{section4} we analyse the correlation functions involving bosonic conserved currents, commenting on some of the general features. Many of these results are known in the literature \cite{Zhiboedov:2012bm,Giombi:2011rz}, however, they have not been derived explicitly using this construction based on the conformal inversion tensor. In addition, within the framework of the generating function methods used in \cite{Zhiboedov:2012bm,Stanev:2012nq}, it is unclear how the generating functions in these works are derived and how they produce an exhaustive list of independent structures. It is in this regard that our analysis is very explicit, as we find all possible structures for a given set of spins and systematically apply linear dependence relations to them. In section \ref{section5} we analyse the mixed three-point functions involving fermionic conserved currents; these results are new and are naturally of interest within the context of superconformal field theories. Finally, in section \ref{section6}, we analyse correlation functions involving combinations of higher-spin currents and fundamental scalars/spinors. We stress that our analysis is based only on symmetries and conservation equations and does not take into account any other features of local field theory. The results are completely analytic and we present explicit formula for $\cH(X; u,v,w)$ in all cases, the results are copied directly from the Mathematica code.



\section{Correlation functions involving bosonic currents}\label{section4}

Three-point correlation functions of conserved bosonic currents have been extensively studied in 3D CFT. In particular, it has been shown that the general structure of the three-point correlation function $\langle J^{}_{s_{1}} J'_{s_{2}} J''_{s_{3}} \rangle$ is fixed up to the following form \cite{Maldacena:2011jn,Giombi:2011rz,Zhiboedov:2012bm}:
\begin{equation}
	\langle J^{}_{s_{1}} J'_{s_{2}} J''_{s_{3}} \rangle = a_{1} \, \langle J^{}_{s_{1}} J'_{s_{2}} J''_{s_{3}} \rangle_{B} + a_{2} \, \langle J^{}_{s_{1}} J'_{s_{2}} J''_{s_{3}} \rangle_{F} + b \, \langle J^{}_{s_{1}} J'_{s_{2}} J''_{s_{3}} \rangle_{odd} \, .
\end{equation}
The solutions $\langle J^{}_{s_{1}} J'_{s_{2}} J''_{s_{3}} \rangle_{B}$, $\langle J^{}_{s_{1}} J'_{s_{2}} J''_{s_{3}} \rangle_{F}$ are generated by theories of a free-boson and free-fermion respectively, while the ``odd" structure, $\langle J^{}_{s_{1}} J'_{s_{2}} J''_{s_{3}} \rangle_{odd}$, is not generated by a free CFT; instead it is generated by a Chern-Simons theory interacting with parity-violating matter \cite{Aharony:2011jz, Giombi:2011kc, Maldacena:2012sf, GurAri:2012is, Aharony:2012nh, Jain:2012qi, Giombi:2016zwa, Chowdhury:2017vel, Sezgin:2017jgm, Skvortsov:2018uru, Inbasekar:2019wdw}. Furthermore, the existence of the odd solution depends on the following set of triangle inequalities:
\begin{align} \label{Triangle inequalities}
	s_{1} &\leq s_{2} + s_{3} \, , & s_{2} &\leq s_{1} + s_{3} \, , & s_{3} &\leq s_{1} + s_{2} \, .
\end{align}
When the triangle inequalities are simultaneously satisfied, there are two even solutions and one odd solution, however, if any of the inequalities above are not satisfied then the odd solution is incompatible with current conservation.\footnote{Existence and uniqueness of the parity-odd solution (inside and outside triangle inequalities) has been proven in the ``light-like" limit in \cite{Giombi:2016zwa}. Similar arguments can be made to show there are only two forms for the parity-even solutions, these are sketched \cite{Maldacena:2011jn}.} Further, if any of the $J$, $J'$, $J''$ coincide (i.e. in cases where the currents are unique and have the same spin), then the resulting point-switch symmetries can kill off the remaining structures. 
Our comments on the general results for three-point functions of bosonic currents are summarised below:
\begin{itemize}
	\item When the triangle inequalities are simultaneously satisfied, each polynomial structure in the solution for all three-point functions can be written as a product of at most 5 of the $P_{i}$, $Q_{i}$, with the $Z_{i}$ completely eliminated. 
	\item For the three-point functions $\langle J^{}_{s_{1}} J'_{s_{1}} J''_{s_{2}} \rangle$, for arbitrary integer $s_{1}$ and $s_{2}$: when the triangle inequalities are satisfied there are two even solutions and one odd solution, otherwise there are only two even solutions. After imposing $J=J'$ the solutions exist only when $s_{2}$ is an even integer. Note that for $s_{1} > s_{2}$ the triangle inequalities are always satisfied.
	\item For the three-point functions $\langle J_{s} \, J_{s} \, J_{s} \rangle$, with $s$ an integer, there are two even solutions and one odd solution, however they exist only for $s$ even. For $s$ odd the solutions survive only if the currents carry a flavour index associated with a non-Abelian symmetry group.
\end{itemize}
Another observation is that the triangle inequalities can be encoded in a discriminant, $\s$, which we define as follows:
\begin{align} \label{Discriminant}
	\s(s_{1}, s_{2}, s_{3}) = q_{1} q_{2} q_{3} \, , \hspace{10mm} q_{i} = s_{i} - s_{j} - s_{k} - 1 \, ,
\end{align}
where $(i,j,k)$ is a cyclic permutation of $(1,2,3)$. For $\s(s_{1}, s_{2}, s_{3}) < 0$, there are two even solutions and one odd solution, while for $\s(s_{1}, s_{2}, s_{3}) \geq 0$ there are only two even solutions. 
The origin of this discriminant equation is actually quite simple within the framework of this formalism: recall that the correlation function can be encoded in a tensor $\cH$, which is a function of a single three-point covariant, $X$. There are three different (but equivalent) representations of a given correlation function, 
call them $\cH^{(i)}$, where the superscript $i$ denotes which point we set to act as the ``third point" in the ansatz \eqref{H ansatz}. As shown in subsection \ref{subsubsection3.2.1}, 
the representations are related by the intertwining operator $\cI$, with each $\cH^{(i)}$ being homogeneous degree $q_{i}$. After exhaustive analysis of the three-point functions with $s_{i} \leq 20$, 
a clear pattern emerges: the odd structure survives if and only if $\forall i$, $q_{i} < 0$. In other words, each $\cH^{(i)}$ must be a rational function of $X$ with homogeneity $q_{i} < 0$. The discriminant \eqref{Discriminant} simply encodes information about whether the $\cH^{(i)}$ are simultaneously of negative homogeneity. 
 
In the next subsections we analyse the structure of three-point functions involving conserved bosonic currents. As a test our approach we begin with an analysis of correlation functions involving low-spin currents such as the energy-momentum tensor and vector current. 

\subsection{Energy-momentum tensor and vector current correlators}\label{subsection4.1}

The conserved currents which are fundamental in any conformal field theory are the conserved vector current, $V_{m}$, and the symmetric, traceless energy-momentum tensor, $T_{mn}$. The vector current has scale dimension $\Delta_{V} = 2$ and satisfies $\pa^{m} V_{m} = 0$, while the energy-momentum tensor has scale dimension $\Delta_{T} = 3$ and satisfies the conservation equation $\pa^{m} T_{mn} = 0$. Converting to spinor notation we have:
\begin{align}
	V_{\a_{1} \a_{2}}(x) = (\g^{m})_{\a_{1} \a_{2}} V_{m}(x) \, , && T_{\a_{1} \a_{2} \a_{3} \a_{4}}(x) = (\g^{m})_{(\a_{1} \a_{2}} (\g^{n})_{\a_{3} \a_{4})} T_{mn}(x) \, .
\end{align}
These objects possess fundamental information associated with internal and spacetime symmetries, hence, analysis of their three-point functions is of great importance. The general structure of correlation functions involving these fields have been widely studied throughout the literature of conformal field theory; here we present the solutions for them using our formalism. The possible three-point functions involving the conserved vector current and the energy-momentum tensor are:
\begin{align} \label{Low-spin component correlators}
	\langle V_{\a(2)}(x_{1}) \, V_{\b(2)}(x_{2}) \, V_{\g(2)}(x_{3}) \rangle \, , &&  \langle V_{\a(2)}(x_{1}) \, V_{\b(2)}(x_{2}) \, T_{\g(4)}(x_{3}) \rangle \, , \\
	\langle T_{\a(4)}(x_{1}) \, T_{\b(4)}(x_{2}) \, V_{\g(2)}(x_{3}) \rangle \, , &&  \langle T_{\a(4)}(x_{1}) \, T_{\b(4)}(x_{2}) \, T_{\g(4)}(x_{3}) \rangle \, .
\end{align}
In all cases, we note that the triangle inequalities \eqref{Triangle inequalities} are simultaneously satisfied, hence, we expect that each of these correlation functions should possess a parity-odd solution after imposing conservation on all three points. The analysis of these three-point functions is quite simple using our computational approach. Let us first consider $\langle V V V \rangle$; within the framework of our formalism we study the three-point function $\langle J^{}_{1} J'_{1} J''_{1} \rangle$.\\[5mm]
\textbf{Correlation function} $\langle J^{}_{1} J'_{1} J''_{1} \rangle$\textbf{:}\\[2mm]
The general ansatz for this correlation function, according to \eqref{Conserved correlator ansatz} is
\begin{align}
	\langle J^{}_{\a(2)}(x_{1}) \, J'_{\b(2)}(x_{2}) \, J''_{\g(2)}(x_{3}) \rangle = \frac{ \cI_{\a(2)}{}^{\a'(2)}(x_{13}) \,  \cI_{\b(2)}{}^{\b'(2)}(x_{23}) }{(x_{13}^{2})^{2} (x_{23}^{2})^{2}}
	\; \cH_{\a'(2) \b'(2) \g(2)}(X_{12}) \, .
\end{align} 
Using the formalism outlined in subsection \ref{subsection3.2}, all information about this correlation function is encoded in the following polynomial:
\begin{align}
	\cH(X; u(2), v(2), w(2)) = \cH_{ \a(2) \b(2) \g(2) }(X) \, \mathbf{U}^{\a(2)}  \mathbf{V}^{\b(2)}  \mathbf{W}^{\g(2)} \, .
\end{align}
Using Mathematica we solve \eqref{Diophantine equations} for the chosen spins and substitute each solution into the generating function \eqref{Generating function}. This provides us with the following list of (linearly dependent) polynomial structures in the even and odd sectors respectively:
\begin{align}
	\includegraphics[width=0.9\textwidth, valign=c]{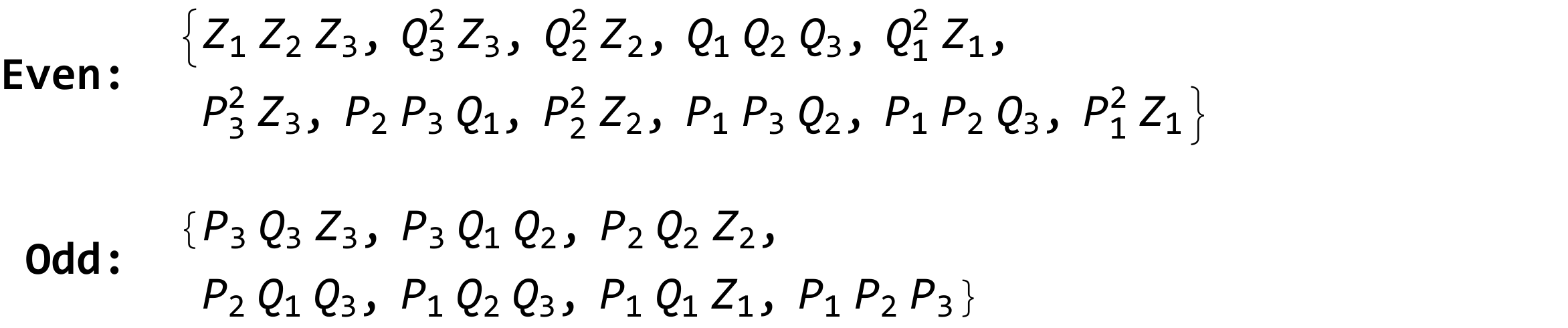}
\end{align}
Next, we systematically apply the linear dependence relations \eqref{Linear dependence 1} to these lists, reducing them to the following sets of linearly independent structures:
\begin{align}
	\includegraphics[width=0.88\textwidth, valign=c]{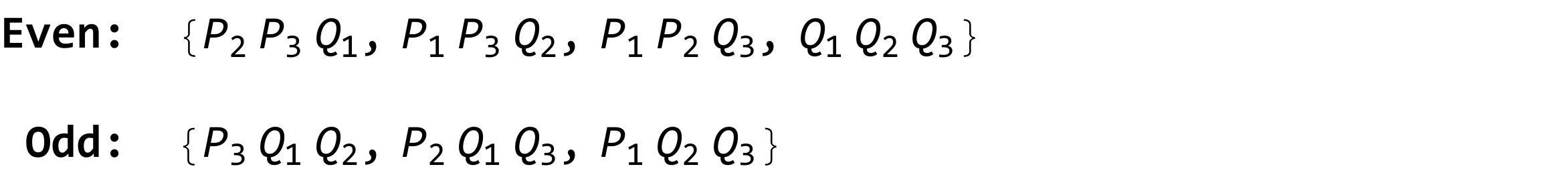}
\end{align}
Note that application of the linear-dependence relations eliminates all terms involving $Z_{i}$ in this case. Next we construct an ansatz out of the linearly independent structures, see \eqref{H decomposition}, where $\cH^{(\pm)}_{i}$ denotes a structure at position `$i$' in the even/odd list respectively. After imposing conservation on all three points using the methods outlined in \ref{subsubsection3.2.3}, we obtain the following relations between the coefficients:
\begin{align}
	\includegraphics[width=0.88\textwidth, valign=c]{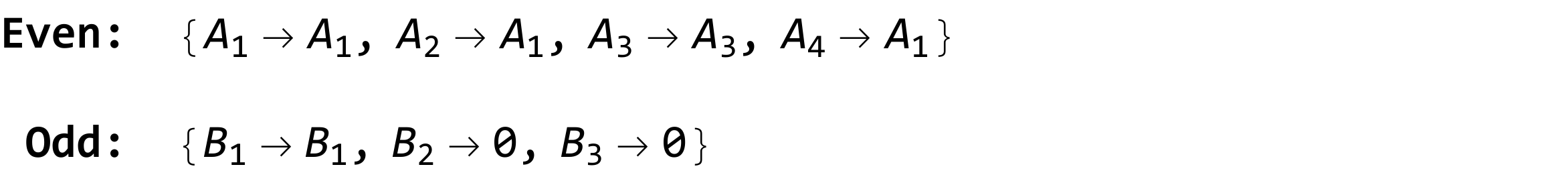}
\end{align}
Hence, the final solutions for the even and odd sectors are
\begin{align}\label{1-1-1}
	\includegraphics[width=0.88\textwidth, valign=c]{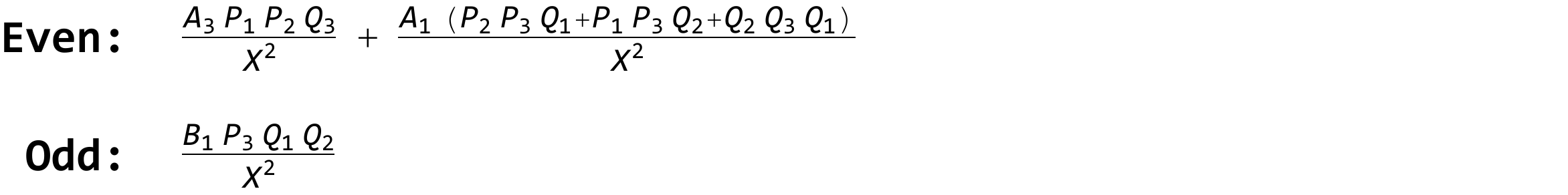}
\end{align}
After imposing symmetries under permutation of spacetime points, e.g. $J=J'=J''$, the remaining structures vanish unless the currents possess a flavour index associated with a non-Abelian symmetry group, in which case all three structures survive. The next example to consider is the mixed correlator $\langle V V T \rangle$. To study this case we may examine the correlation function $\langle J^{}_{1} J'_{1} J''_{2} \rangle$. \\[5mm]
\noindent
\textbf{Correlation function} $\langle J^{}_{1} J'_{1} J''_{2} \rangle$\textbf{:}\\[2mm]
Using the general formula, the ansatz for this three-point function:
\begin{align}
	\langle J^{}_{\a(2)}(x_{1}) \, J'_{\b(2)}(x_{2}) \, J''_{\g(4)}(x_{3}) \rangle = \frac{ \cI_{\a(2)}{}^{\a'(2)}(x_{13}) \,  \cI_{\b(2)}{}^{\b'(2)}(x_{23}) }{(x_{13}^{2})^{2} (x_{23}^{2})^{2}}
	\; \cH_{\a'(2) \b'(2) \g(4)}(X_{12}) \, .
\end{align} 
Using the formalism outlined in \ref{subsection3.2}, all information about this correlation function is encoded in the following polynomial:
\begin{align}
	\cH(X; u(2), v(2), w(4)) = \cH_{ \a(2) \b(2) \g(4) }(X) \, \mathbf{U}^{\a(2)}  \mathbf{V}^{\b(2)}  \mathbf{W}^{\g(4)} \, .
\end{align}
After solving \eqref{Diophantine equations}, we find the following linearly dependent polynomial structures in the even and odd sectors respectively:
\begin{align}
	\includegraphics[width=0.9\textwidth, valign=c]{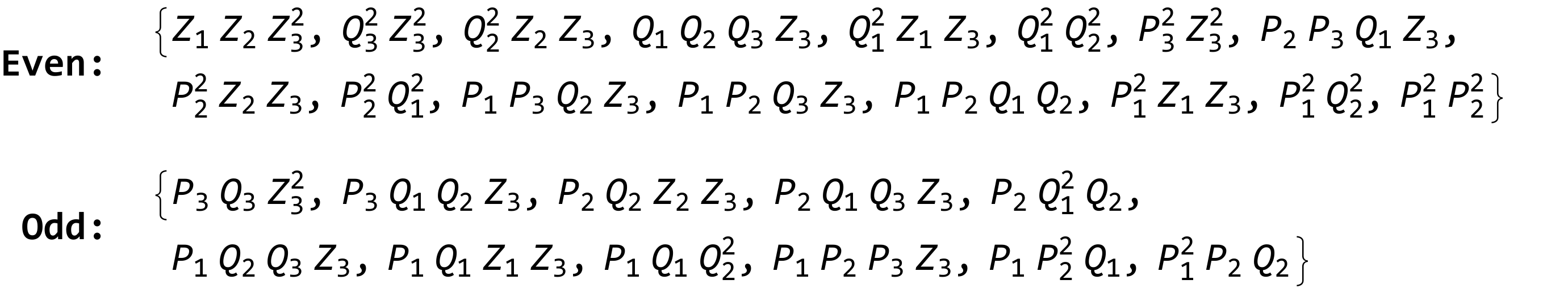}
\end{align}
Next we systematically apply the linear dependence relations \eqref{Linear dependence 1} to these lists, reducing them to the following linearly independent structures:
\begin{align}
	\includegraphics[width=0.9\textwidth, valign=c]{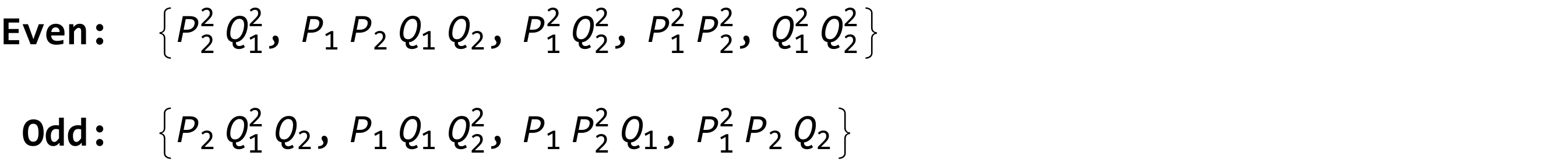}
\end{align}
After constructing an appropriate ansatz for each sector, we then impose conservation on all three points using the methods outlined in \ref{subsubsection3.2.3} and obtain the following relations between the coefficients:
\begin{align}
	\includegraphics[width=0.9\textwidth, valign=c]{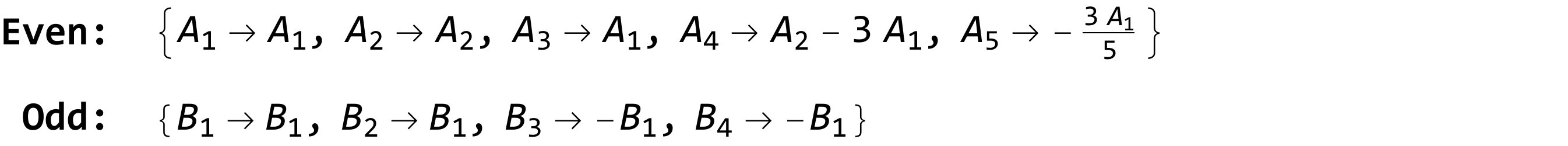}
\end{align}
Hence, the final solutions for the even and odd sectors are
\begin{align}\label{1-1-2}
	\includegraphics[width=0.9\textwidth, valign=c]{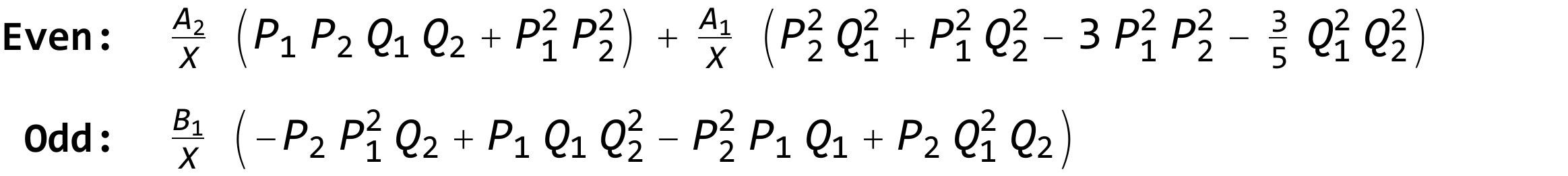}
\end{align}
All structures survive after setting $J = J'$. Hence, this correlation function is fixed up to two independent even structures, and one odd structure.

Since the number of tensor structures rapidly increases with spin, we will skip the technical details for the other correlation functions and present only the final results after imposing conservation. For $\langle T T V \rangle$ we may consider the correlation function $\langle J^{}_{2} J'_{2} J''_{1} \rangle$, for which we find the solution:
\begin{align}\label{2-2-1}
	\includegraphics[width=0.9\textwidth, valign=c]{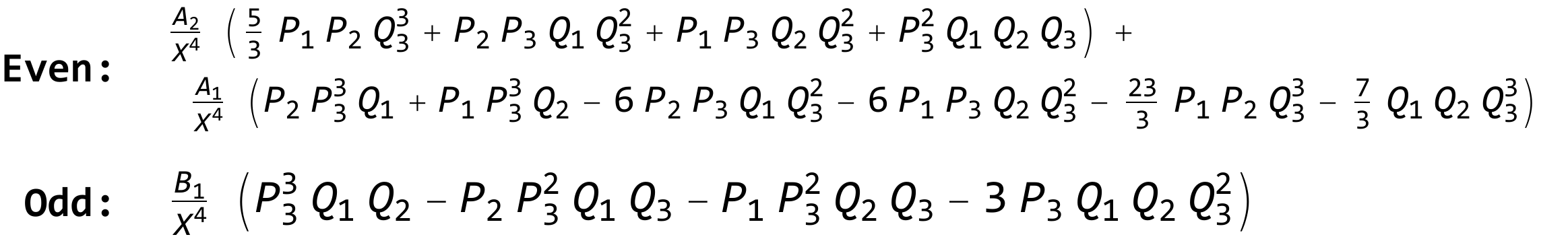}
\end{align} 
In this case, all structures vanish after setting $J = J'$ and imposing the required symmetries under the exchange of $x_{1}$ and $x_{2}$. Hence, the correlation function $\langle T T V \rangle$ vanishes in any CFT. Finally, to study $\langle T T T \rangle$ we can analyse the correlation function $\langle J^{}_{2} J'_{2} J''_{2} \rangle$, which has the solution:
\begin{align}\label{2-2-2}
	\includegraphics[width=0.9\textwidth, valign=c]{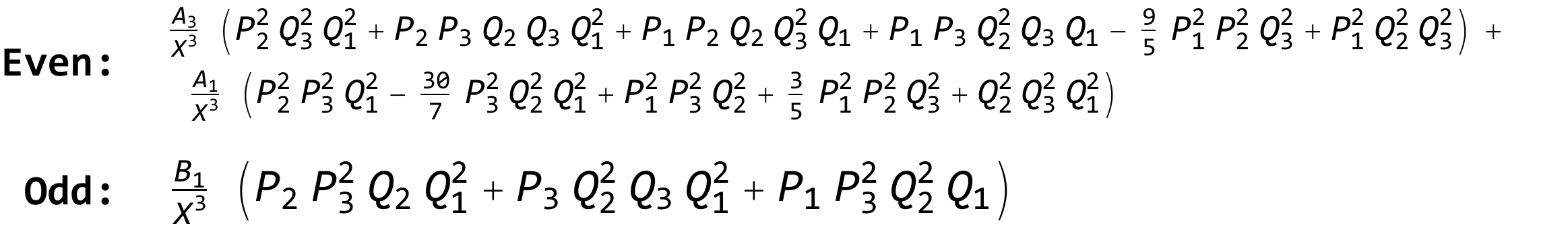}
\end{align}
In all cases, it is clear that the general solutions are determined up to two independent even structures, and one odd structure. These results are consistent with \cite{Giombi:2011rz} in terms of the number of independent polynomial structures, however it is difficult to make a direct comparison.

\subsection{Higher-spin correlators}\label{subsection4.2}
In this subsection we obtain explicit results for three-point functions involving higher-spin currents. We present the final results after imposing conservation on all three-points. \\[3mm]
\noindent
\textbf{Correlation function} $\langle J^{}_{1} J'_{1} J''_{3} \rangle$\textbf{:} \hspace{3mm} $\sigma = 0$
\begin{align}\label{1-1-3}
	\includegraphics[width=0.9\textwidth, valign=c]{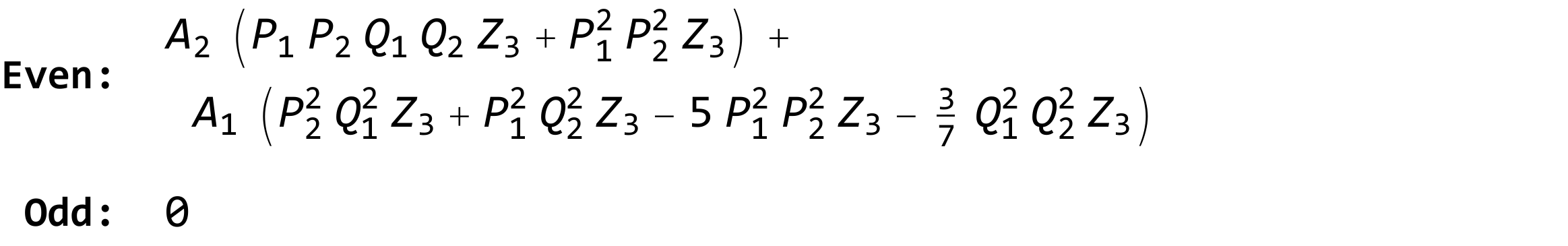}
\end{align}
This is an instance in which one of the triangle inequalities is not satisfied, and we can see here that the odd solution vanishes. The even structures vanish after imposing $J = J'$. \\[4mm]
\textbf{Correlation function} $\langle J^{}_{1} J'_{1} J''_{4} \rangle$\textbf{:} \hspace{3mm} $\sigma > 0$
\begin{align}\label{1-1-4}
	\includegraphics[width=0.9\textwidth, valign=c]{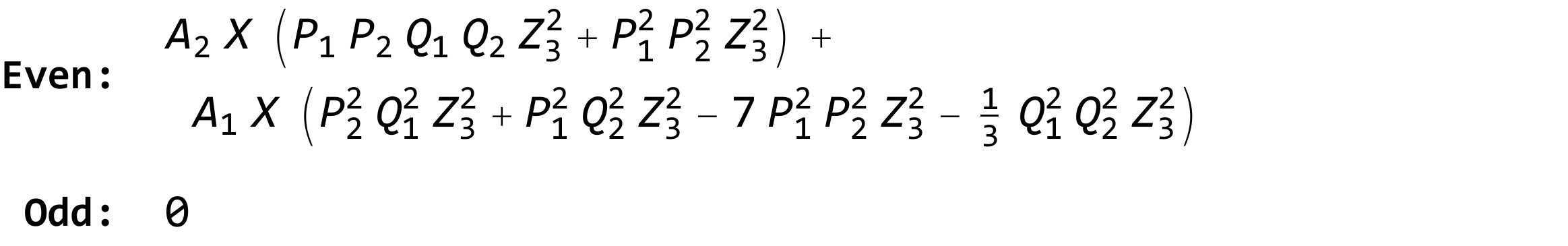}
\end{align} \\
\textbf{Correlation function} $\langle J^{}_{1} J'_{2} J''_{3} \rangle$\textbf{:} \hspace{3mm} $\sigma < 0$
\begin{align}\label{1-2-3}
	\includegraphics[width=0.9\textwidth, valign=c]{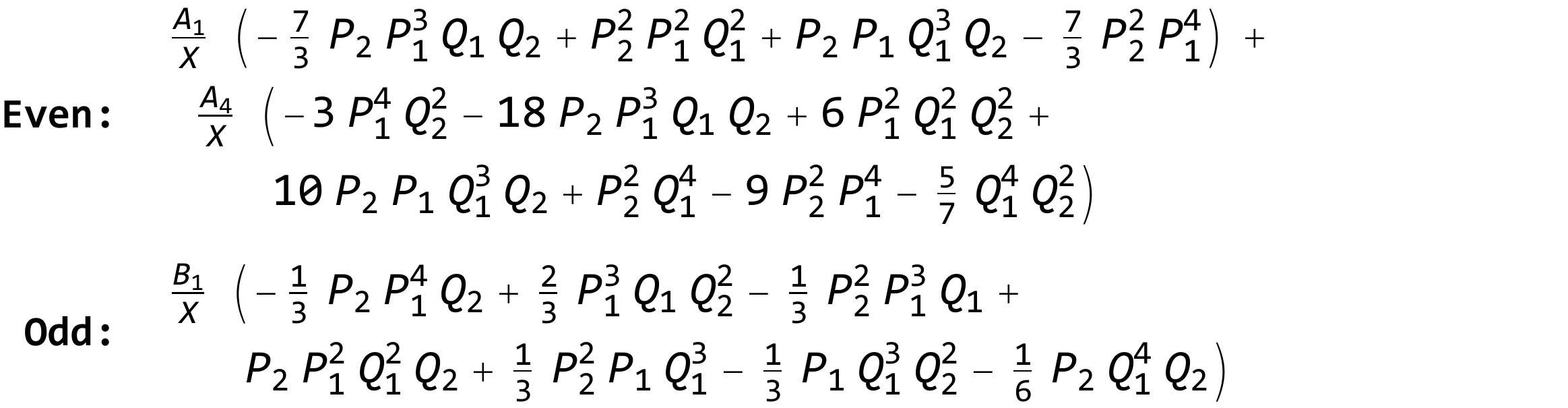}
\end{align} \\
\textbf{Correlation function} $\langle J^{}_{2} J'_{2} J''_{3} \rangle$\textbf{:} \hspace{3mm} $\sigma < 0$
\begin{align}\label{2-2-3}
	\includegraphics[width=0.9\textwidth, valign=c]{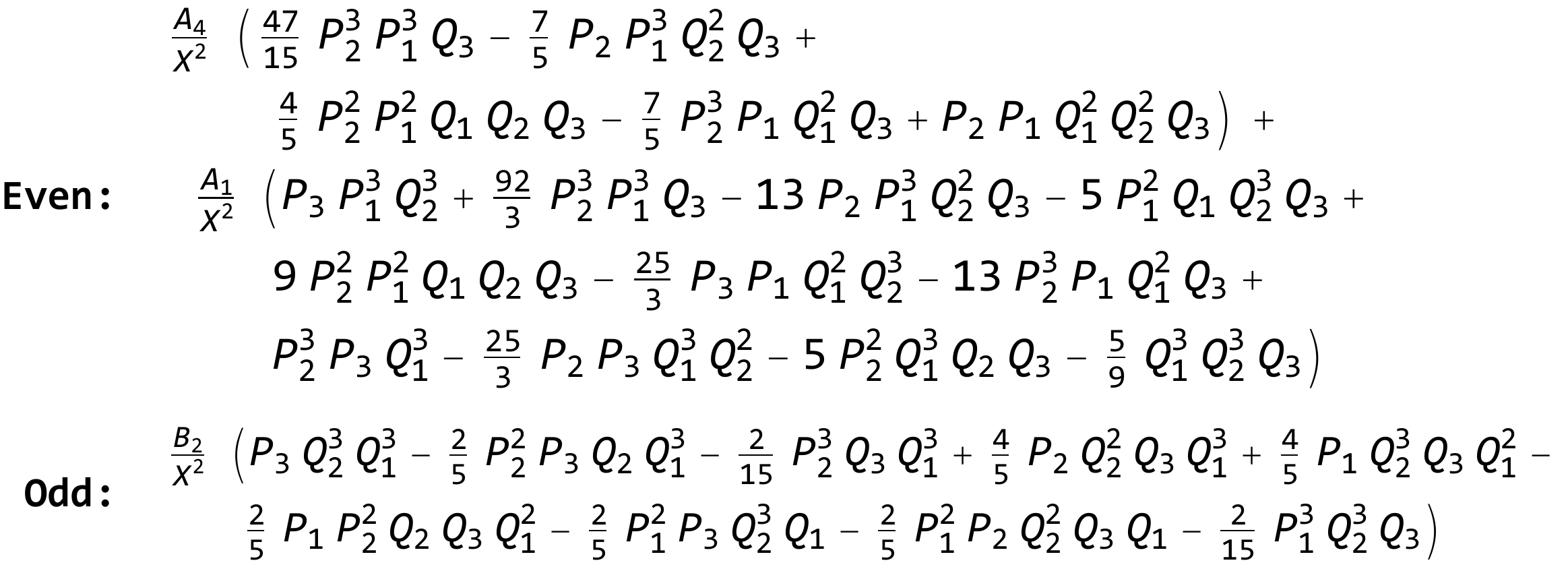}
\end{align} \\
\textbf{Correlation function} $\langle J^{}_{2} J'_{2} J''_{4} \rangle$\textbf{:} \hspace{3mm} $\sigma < 0$
\begin{align}\label{2-2-4}
	\includegraphics[width=0.9\textwidth, valign=c]{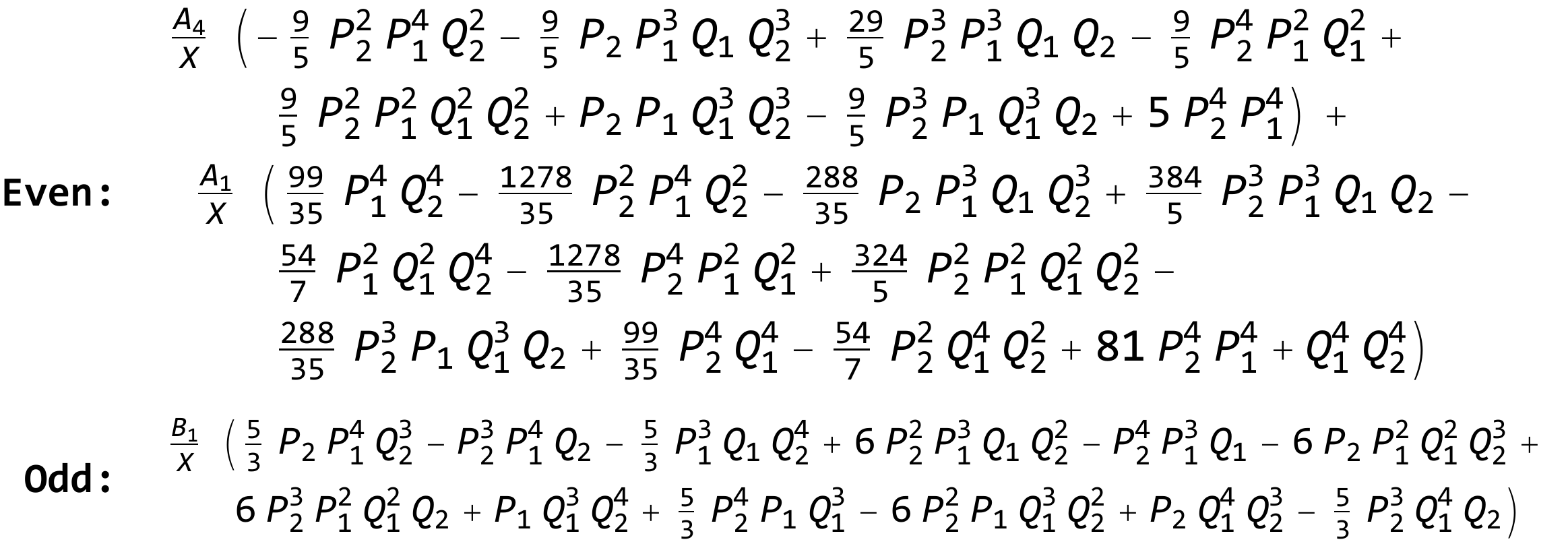}
\end{align} \\
\textbf{Correlation function} $\langle J^{}_{2} J'_{2} J''_{5} \rangle$\textbf{:} \hspace{3mm} $\sigma = 0$
\begin{align}\label{2-2-5}
	\includegraphics[width=0.9\textwidth, valign=c]{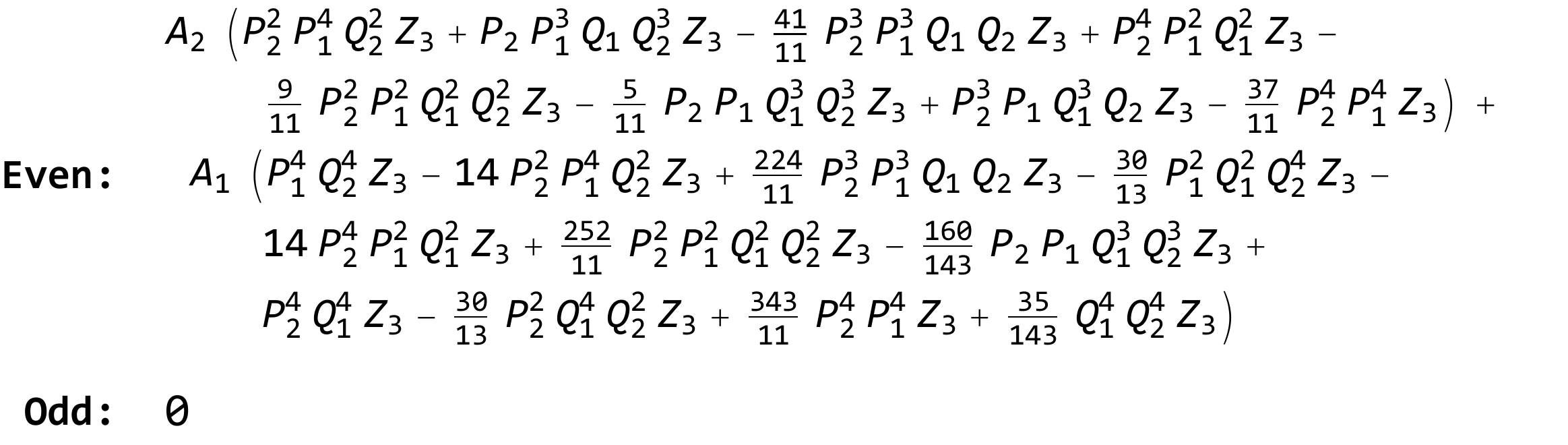}
\end{align} \\
\textbf{Correlation function} $\langle J^{}_{3} J'_{3} J''_{3} \rangle$\textbf{:} \hspace{3mm} $\sigma < 0$
\begin{align}\label{3-3-3}
	\includegraphics[width=0.9\textwidth, valign=c]{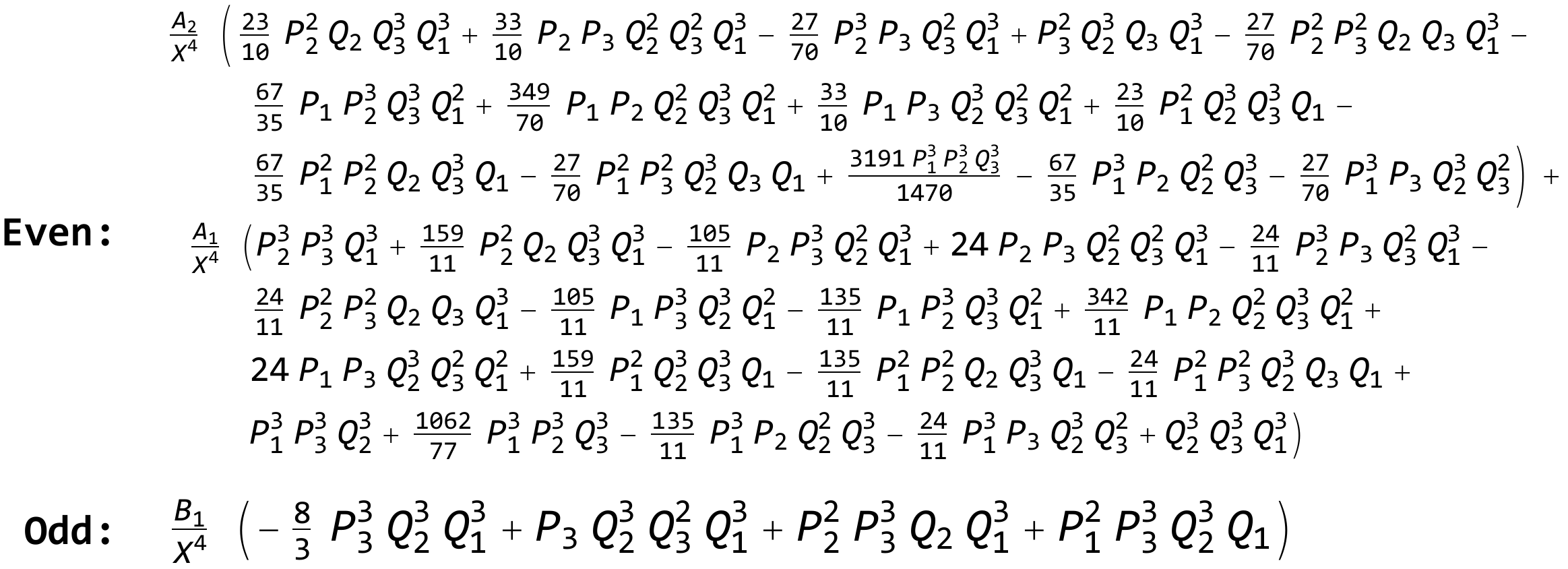}
\end{align} \\
\textbf{Correlation function} $\langle J^{}_{4} J'_{4} J''_{4} \rangle$\textbf{:} \hspace{3mm} $\sigma < 0$
\begin{align}\label{4-4-4}
	\includegraphics[width=0.9\textwidth, valign=c]{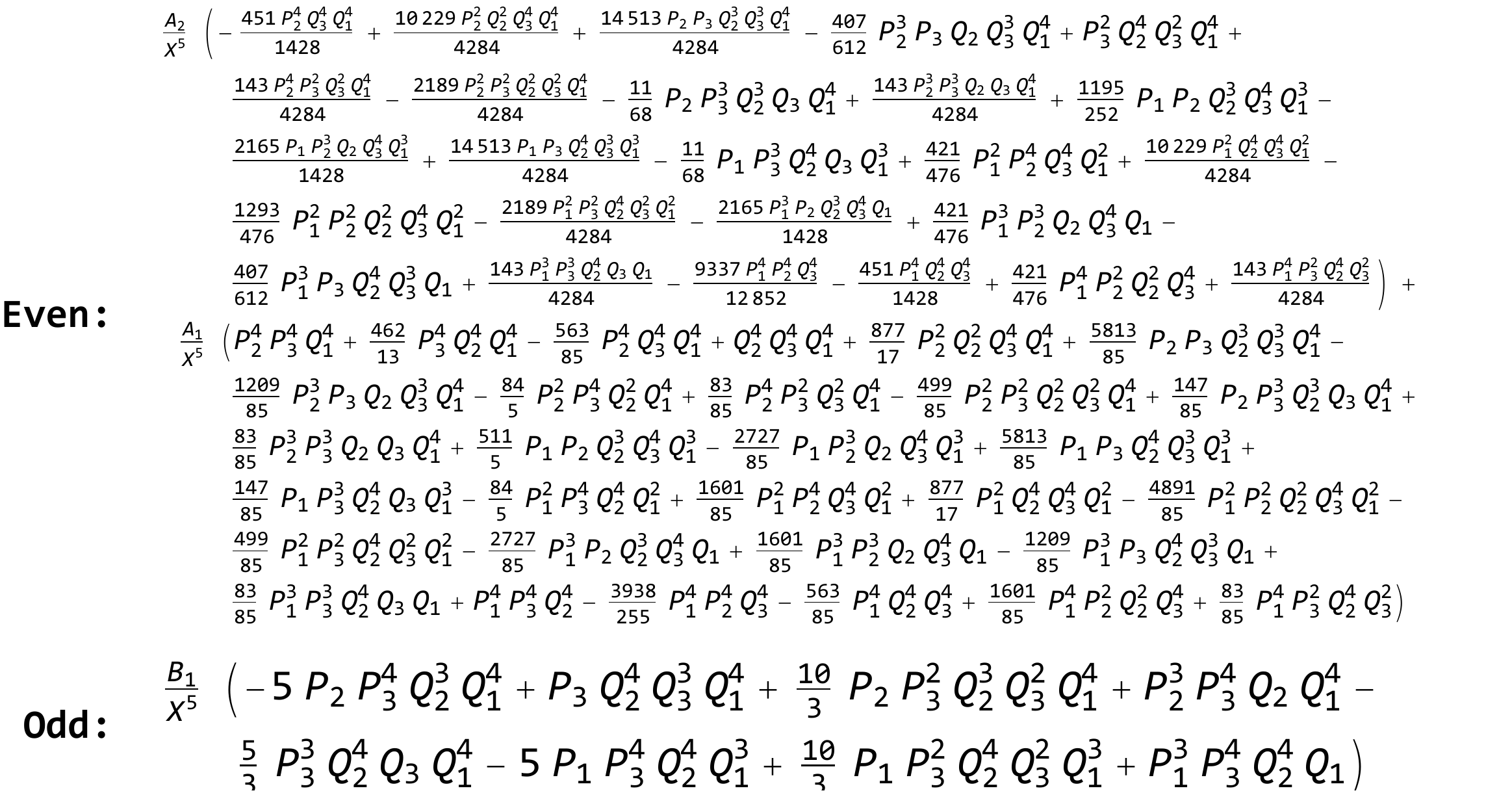}
\end{align} \\
The solutions quickly become cumbersome to present beyond these cases, however, our method effectively produces explicit results for any chosen spins within the confines of our computational limitations. We have explicitly tested our approach up to $s_{i} = 20$, and we present additional results in appendix \ref{AppB}.

\section{Mixed correlators involving fermionic currents}\label{section5}

In this section we will evaluate three-point functions involving conserved fermionic currents. 
To the best of our knowledge, these correlation functions have not been studied in much detail in the literature, particularly in three dimensions. The most important example of a fermionic conserved current is the supersymmetry current, $Q_{m,\a}$, which is prevalent in $\cN$-extended superconformal field theories. Such a field is primary with dimension $\D_{Q} = 5/2$, and satisfies the conservation equation $\pa^{m} Q_{m, \a } = 0$. In spinor notation, we have:
\begin{equation}
	Q_{\a(3)}(x) = (\g^{m})_{(\a_{1} \a_{2}} Q_{m, \a_{3})}(x) \, .
\end{equation}
Recall that in three-dimensional superconformal field theory, the supersymmetry current and the energy-momentum tensor are contained in the supercurrent multiplet, $\mathbf{J}_{\a(3)}(z)$, where $z^{\cA} = (x^{a}, \q^{\a})$ is a point in $\cN = 1$ superspace. The supersymmetry current, $Q_{\a(3)}$, and the energy-momentum tensor, $T_{\a(4)}$, are extracted through bar-projection as follows:
\begin{align}
	Q_{\a(3)}(x) = \mathbf{J}_{\a(3)}(z)|_{\q = 0} \, ,  && T_{\a(4)}(x) = D_{(\a_{1}} \mathbf{J}_{\a_{2} \a_{3} \a_{4} )}(z)|_{\q = 0} \, ,
\end{align}
where 
\begin{equation}
	D_{\a} = \frac{\pa}{\pa \q^{\a}} + \text{i} (\g^{m})_{\a \b} \q^{\b} \frac{\pa}{\pa x^{m}} \, ,
\end{equation}
is the standard spinor-covariant derivative \cite{Buchbinder:1998qv}. Likewise, the conserved vector current, $V_{\a(2)}$, is contained within the flavour current multiplet, $\mathbf{L}_{\a}(z)$, and is extracted as follows:
\begin{align}
	V_{\a(2)}(x) = D_{(\a_{1}} \mathbf{L}_{\a_{2})}(z)|_{\q = 0} \, .
\end{align}
Since the supersymmetry current is a conserved current associated with supersymmetry transformations, it is interesting to study the correlation functions involving the supersymmetry current, the vector current, and the energy-momentum tensor. The two three-point functions involving $Q$, $V$ and $T$ which are of interest in $\cN=1$ superconformal field theories are:
\begin{align} \label{Susy current correlators - 1}
	\langle Q_{\a(3)}(x_{1}) \, Q_{\b(3)}(x_{2}) \, V_{\g(2)}(x_{3}) \rangle \, , && \langle Q_{\a(3)}(x_{1}) \,  Q_{\b(3)}(x_{2}) \, T_{\g(4)}(x_{3}) \rangle \, .
\end{align}
These correlation functions are naturally contained in the following supersymmetric three-point functions
\begin{align}
	\langle \, \mathbf{J}_{\a(3)}(z_{1}) \, \mathbf{J}_{\b(3)}(z_{2}) \, \mathbf{L}_{\g}(z_{3}) \rangle \, , &&  \langle \, \mathbf{J}_{\a(3)}(z_{1}) \, \mathbf{J}_{\b(3)}(z_{2}) \, \mathbf{J}_{\g(3)}(z_{3}) \rangle \, .
\end{align}
In three-dimensions, $\langle \mathbf{J} \mathbf{J} \mathbf{L} \rangle$ vanishes, while $\langle \mathbf{J} \mathbf{J} \mathbf{J} \rangle$ is fixed up to a single tensor 
structure \cite{Nizami:2013tpa,Buchbinder:2015qsa, Buchbinder:2021gwu}. Therefore, since the component correlators \eqref{Susy current correlators - 1} are obtained by 
bar-projecting the supersymmetric correlation functions, we find $\langle Q Q V \rangle = 0$, while $\langle Q Q T \rangle$ is fixed up to a single tensor structure. 

However, in this paper we do not assume supersymmetry. Our goal is to find the most general structure of the correlation functions consistent with only conformal symmetry.
Hence, in the next subsection we will evaluate the correlation functions
\begin{align} \label{Susy current correlators}
	\langle \tilde{Q}_{\a(3)}(x_{1}) \, \tilde{Q}'_{\b(3)}(x_{2}) \, V_{\g(2)}(x_{3}) \rangle \, , && \langle \tilde{Q}_{\a(3)}(x_{1}) \,  \tilde{Q}'_{\b(3)}(x_{2}) \, T_{\g(4)}(x_{3}) \rangle \, ,
\end{align}
where in this case $\tilde{Q}_{\a(3)}$ and $\tilde{Q}'_{\b(3)}$ now denote ``supersymmetry-like" currents; that is, they possess identical properties to supersymmetry currents but are not necessarily 
equal to them. 
It is of interest to us to examine the number of independent tensor structures contained within these three-point functions to see if they are consistent with the supersymmetric results. A similar analysis was recently carried out in four-dimensions \cite{Buchbinder:2022cqp}, where it was found that the number of independent structures is, in general, inconsistent with supersymmetry.

In the next subsections we use our formalism to constrain the general form of correlation functions involving ``supersymmetry-like" currents consistent with conservation equations and point-switch symmetries. This is followed by an analysis of correlation functions involving conserved fermionic higher-spin currents. Our comments on the results for correlation functions involving fermionic symmetry currents for general spins are summarised below:
\begin{itemize}
	\item In general, the structure of the three-point function $\langle J^{}_{s_{1}} J'_{s_{2}} J''_{s_{3}} \rangle$, for arbitrary half-integer $s_{1}, s_{2}$ and integer $s_{3}$, 
	adheres to the triangle inequalities \eqref{Triangle inequalities}, the same as the bosonic case. That is, if the triangle inequalities are satisfied we obtain two even structures and one odd structure. Otherwise, there are just two even structures.
	\item For the three-point functions $\langle J^{}_{s_{1}} J'_{s_{1}} J''_{s_{2}} \rangle$, for arbitrary half-integer $s_{1}$ and integer $s_{2}$: when the triangle inequalities are satisfied there are two even solutions and one odd solution, otherwise there are only two even solutions. After imposing $J=J'$ the solutions exist only when $s_{2}$ is an even integer. Note that for $s_{1} > s_{2}$ the triangle inequalities are always satisfied.
\end{itemize}

\subsection{Spin - 3/2 current correlators}

In this subsection we present an explicit analysis of the general structure of the correlation functions involving $\tilde{Q}$, $V$ and $T$ that are compatible with the constraints of conformal symmetry and conservation equations. Let us first consider $\langle \tilde{Q} \tilde{Q}' V \rangle$, for which we may analyse the general structure of the correlation function $\langle J^{}_{3/2} J'_{3/2} J''_{1} \rangle$. \\[5mm]
\noindent
\textbf{Correlation function} $\langle J^{}_{3/2} J'_{3/2} J''_{1} \rangle$\textbf{:}\\[2mm]
Using the general formula, the ansatz for this three-point function:
\begin{align}
	\langle J^{}_{\a(3)}(x_{1}) \, J'_{\b(3)}(x_{2}) \, J''_{\g(2)}(x_{3}) \rangle = \frac{ \cI_{\a(3)}{}^{\a'(3)}(x_{13}) \,  \cI_{\b(3)}{}^{\b'(3)}(x_{23}) }{(x_{13}^{2})^{5/2} (x_{23}^{2})^{5/2}}
	\; \cH_{\a'(3) \b'(3) \g(2)}(X_{12}) \, .
\end{align} 
Using the formalism outlined in \ref{subsection3.2}, all information about this correlation function is encoded in the following polynomial:
\begin{align}
	\cH(X; u(3), v(3), w(2)) = \cH_{ \a(3) \b(3) \g(2) }(X) \, \mathbf{U}^{\a(3)}  \mathbf{V}^{\b(3)}  \mathbf{W}^{\g(2)} \, .
\end{align}
After solving \eqref{Diophantine equations}, we find the following linearly dependent polynomial structures in the even and odd sectors respectively:
\begin{align}
	\includegraphics[width=0.9\textwidth, valign=c]{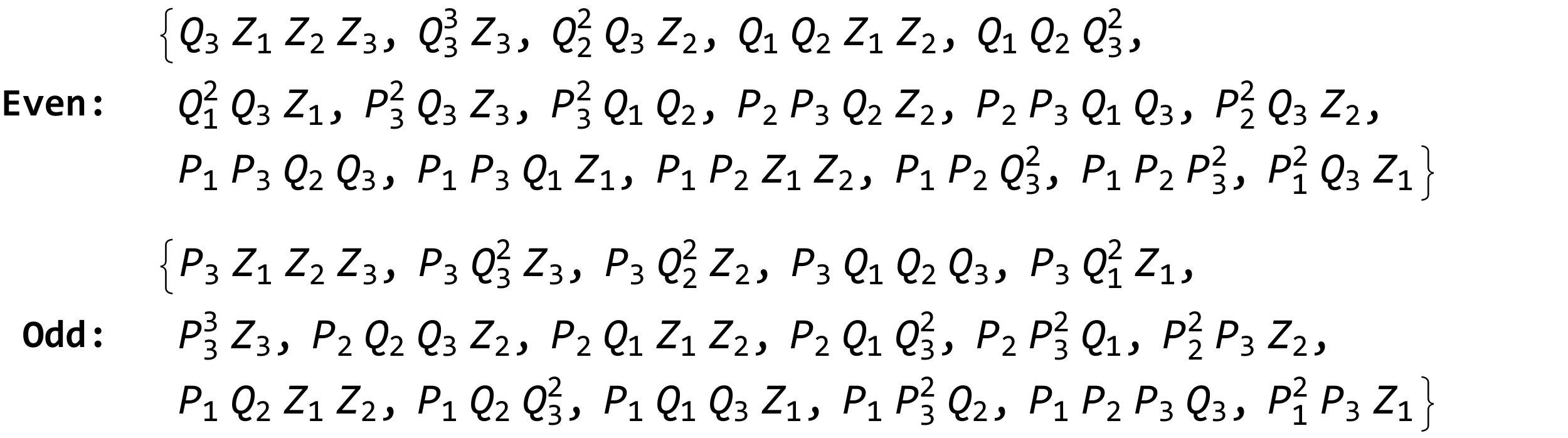}
\end{align}
Next we systematically apply the linear dependence relations \eqref{Linear dependence 1} to these lists, reducing them to the following linearly independent structures:
\begin{align}
	\includegraphics[width=0.9\textwidth, valign=c]{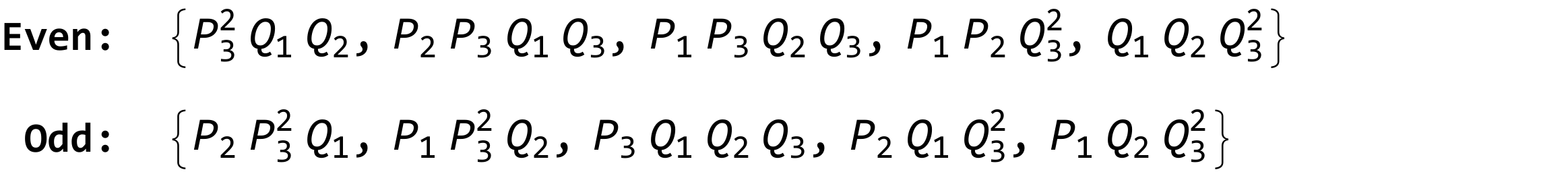}
\end{align}
After constructing an appropriate ansatz for each sector, we obtain the following relations between the coefficients after imposing conservation on all three points:
\begin{align}
	\includegraphics[width=0.9\textwidth, valign=c]{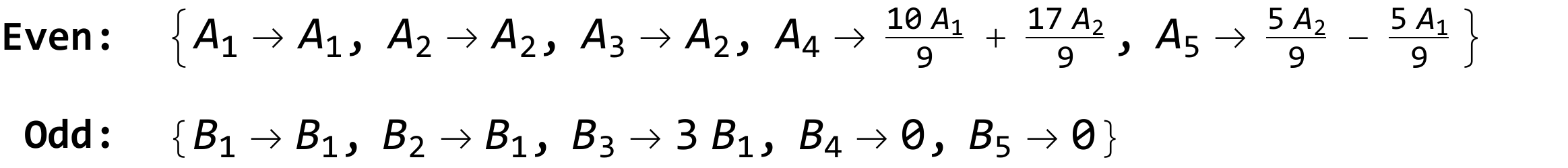}
\end{align}
The final solutions for the even and odd sectors are
\begin{align}\label{3/2-3/2-1}
	\includegraphics[width=0.9\textwidth, valign=c]{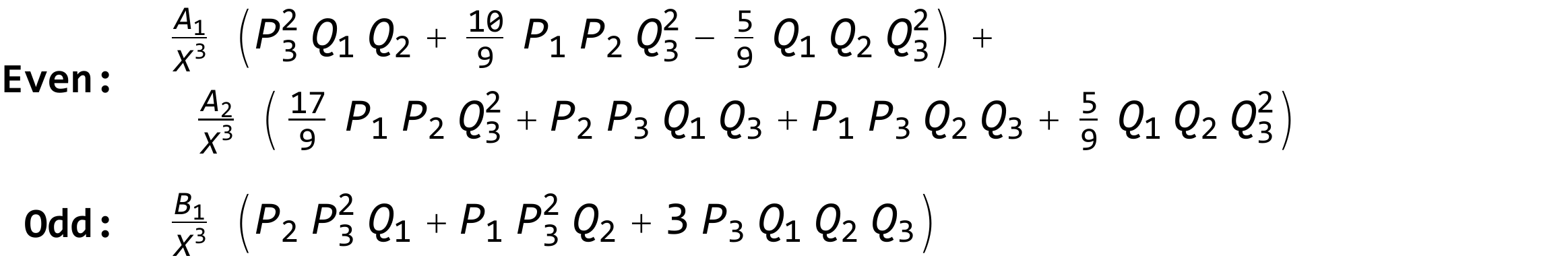}
\end{align} \\
Hence, we see that the correlation function $\langle J^{}_{3/2} J'_{3/2} J''_{1} \rangle$, and therefore $\langle \tilde{Q} \tilde{Q}' V \rangle$, is fixed up to two even structures and one odd structure. It may be show that all structures vanish for $J=J'$ as they do not-possess the correct symmetry under permutation of points $x_{1}$ and $x_{2}$, therefore we find that the correlation function $\langle \tilde{Q} \tilde{Q} V \rangle$ vanishes. 

Next we will analyse the general structure of $\langle \tilde{Q} \tilde{Q} T \rangle$, which is associated with the correlation function $\langle J^{}_{3/2} J'_{3/2} J''_{2} \rangle$ using the ansatz \eqref{Conserved correlator ansatz}.\\

\noindent
\textbf{Correlation function} $\langle J^{}_{3/2} J'_{3/2} J''_{2} \rangle$\textbf{:}\\[2mm]
According to the general formula \eqref{Conserved correlator ansatz}, the ansatz for this three-point function is:
\begin{align}
	\langle J^{}_{\a(3)}(x_{1}) \, J'_{\b(3)}(x_{2}) \, J''_{\g(4)}(x_{3}) \rangle = \frac{ \cI_{\a(3)}{}^{\a'(3)}(x_{13}) \,  \cI_{\b(3)}{}^{\b'(3)}(x_{23}) }{(x_{13}^{2})^{5/2} (x_{23}^{2})^{5/2}}
	\; \cH_{\a'(3) \b'(3) \g(4)}(X_{12}) \, .
\end{align} 
Using the formalism outlined in \ref{subsection3.2}, all information about this correlation function is encoded in the following polynomial:
\begin{align}
	\cH(X; u(3), v(3), w(4)) = \cH_{ \a(3) \b(3) \g(4) }(X) \, \mathbf{U}^{\a(3)}  \mathbf{V}^{\b(3)}  \mathbf{W}^{\g(4)} \, .
\end{align}
After solving \eqref{Diophantine equations}, we find the following linearly dependent polynomial structures in the even and odd sectors respectively:
\begin{align}
	\includegraphics[width=0.9\textwidth, valign=c]{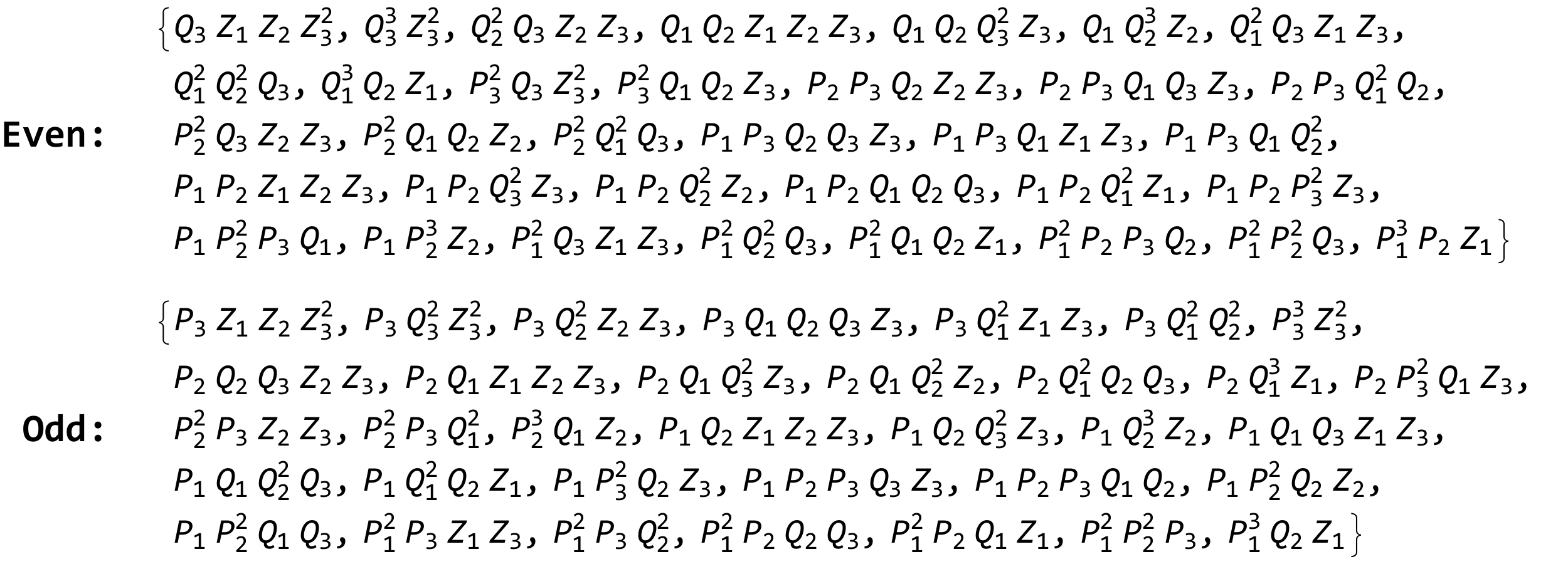}
\end{align}
Next we systematically apply the linear dependence relations \eqref{Linear dependence 1} to these lists, reducing them to the following linearly independent structures:
\begin{align}
	\includegraphics[width=0.9\textwidth, valign=c]{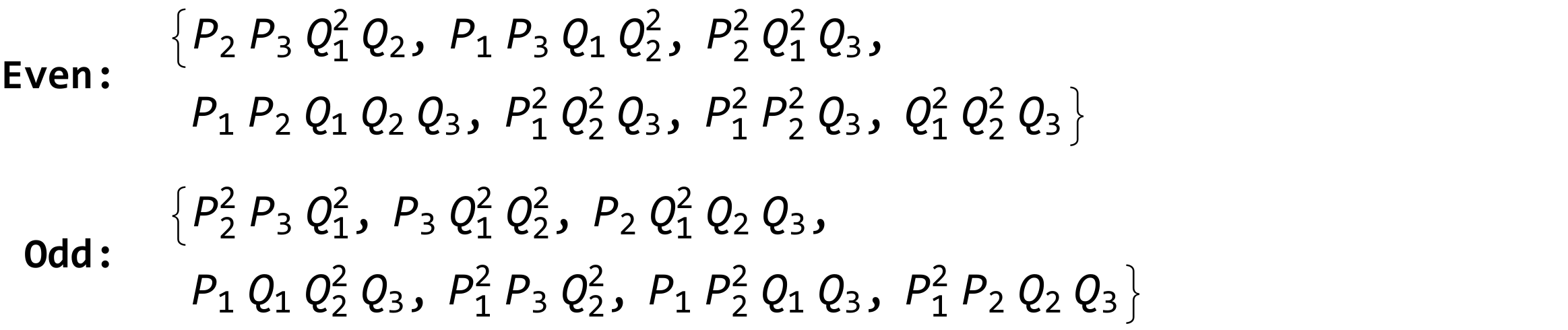}
\end{align}
After constructing an appropriate ansatz for each sector, we obtain the following relations between the coefficients after imposing conservation on all three points:
\begin{align}
	\includegraphics[width=0.9\textwidth, valign=c]{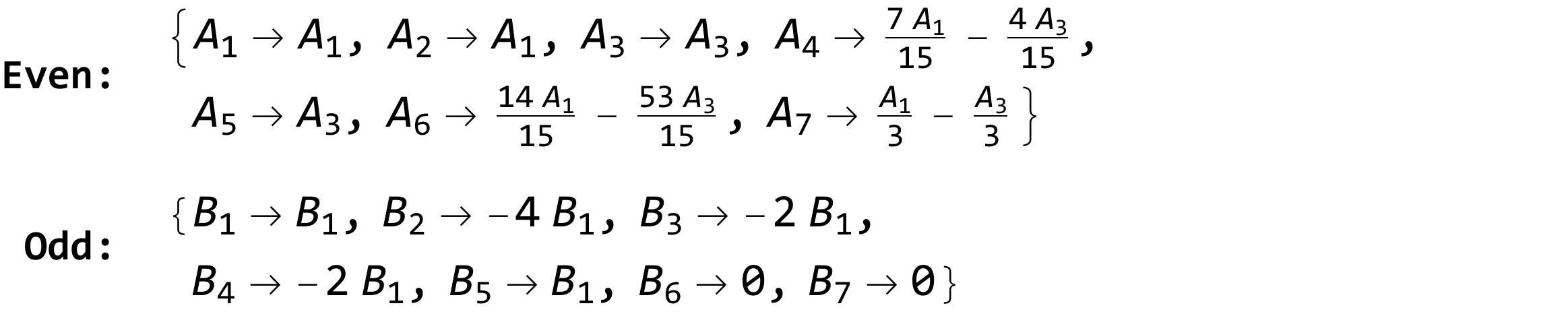}
\end{align}
Therefore the final solutions for the even and odd sectors are
\begin{align}\label{3/2-3/2-2}
	\includegraphics[width=0.9\textwidth, valign=c]{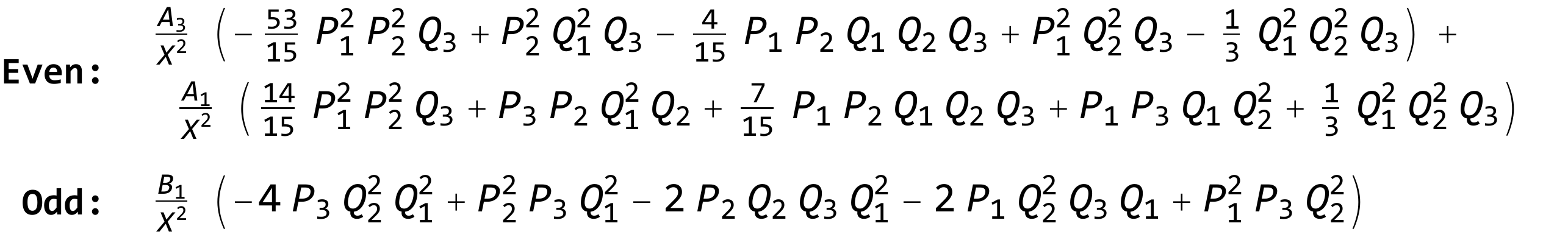}
\end{align}
Hence, we note that the three-point function $\langle J^{}_{3/2} J'_{3/2} J''_{2} \rangle$, and therefore $\langle \tilde{Q} \tilde{Q}' T \rangle$, is fixed up to two independent ``even" structures and one ``odd" structure. In this case, both structures survive after imposing the symmetry under the exchange of $x_{1}$ and $x_{2}$, that is, when $J = J'$. Hence, the correlation function $\langle \tilde{Q} \tilde{Q} T \rangle$ is also fixed up to two even structures and a single odd structure.

\subsection{Higher-spin correlators}

In this subsection we compile some results for three-point correlation functions involving fermionic higher-spin currents. We present only the final results after imposing conservation on all three points.

\noindent
\textbf{Correlation function} $\langle J^{}_{3/2} J'_{3/2} J''_{3} \rangle$\textbf{:} \hspace{3mm} $\sigma < 0$
\begin{align}\label{3/2-3/2-3}
	\includegraphics[width=0.9\textwidth, valign=c]{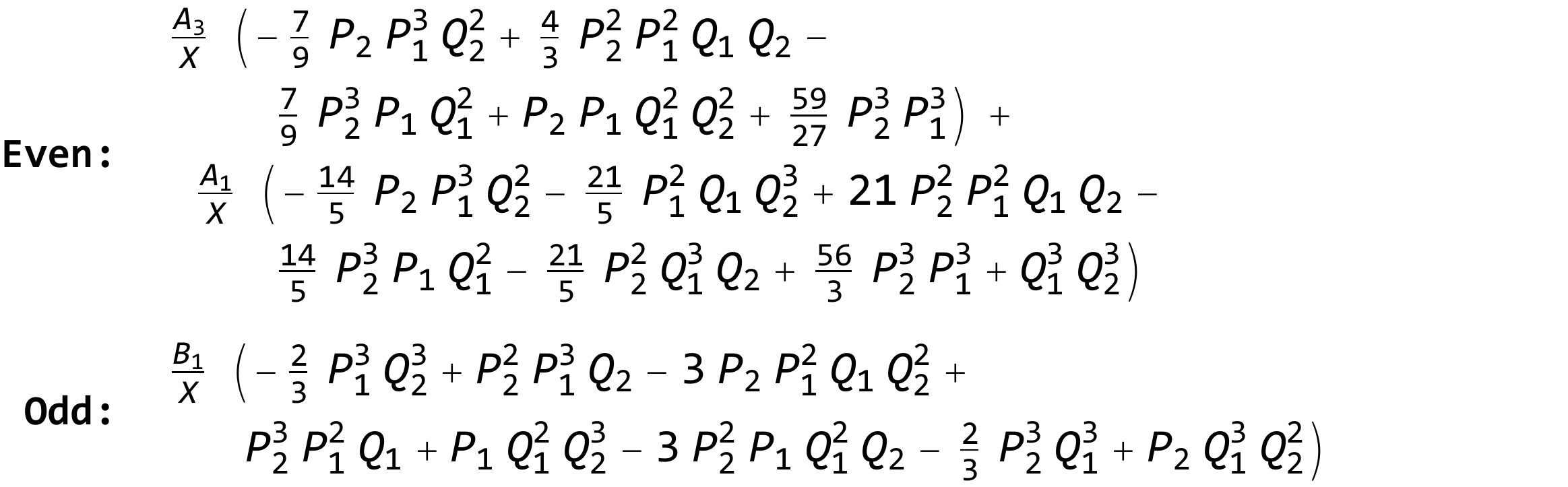}
\end{align} \\
Imposing the symmetry under permutation of spacetimes points $x_{1}$ and $x_{2}$, i.e. when $J = J'$, requires that the three-point function must vanish. \\[3mm]
\noindent\textbf{Correlation function} $\langle J^{}_{3/2} J'_{3/2} J''_{4} \rangle$\textbf{:} \hspace{3mm} $\sigma = 0$
\begin{align}\label{3/2-3/2-4}
	\includegraphics[width=0.9\textwidth, valign=c]{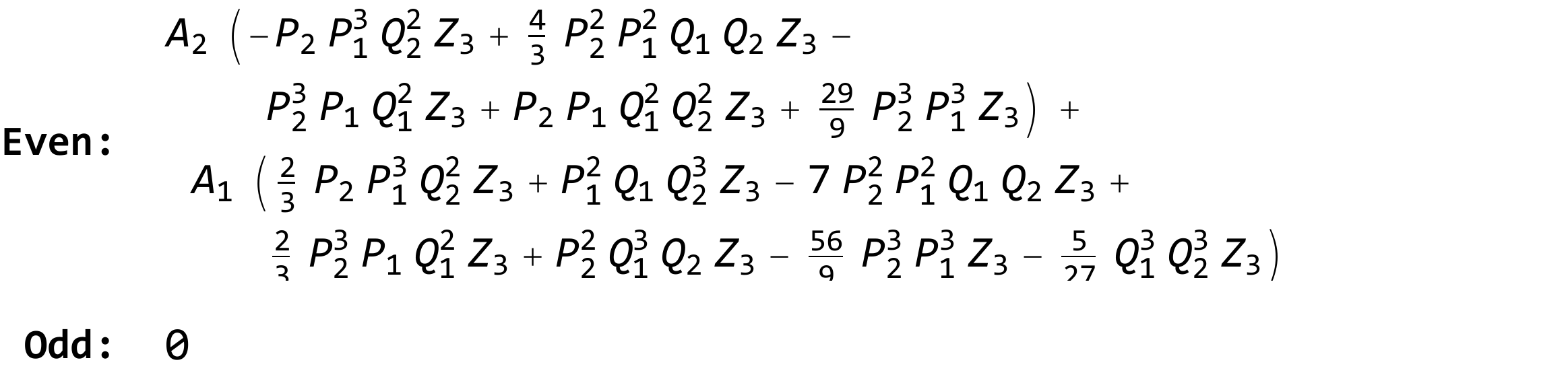}
\end{align} \\
This correlation function is compatible with the symmetry under permutation of spacetimes points $x_{1}$ and $x_{2}$. \\[3mm]
\noindent\textbf{Correlation function} $\langle J^{}_{3/2} J'_{3/2} J''_{5} \rangle$\textbf{:} \hspace{3mm} $\sigma > 0$
\begin{align}\label{3/2-3/2-5}
	\includegraphics[width=0.9\textwidth, valign=c]{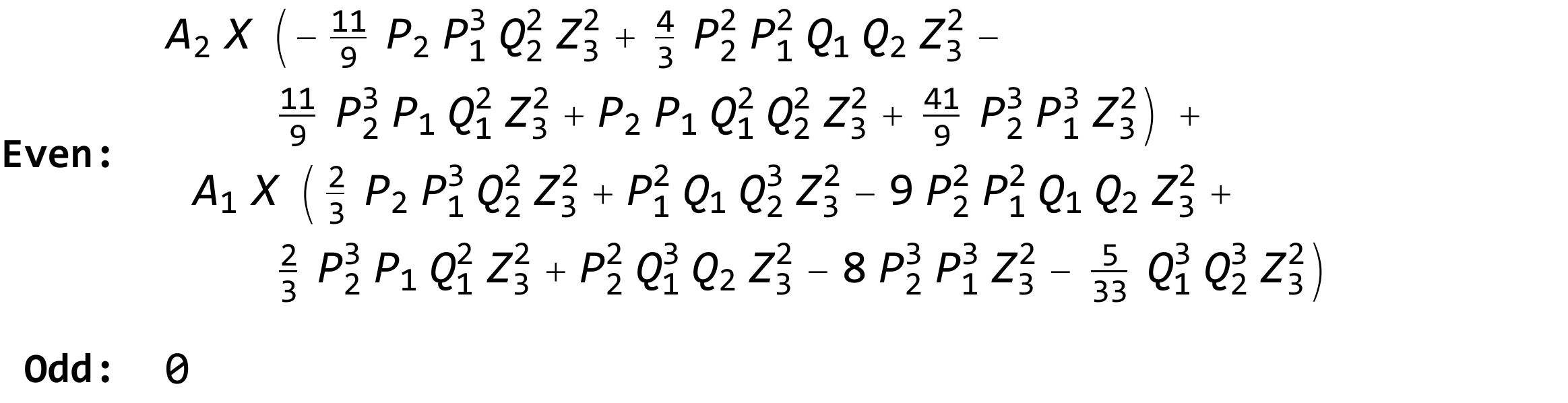}
\end{align} \\
\textbf{Correlation function} $\langle J^{}_{5/2} J'_{3/2} J''_{1} \rangle$\textbf{:} \hspace{3mm} $\sigma < 0$
\begin{align}\label{5/2-3/2-1}
	\includegraphics[width=0.9\textwidth, valign=c]{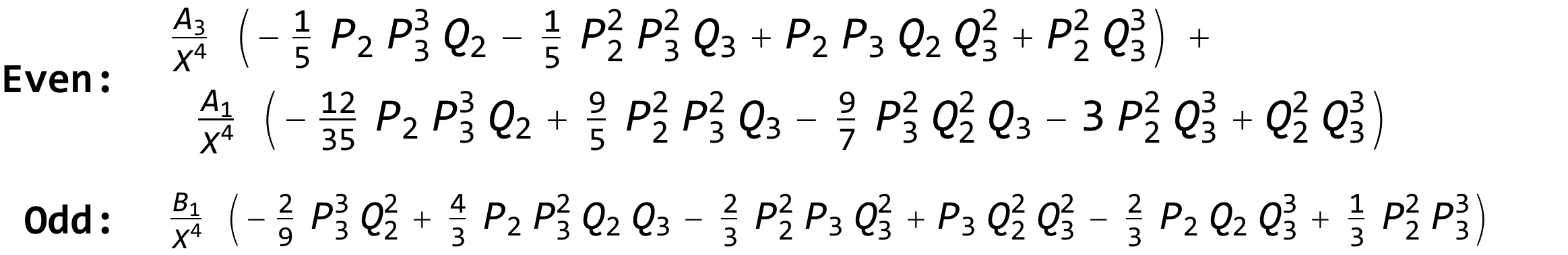}
\end{align} \\
\textbf{Correlation function} $\langle J^{}_{5/2} J'_{3/2} J''_{2} \rangle$\textbf{:} \hspace{3mm} $\sigma < 0$
\begin{align}\label{5/2-3/2-2}
	\includegraphics[width=0.9\textwidth, valign=c]{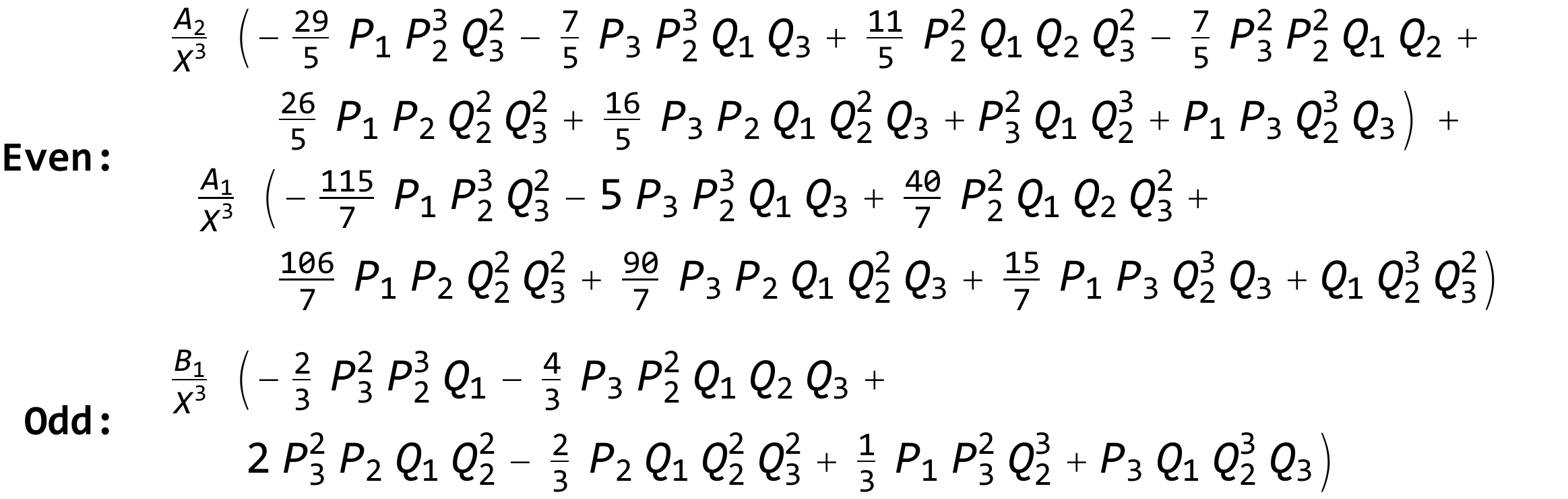}
\end{align} \\
\textbf{Correlation function} $\langle J^{}_{5/2} J'_{3/2} J''_{3} \rangle$\textbf{:} \hspace{3mm} $\sigma < 0$
\begin{align}\label{5/2-3/2-3}
	\includegraphics[width=0.9\textwidth, valign=c]{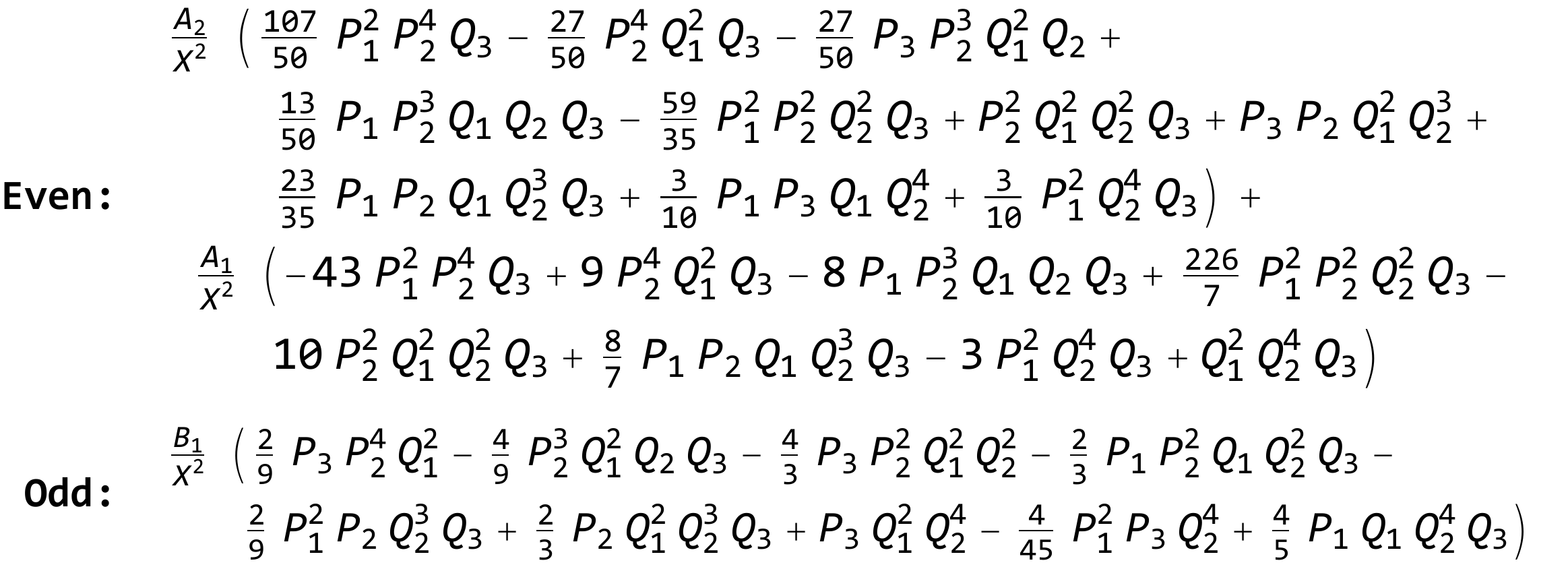}
\end{align} \\
\textbf{Correlation function} $\langle J^{}_{5/2} J'_{3/2} J''_{4} \rangle$\textbf{:} \hspace{3mm} $\sigma < 0$
\begin{align}\label{5/2-3/2-4}
	\includegraphics[width=0.9\textwidth, valign=c]{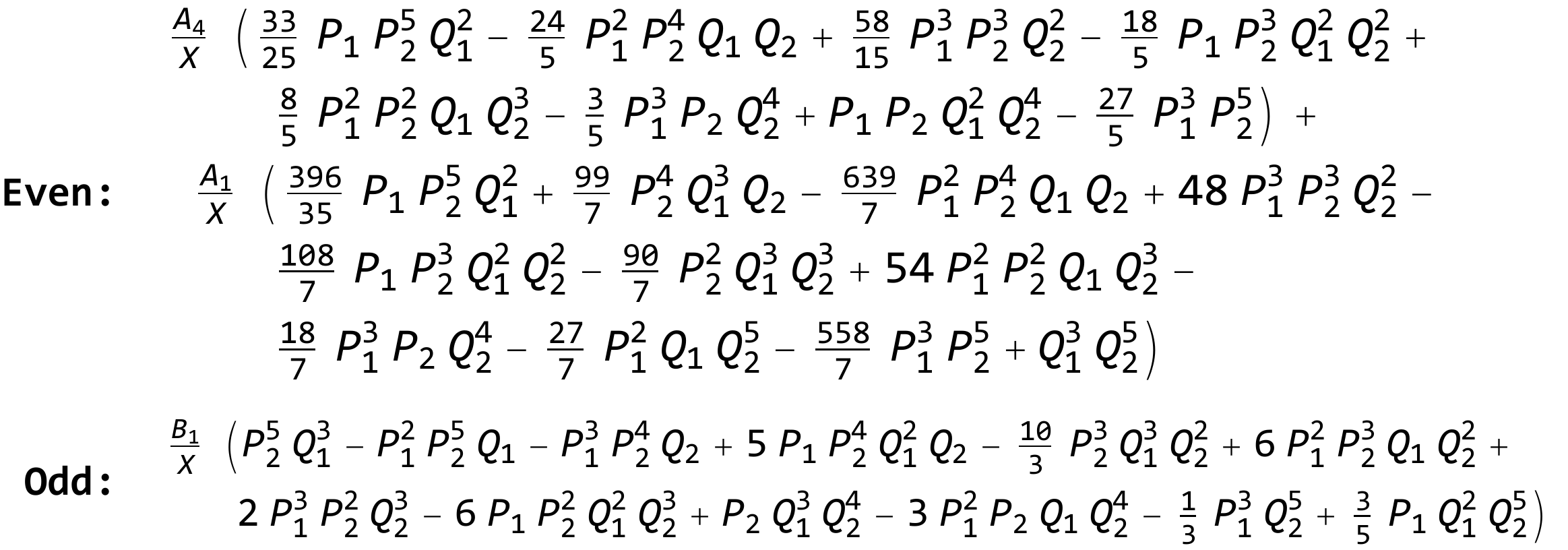}
\end{align} \\
\textbf{Correlation function} $\langle J^{}_{5/2} J'_{3/2} J''_{5} \rangle$\textbf{:} \hspace{3mm} $\sigma = 0$
\begin{align}\label{5/2-3/2-5}
	\includegraphics[width=0.9\textwidth, valign=c]{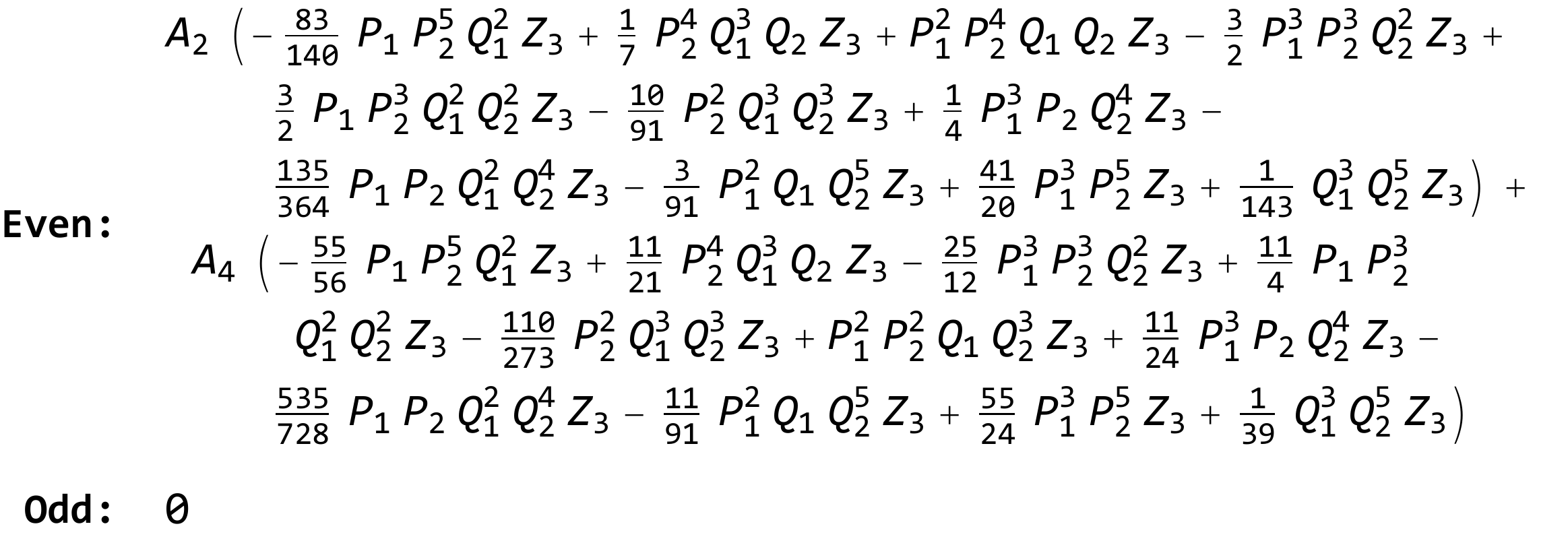}
\end{align} \\
\textbf{Correlation function} $\langle J^{}_{5/2} J'_{5/2} J''_{1} \rangle$\textbf{:} \hspace{3mm} $\sigma < 0$
\begin{align}\label{5/2-5/2-1}
	\includegraphics[width=0.9\textwidth, valign=c]{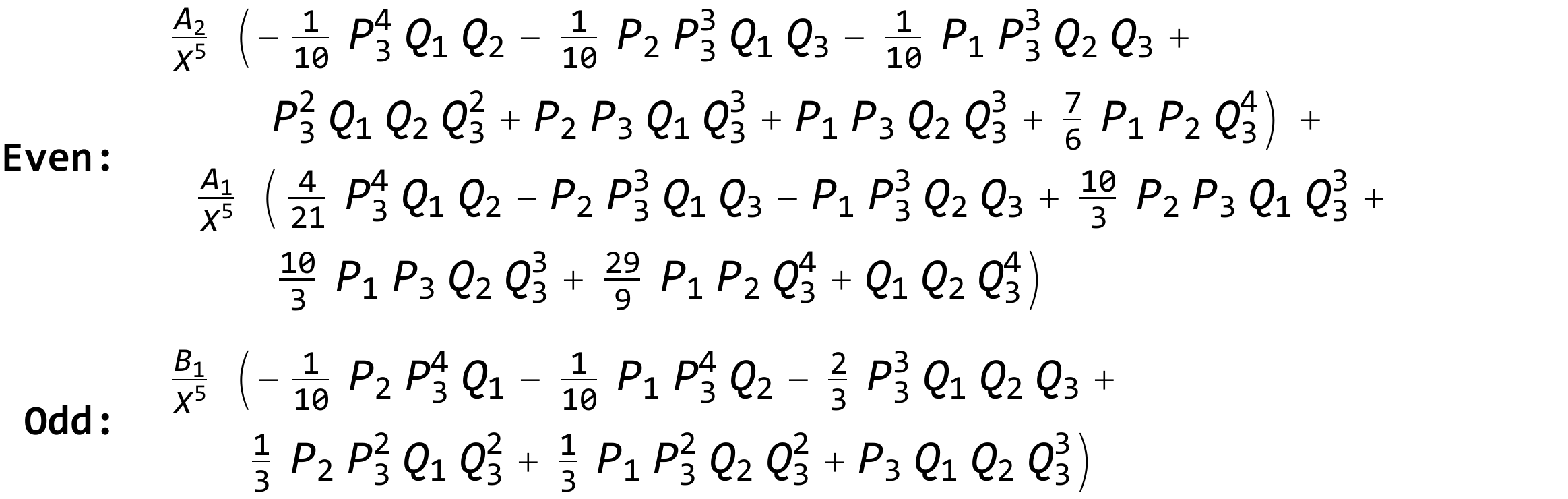}
\end{align} \\
\textbf{Correlation function} $\langle J^{}_{5/2} J'_{5/2} J''_{2} \rangle$\textbf{:} \hspace{3mm} $\sigma < 0$
\begin{align}\label{5/2-5/2-2}
	\includegraphics[width=0.9\textwidth, valign=c]{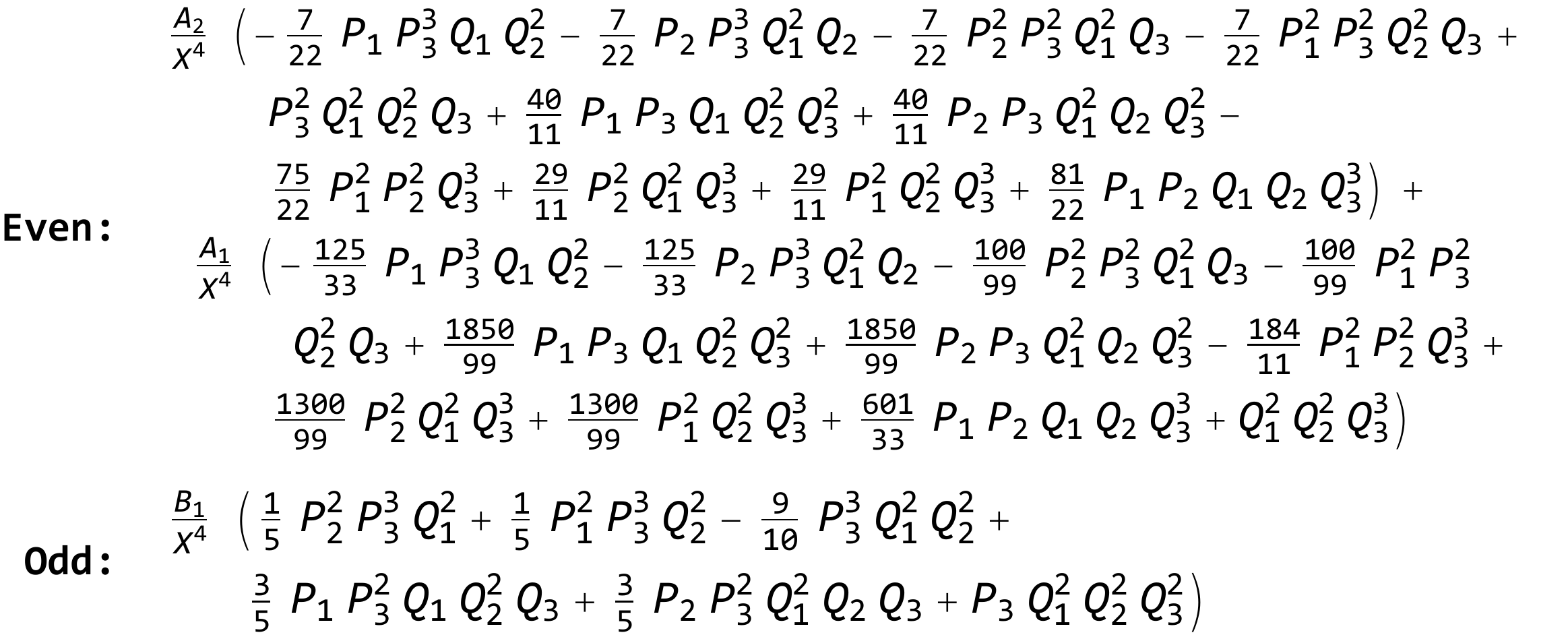}
\end{align} \\
\textbf{Correlation function} $\langle J^{}_{5/2} J'_{5/2} J''_{3} \rangle$\textbf{:} \hspace{3mm} $\sigma < 0$
\begin{align}\label{5/2-5/2-3}
	\includegraphics[width=0.9\textwidth, valign=c]{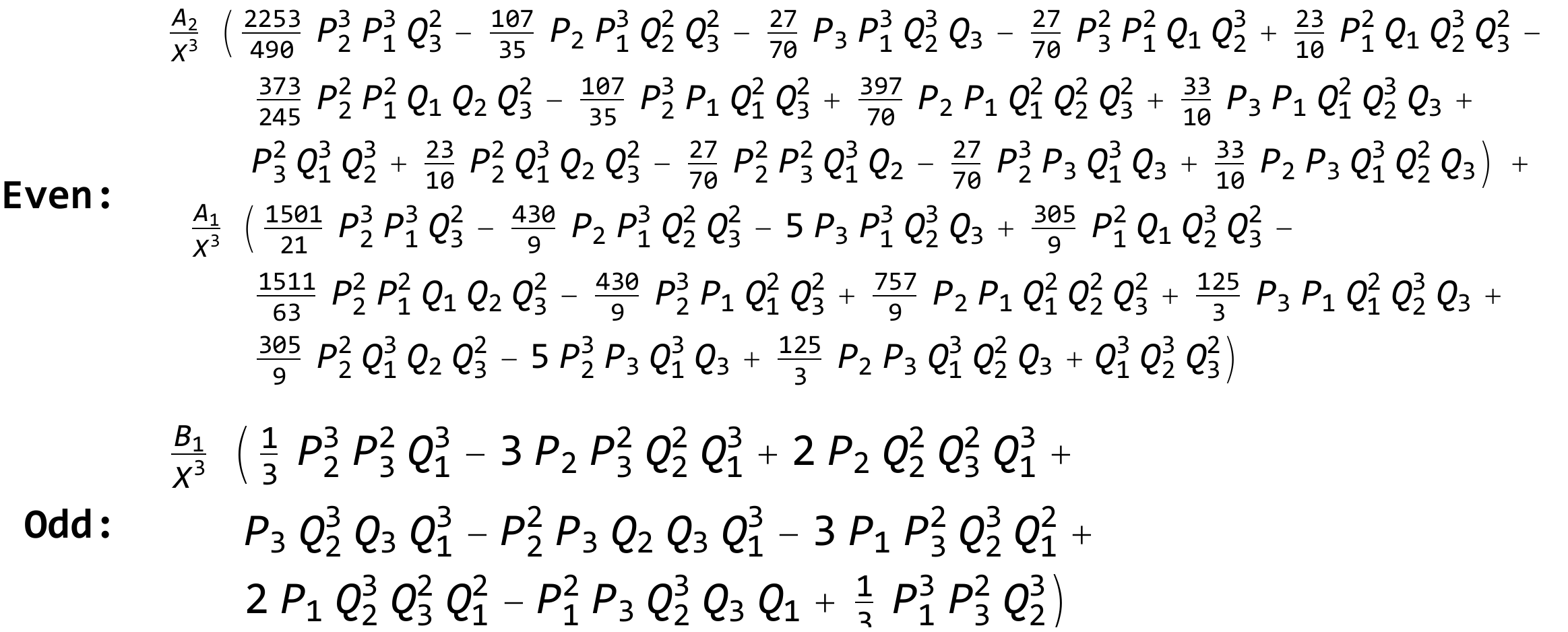}
\end{align} \\
\textbf{Correlation function} $\langle J^{}_{5/2} J'_{5/2} J''_{4} \rangle$\textbf{:} \hspace{3mm} $\sigma < 0$
\begin{align}\label{5/2-5/2-4}
	\includegraphics[width=0.9\textwidth, valign=c]{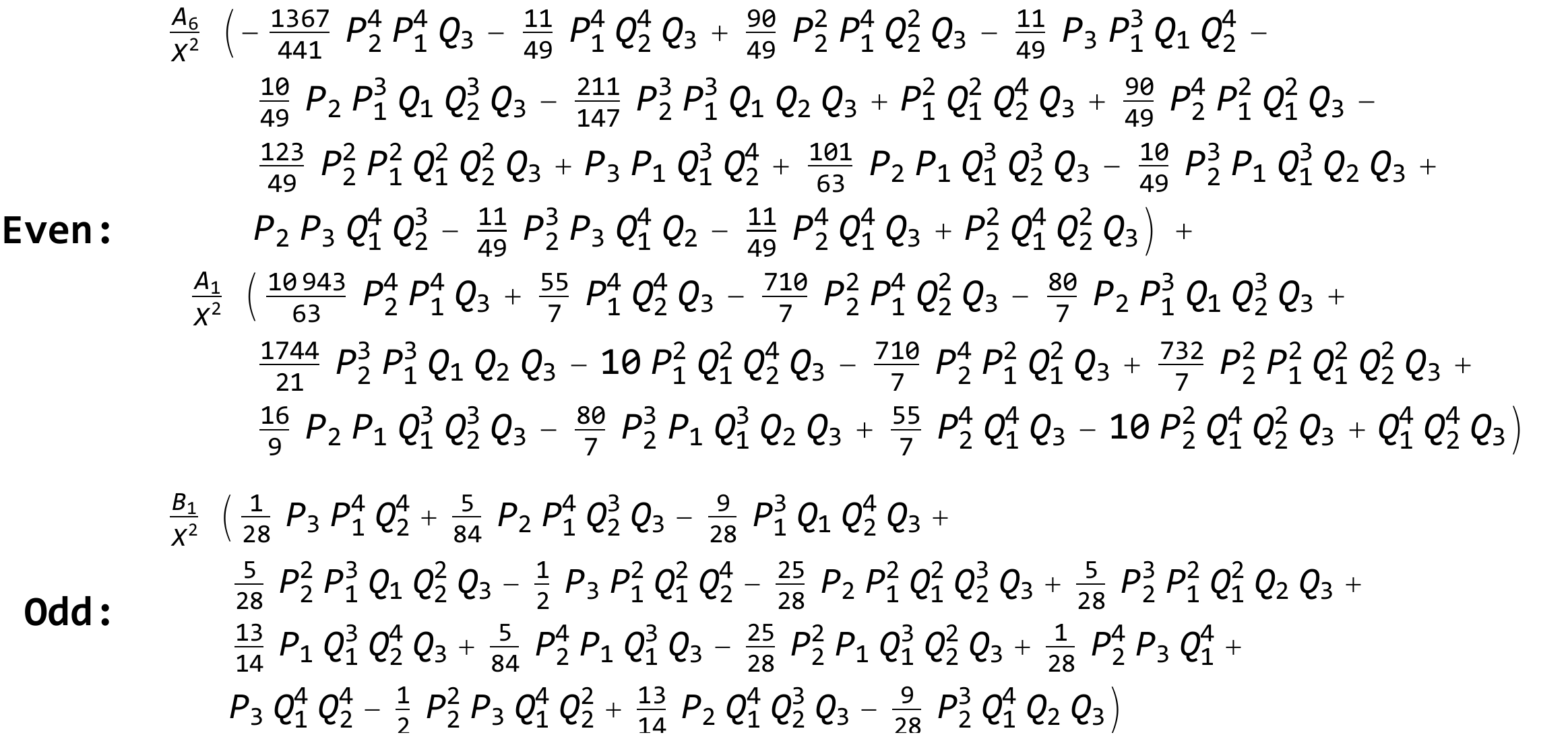}
\end{align} \\
\textbf{Correlation function} $\langle J^{}_{5/2} J'_{5/2} J''_{5} \rangle$\textbf{:} \hspace{3mm} $\sigma < 0$
\begin{align}\label{5/2-5/2-5}
	\includegraphics[width=0.9\textwidth, valign=c]{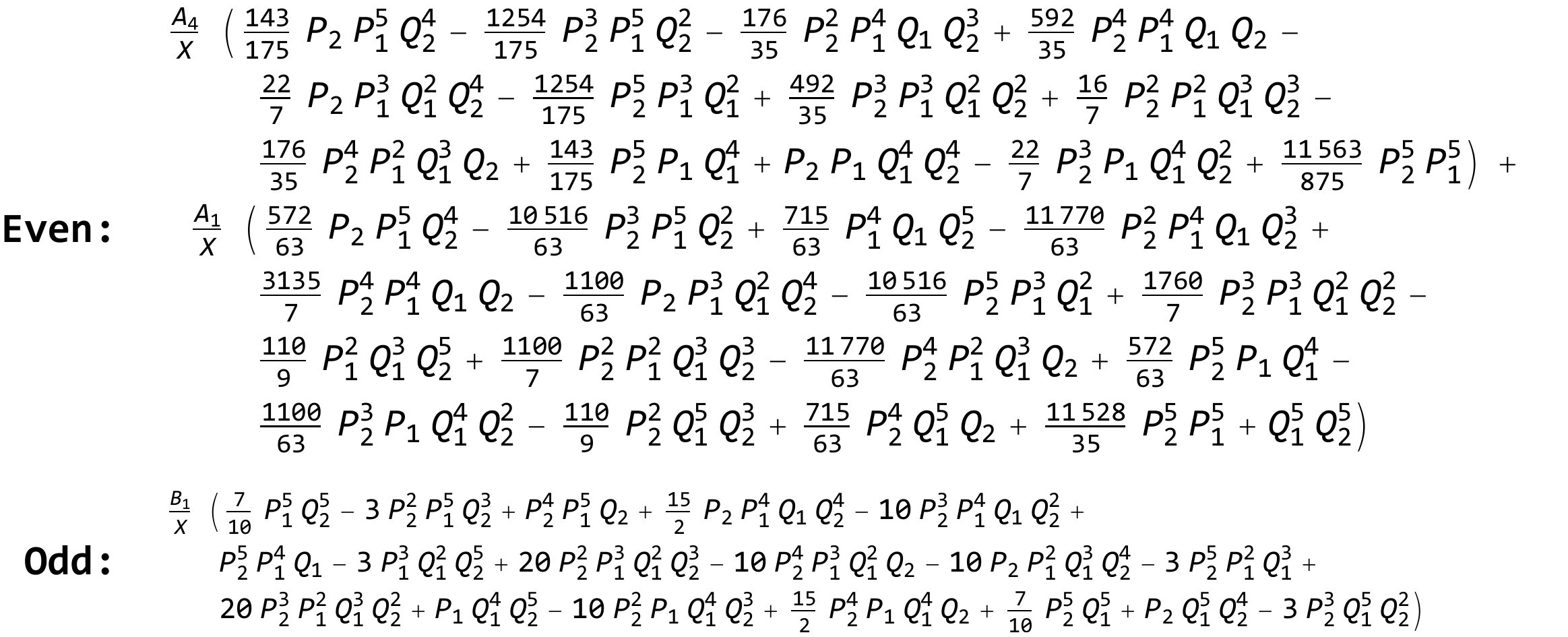}
\end{align} \\
\textbf{Correlation function} $\langle J^{}_{5/2} J'_{5/2} J''_{6} \rangle$\textbf{:} \hspace{3mm} $\sigma = 0$
\begin{align}\label{5/2-5/2-6}
	\includegraphics[width=0.9\textwidth, valign=c]{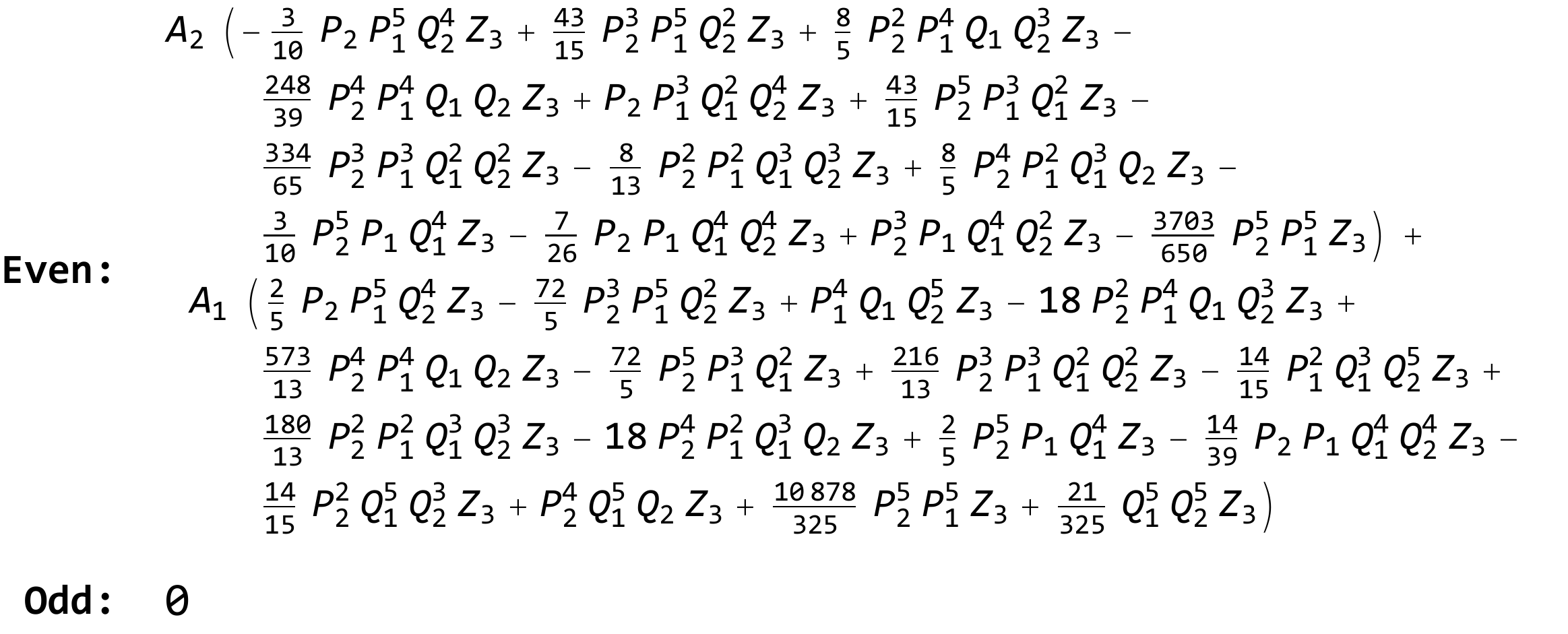}
\end{align}
Additional results for three-point functions involving fermionic higher-spin currents are contained in appendix \ref{AppB}.

\section{Correlators involving scalars and spinors}\label{section6}

In this section, for completeness, we analyse some of the important three-point correlation functions involving scalar and spinor fields. The results are interesting because the correlation functions can contain parity-odd solutions, with their existence depending on both triangle inequalities and the scale dimensions of the scalars/spinors. The correlation functions are analysed using the same methods as in the previous sections; the full classification of the results is presented below:
\begin{itemize}
	\item The three-point function $\langle \psi \, \psi' \, O \rangle$, where $\psi, \psi'$ are fundamental fermions and $O$ is a fundamental scalar, is fixed up to one even structure and one odd structure. All structures remain after imposing $\psi = \psi'$.
	\item The three-point function $\langle O \, O' J_{s} \rangle$, where $O,O'$ are fundamental scalars with dimension $\delta,\delta'$ respectively: for $\delta = \delta'$, there is a single even solution compatible with conservation which survives after imposing $O = O'$ only for even $s$. For $\delta \neq \delta'$ there are no solutions.
	\item The three-point function $\langle \psi \, \psi' J_{s} \rangle$, where $\psi,\psi$ are fundamental fermions with dimension $\delta,\delta'$ respectively: when $s=1$, the triangle inequalities are satisfied, and for $\delta = \delta'$ there are two even solutions and one odd solution. For $\delta \neq \delta'$ there is one even solution and one odd solution. In both cases, the three-point function vanishes after imposing $\psi = \psi'$. For $s >1$, the triangle inequalities are not satisfied, and for $\d = \d'$ there are two even solutions which survive after imposing $\psi = \psi'$ provided that $s$ is even. For $\d \neq \d'$ there are no solutions for general $s$.
	\item The three-point function $\langle \psi \, J^{}_{s} \, O \rangle$, for half-integer $s \geq 3/2$, where $\psi$ is a fundamental fermion with dimension $\delta$, and $O$ is a scalar with dimension $\delta'$: the triangle inequalities are not satisfied for any $s$, and in general there are no solutions after imposing conservation for arbitrary $\d, \d'$. However, there are two special cases; for $\delta = 3/2$ there is an even solution for $\delta' = 1$ and an odd solution for $\delta' = 2$.
	\item For three-point functions of the form $\langle J^{}_{s_{1}} J'^{}_{s_{2}} \, O \rangle$, where $O$ is a scalar field with dimension $\delta$, $s_{1}$ and $s_{2}$ must be simultaneously integer/half-integer for there to be a solution. For $s_{1} > s_{2}$, the triangle inequalities are not satisfied and there is no solution for general $\delta$, however, there is an even solution for $\delta = 1$, and an odd solution for $\delta = 2$. For $s_{1} = s_{2}$, the triangle inequalities are satisfied and there exists an even and odd solution for general $\delta$. The solutions also survive after imposing the symmetry $J=J'$.
	\item For three-point functions of the form $\langle \psi \, J^{}_{s_{1}} J'_{s_{2}} \rangle$, for half-integer $s_{1} \geq 3/2$ and integer $s_{2}$, where $\psi$ is a fundamental fermion with dimension $\delta$: for $\delta = 3/2$ there are two even solutions and one odd solution provided that the triangle inequalities are satisfied, otherwise, there are only two even solutions. In addition, for $\delta = 5/2$ there is one even solution and one odd solution when the triangle inequalities are satisfied, otherwise, there is a single odd solution. For general $\delta$, an even and odd solution exists if the triangle inequalities are satisfied, otherwise, there are no solutions. 
\end{itemize}
In the next subsections we present explicit solutions for some of the above cases.

\subsection{Low-spin correlators}
\noindent
\textbf{Correlation function} $\langle \psi \, \psi' \, O \rangle$\textbf{:}\\[2mm]
For $\Delta_{\psi} = \delta_{1}$, $\Delta_{\psi'} = \d'_{1}$, $\Delta_{O} = \d_{2}$, there is always one even and one odd solution
\begin{align}\label{1/2-1/2-0}
	\includegraphics[width=0.9\textwidth, valign=c]{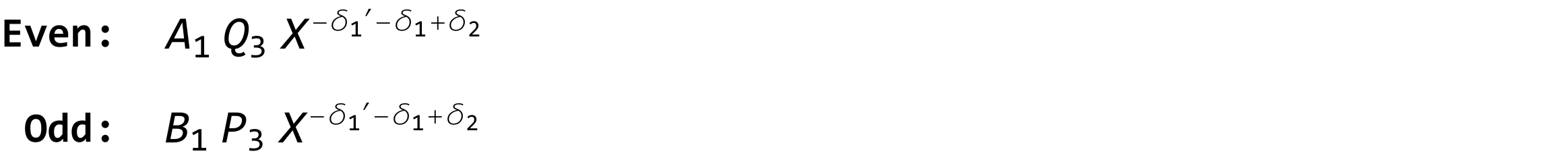}
\end{align} \\
\textbf{Correlation function} $\langle O \, O' J_{1} \rangle$\textbf{:}\\[2mm]
An even solution exists for $\Delta_{O} = \Delta_{O'} = \delta$:
\begin{align}\label{0-0-1}
	\includegraphics[width=0.9\textwidth, valign=c]{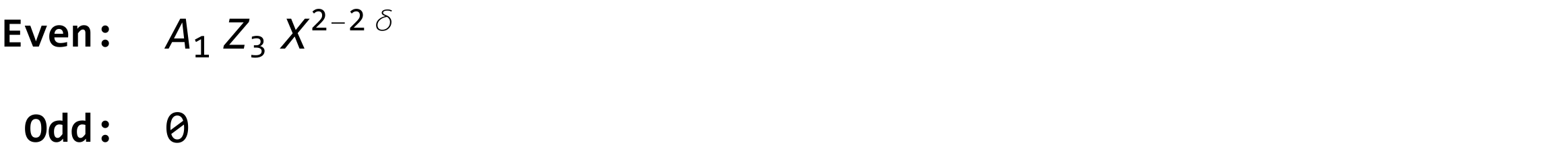}
\end{align}
The solution vanishes upon imposing the symmetry between $x_{1}$ and $x_{2}$, i.e, when the fields $O$, $O'$ coincide. \\[3mm]
\textbf{Correlation function} $\langle O \, O' J_{2} \rangle$\textbf{:}\\[2mm]
An even solution exists for $\Delta_{O} = \Delta_{O'} = \delta$:
\begin{align}\label{0-0-2}
	\includegraphics[width=0.9\textwidth, valign=c]{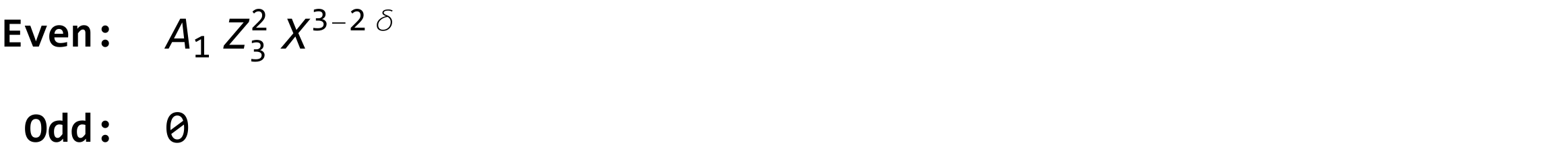}
\end{align}
In general, for $\langle O \, O' J_{s} \rangle$ there is always an even solution, however, it only survives the $O = O'$ point-switch symmetry for even $s$. \\[3mm]
\textbf{Correlation function} $\langle \psi \, \psi' J_{1} \rangle$\textbf{:}
\begin{align}\label{1/2-1/2-1}
	\includegraphics[width=0.9\textwidth, valign=c]{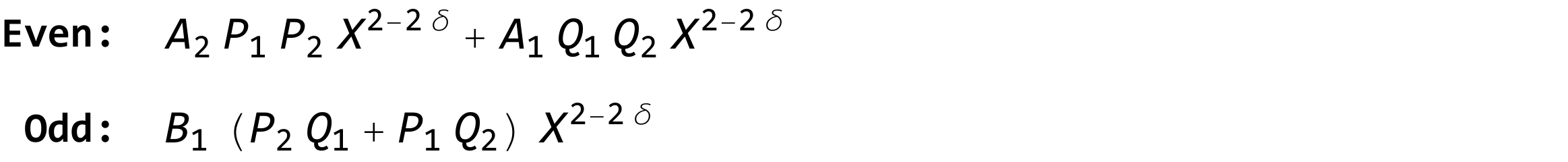}
\end{align} 
In this case the triangle inequalities are satisfied and there are two even solutions and one odd solution. All structures vanish after imposing $\psi = \psi'$. \\[3mm]
\textbf{Correlation function} $\langle \psi \, \psi' J_{2} \rangle$\textbf{:}
\begin{align}\label{1/2-1/2-2}
	\includegraphics[width=0.9\textwidth, valign=c]{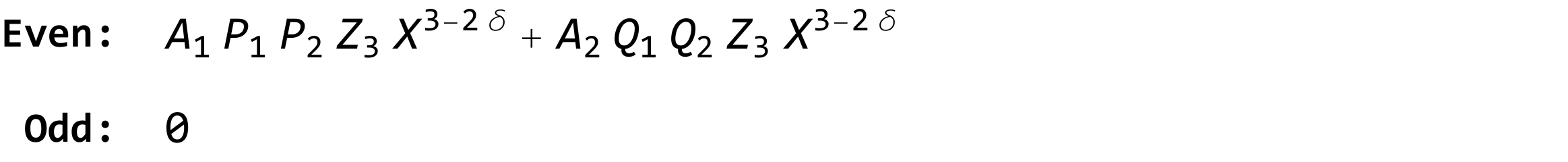}
\end{align}
In this case the triangle inequalities are not satisfied and there are two even solutions. All structures survive after imposing $\psi = \psi'$. \\[3mm]
\textbf{Correlation function} $\langle J^{}_{1} J'_{1} \, O \rangle$\textbf{:}
\begin{align}\label{1-1-0}
	\includegraphics[width=0.9\textwidth, valign=c]{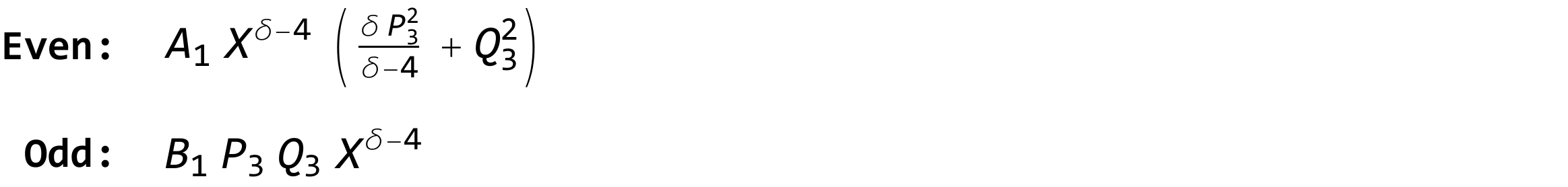}
\end{align}
In this case the triangle inequalities are satisfied and there is one even solution and one odd solution. The structures survive after imposing $J = J'$. \\[3mm]
\textbf{Correlation function} $\langle J^{}_{2} J'_{1} \, O \rangle$\textbf{:}\\[2mm]
In this case there is a single even solution for $\D_{O} = 1$:
\begin{align}\label{2-1-0-A}
	\includegraphics[width=0.9\textwidth, valign=c]{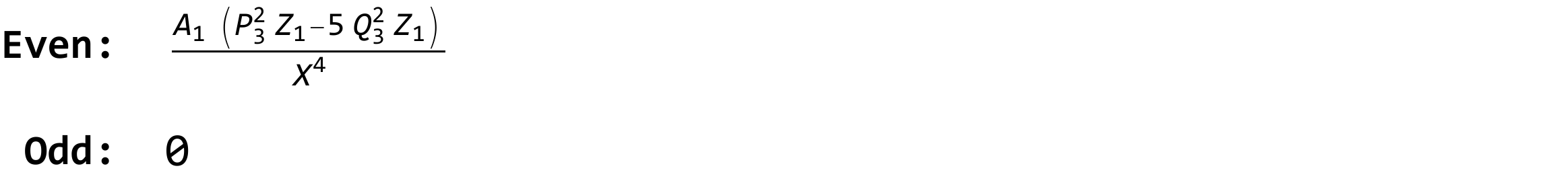}
\end{align}
There is also a single odd solution for $\D_{O} = 2$:
\begin{align}\label{2-1-0-B}
	\includegraphics[width=0.9\textwidth, valign=c]{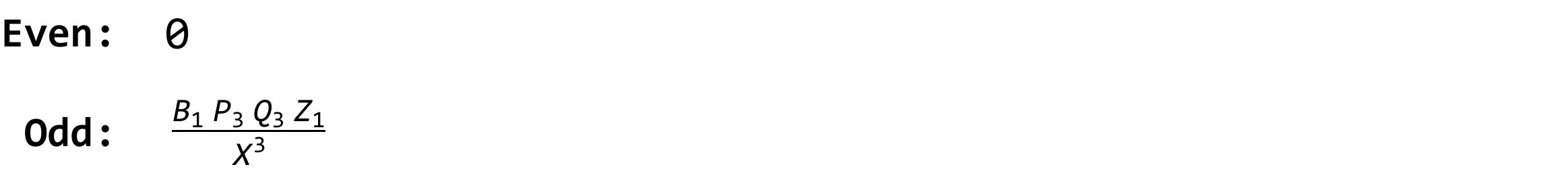}
\end{align}
However, there are no solutions for arbitrary $\D_{O}$. The same results was found in \cite{Giombi:2011rz}. \\[3mm]
\noindent
\textbf{Correlation function} $\langle J^{}_{3/2} J'_{3/2} \, O \rangle$\textbf{:}
\begin{align}\label{3/2-3/2-0}
	\includegraphics[width=0.9\textwidth, valign=c]{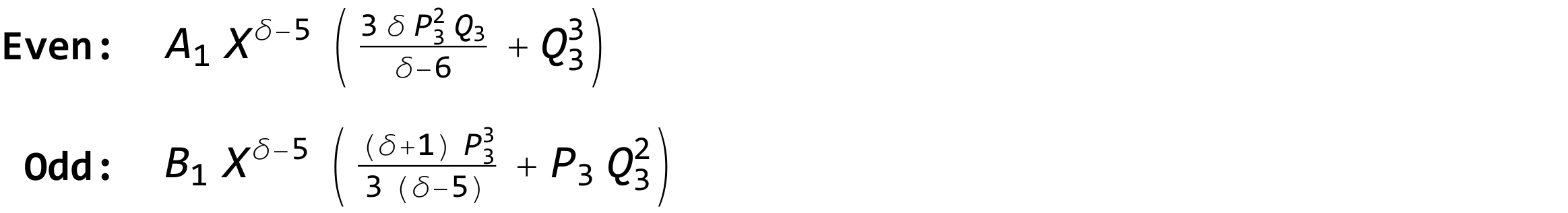}
\end{align} \\
\textbf{Correlation function} $\langle J^{}_{2} J'_{2} \, O \rangle$\textbf{:}
\begin{align}\label{2-2-0}
	\includegraphics[width=0.9\textwidth, valign=c]{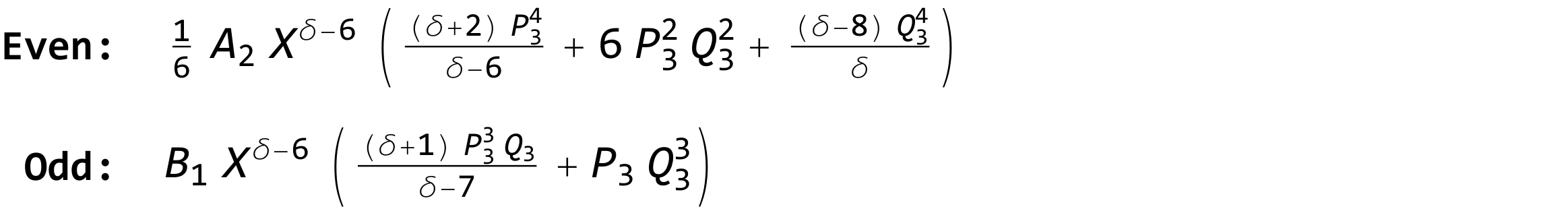}
\end{align} \\
\textbf{Correlation function} $\langle \psi \, J_{3/2} \, O \rangle$\textbf{:}\\[2mm]
For $\Delta_{\psi} = 3/2$, there is an even solution for $\Delta_{O} = 1$:
\begin{align}\label{1/2-3/2-0-A}
	\includegraphics[width=0.9\textwidth, valign=c]{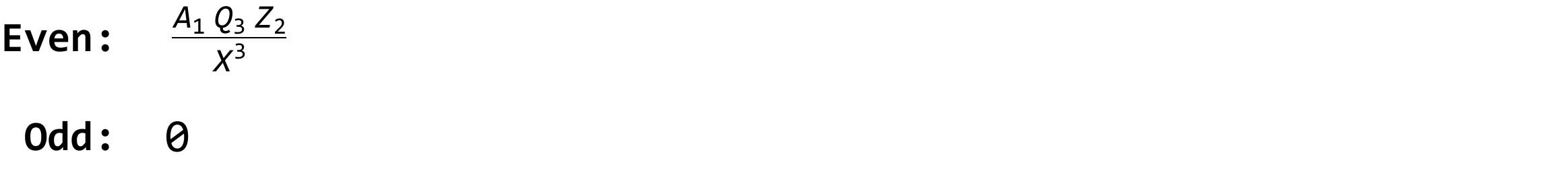}
\end{align}
There is also an odd solution for $\Delta_{O} = 2$:
\begin{align}\label{1/2-3/2-0-B}
	\includegraphics[width=0.9\textwidth, valign=c]{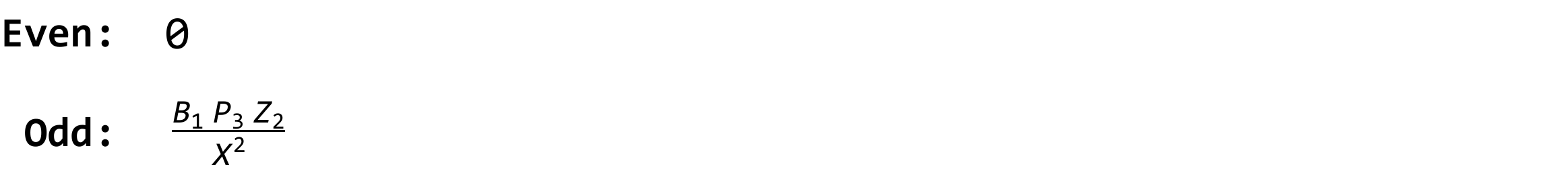}
\end{align} \\
\textbf{Correlation function} $\langle \psi \, J^{}_{3/2} J'_{1} \rangle$\textbf{:}\\[2mm]
In this case there is an even and odd solution for general $\Delta_{\psi}$:
\begin{align}\label{1/2-3/2-1-A}
	\includegraphics[width=0.9\textwidth, valign=c]{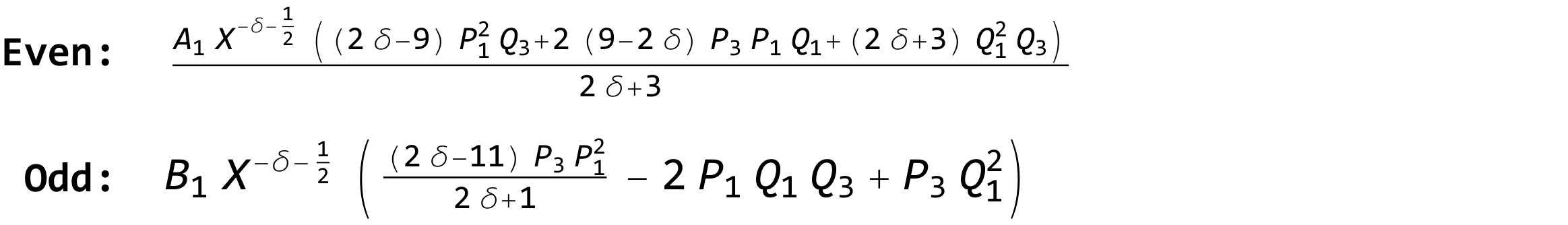}
\end{align} \\
There is also an additional even solution for $\Delta_{\psi} = 3/2$:
\begin{align}\label{1/2-3/2-1-B}
	\includegraphics[width=0.9\textwidth, valign=c]{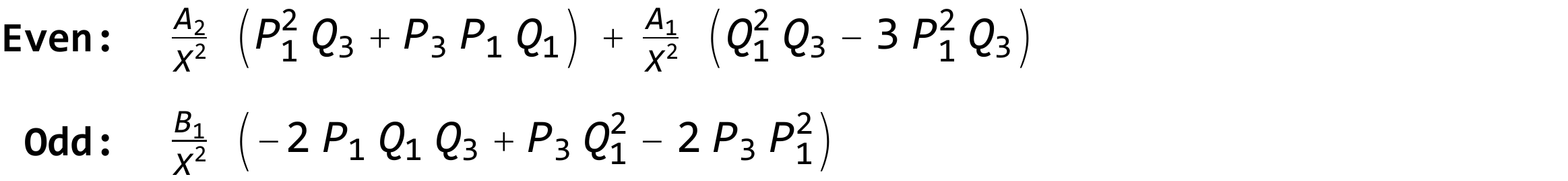}
\end{align} \\
\textbf{Correlation function} $\langle \psi \, J^{}_{3/2} J'_{2} \rangle$\textbf{:}\\[2mm]
In this case there is one even and one odd solution for general $\Delta_{\psi}$:
\begin{align}\label{1/2-3/2-2-A}
	\includegraphics[width=0.9\textwidth, valign=c]{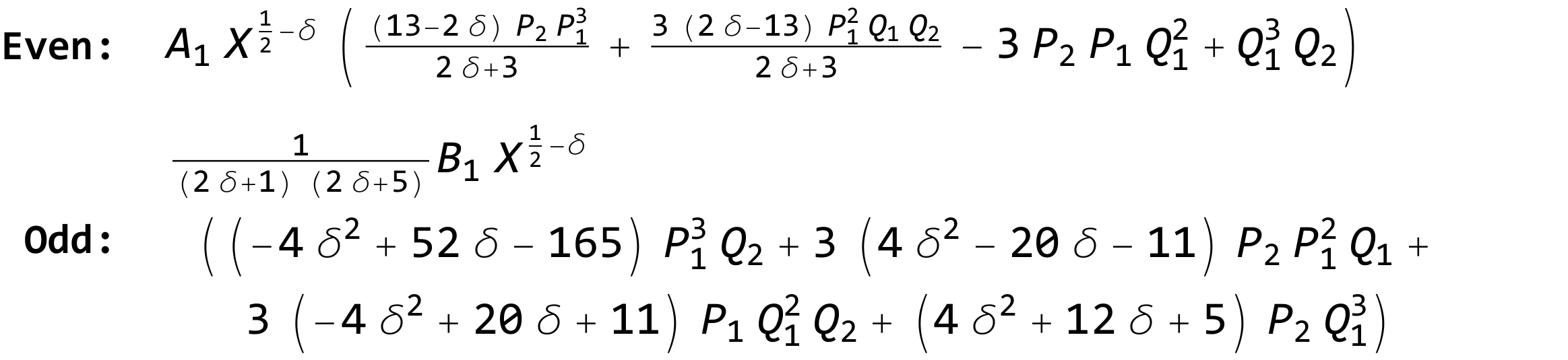}
\end{align}
However, there is an additional even solution for $\Delta_{\psi} = 3/2$:
\begin{align}\label{1/2-3/2-2-B}
	\includegraphics[width=0.9\textwidth, valign=c]{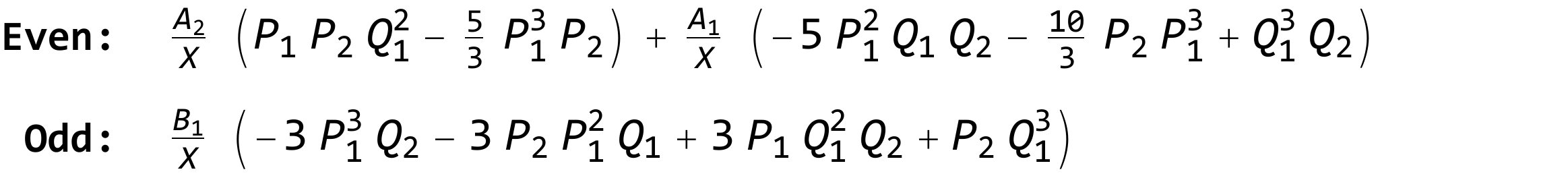}
\end{align}
\subsection{Higher-spin correlators}
In this subsection we provide some more examples of three-point correlation functions involving combinations of scalars, spinors and higher-spin conserved currents. 

\noindent
\textbf{Correlation function} $\langle J^{}_{5/2} J'_{5/2} \, O \rangle$\textbf{:}
\begin{align}\label{5/2-5/2-0}
	\includegraphics[width=0.9\textwidth, valign=c]{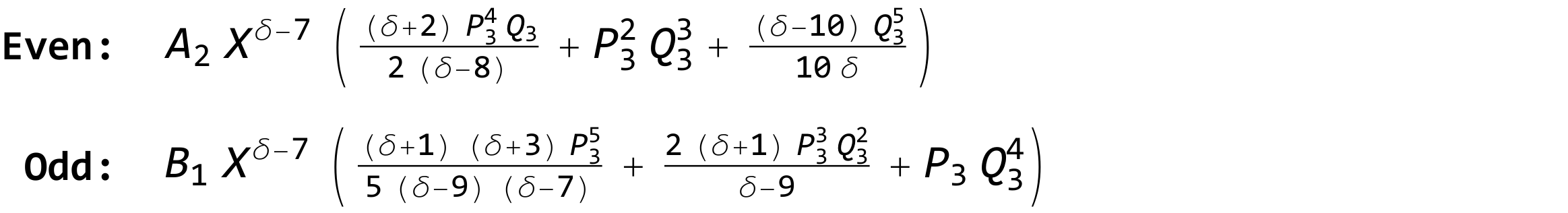}
\end{align} \\
\textbf{Correlation function} $\langle J^{}_{7/2} J'_{7/2} \, O \rangle$\textbf{:}
\begin{align}\label{7/2-7/2-0}
	\includegraphics[width=0.9\textwidth, valign=c]{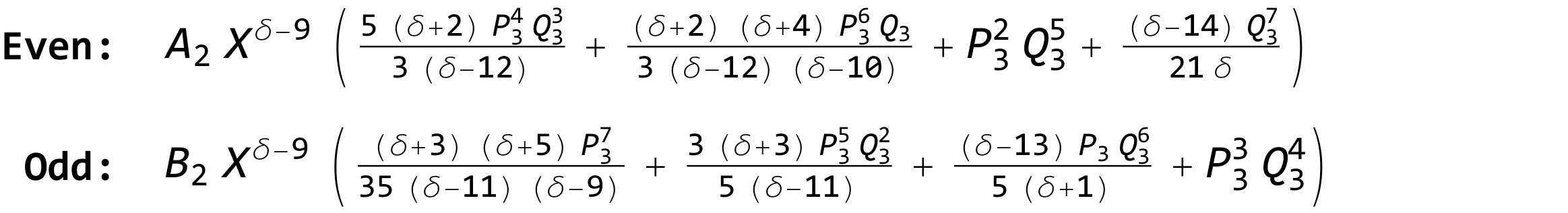}
\end{align} \\
\textbf{Correlation function} $\langle J^{}_{5} J'_{5} \, O \rangle$\textbf{:}
\begin{align}\label{5-5-0}
	\includegraphics[width=0.9\textwidth, valign=c]{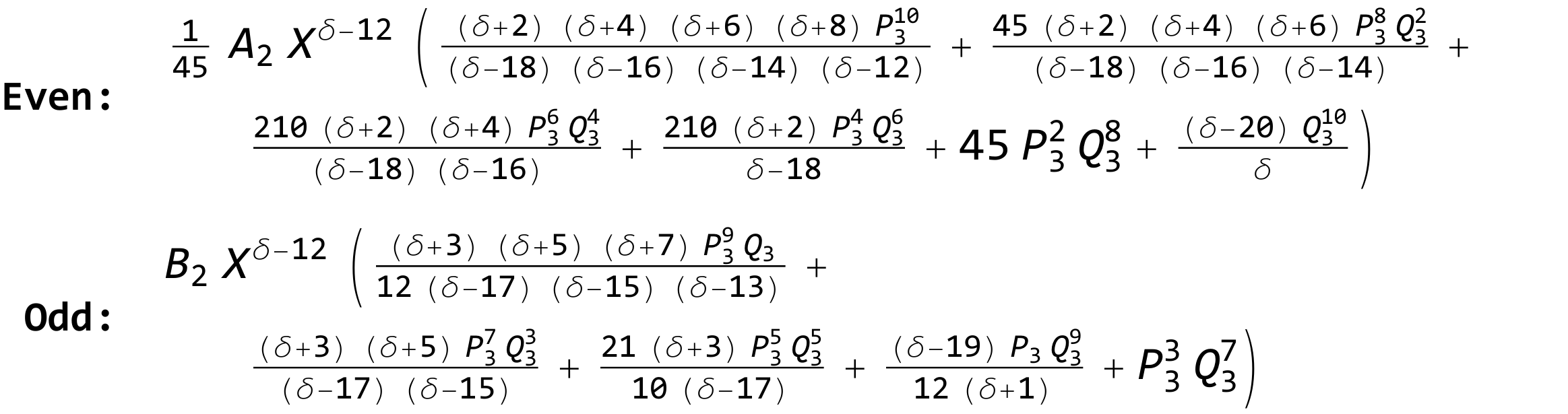}
\end{align} \\
\textbf{Correlation function} $\langle J^{}_{9/2}\, J'_{7/2} \, O \rangle$\textbf{:}\\[2mm]
In this case the triangle inequalities are not satisfied, and there is an even solution for $\Delta_{O} = 1$:
\begin{align}\label{9/2-7/2-0-A}
	\includegraphics[width=0.9\textwidth, valign=c]{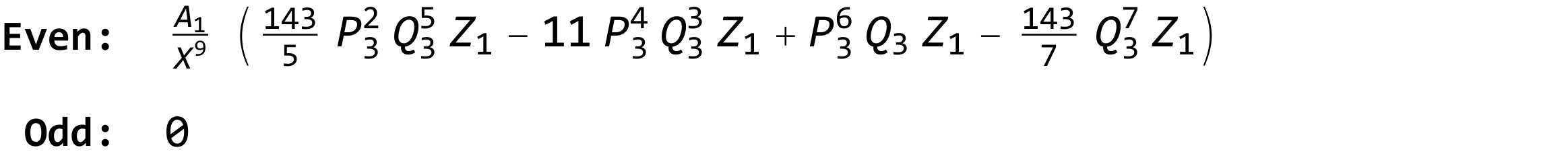}
\end{align}
There is also an odd solution for $\Delta_{O} = 2$:
\begin{align}\label{9/2-7/2-0-B}
	\includegraphics[width=0.9\textwidth, valign=c]{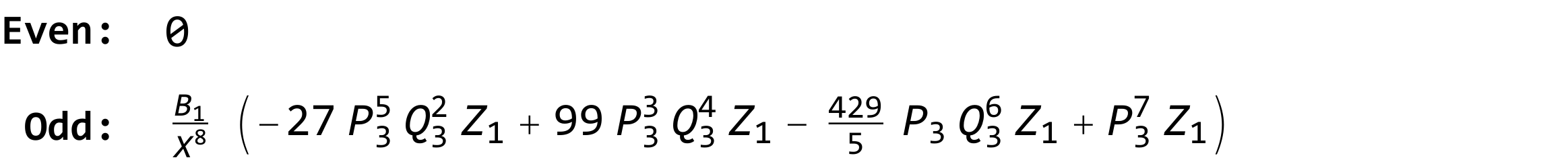}
\end{align} \\
\textbf{Correlation function} $\langle \psi \, J^{}_{7/2} J'_{4} \rangle$\textbf{:}\\[2mm]
In this case the triangle inequalities are satisfied. For $\Delta_{\psi} = 3/2$, we see there are two even solutions and one odd solution:
\begin{align}\label{1/2-7/2-4}
	\includegraphics[width=0.9\textwidth, valign=c]{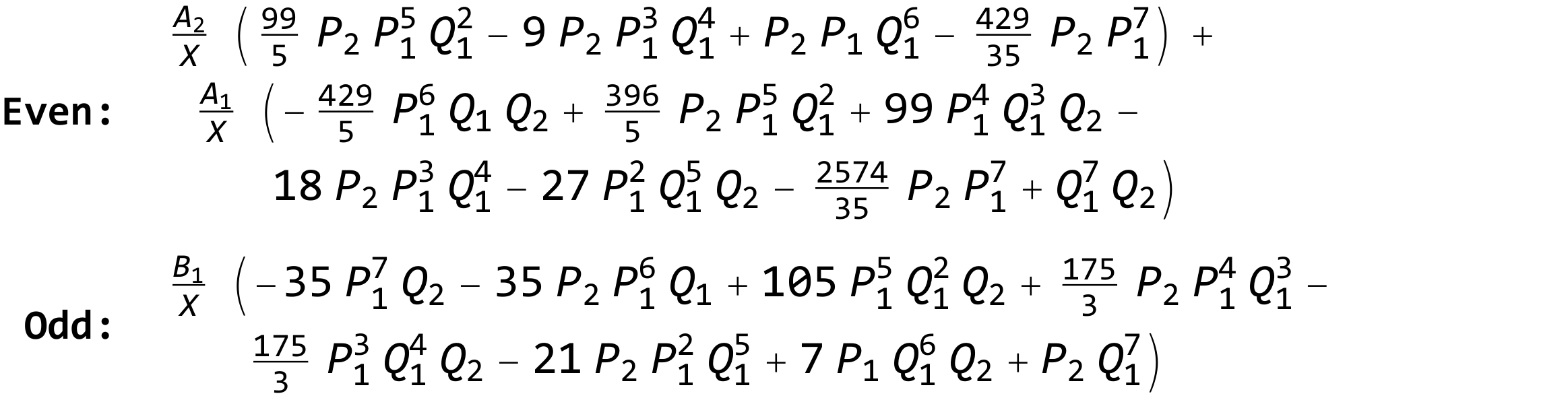}
\end{align}
%

\section{Discussion}\label{section7}

The purpose of this paper is to develop a formalism to determine the general form of the three-point correlation functions of conserved currents with arbitrary spins in three-dimensional conformal field theory. Our method gives explicit results and is limited only by computer power. We managed to find solutions for spins up to $s_{i} = 20$, but the pattern of the number of independent structures is very clear and allows us to conclude that it holds in general. We demonstrate that in all cases where the triangle inequalities are simultaneously satisfied, there are two even solutions and one odd solution for $\langle J^{}_{s_{1}} J'_{s_{2}} J''_{s_{3}} \rangle$, otherwise, there are only two even solutions. Although the results for three-point functions involving bosonic currents have been proposed previously~\cite{Maldacena:2011jn,Giombi:2011rz,Zhiboedov:2012bm}, we believe that our analysis stands on its own merit, as our method for imposing conservation on all three points is very explicit and analytic at every step of the computations. In addition, we construct a discriminant equation which governs the existence of the odd structure, and we extend the scope of our analysis to include correlation functions of conserved fermionic currents. Another benefit of our approach is that it can be directly generalised to four- and higher-dimensional conformal field theories as well as to (extended) superconformal field theories in diverse dimensions. We intend to explore these ideas in future works.

Finally, let us remark on three-point functions of fermionic, spin-$3/2$ currents. These currents naturally appear as supersymmetry currents in (extended) superconformal field theories, hence, it is also interesting to understand the general structure of correlation functions of spin-3/2 currents when supersymmetry is not manifest. In a superconformal theory the correlation functions \eqref{Susy current correlators - 1}
are contained within the following supersymmetric three-point functions:
%
\begin{equation}
	\langle \, \mathbf{J}_{\a(3)}(z_{1}) \, \mathbf{J}_{\b(3)}(z_{2}) \, \mathbf{L}_{\g}(z_{3}) \rangle \, , \qquad  \langle \, 
	\mathbf{J}_{\a(3)}(z_{1}) \, \mathbf{J}_{\b(3)}(z_{2}) \, \mathbf{J}_{\g(3)}(z_{3}) \rangle \,.
	\label{Susy correlators}
\end{equation}
In three-dimensions, $\langle \mathbf{J} \mathbf{J} \mathbf{L} \rangle$ vanishes, while $\langle \mathbf{J} \mathbf{J} \mathbf{J} \rangle$ is fixed up to a single parity-even tensor structure \cite{Nizami:2013tpa, Buchbinder:2015qsa, Buchbinder:2021gwu}.
Hence, it appears that supersymmetry imposes additional restrictions on the general structure of three-point correlation functions. It then follows that the 
general form of the correlation functions \eqref{Susy current correlators} is inconsistent with supersymmetry in the sense that they are fixed up to 
more tensor structures than \eqref{Susy correlators}. 
Similar phenomenon was also found in four-dimensional (super)conformal field theories in \cite{Buchbinder:2021izb, Buchbinder:2022cqp}.




\section*{Acknowledgements}
The authors are grateful to Sergei Kuzenko and Jessica Hutomo for valuable discussions. The work of E.I.B. is supported in part by the Australian Research Council, project No. DP200101944. The work of B.S. is supported by the \textit{Bruce and Betty Green Postgraduate Research Scholarship} under the Australian Government Research Training Program. 



\appendix

\section{3D conventions and notation}\label{AppA}

For the Minkowski metric we use the ``mostly plus'' convention: $\eta_{mn} = \text{diag}(-1,1,1)$. Spinor indices are then raised and lowered with the $\text{SL}(2,\mathbb{R})$ invariant anti-symmetric $\varepsilon$-tensor
\begin{align}
	\ve_{\a \b} = 
	\begingroup
	\setlength\arraycolsep{4pt}
	\begin{pmatrix}
		\, 0 & -1 \, \\
		\, 1 & 0 \,
	\end{pmatrix}
	\endgroup 
	\, , & \hspace{5mm}
	\ve^{\a \b} =
	\begingroup
	\setlength\arraycolsep{4pt}
	\begin{pmatrix}
		\, 0 & 1 \, \\
		\, -1 & 0 \,
	\end{pmatrix}
	\endgroup 
	\, , \hspace{5mm}
	\ve_{\a \g} \ve^{\g \b} = \d_{\a}{}^{\b} \, , \\[4mm]
	& \hspace{-8mm} \f_{\a} = \ve_{\a \b} \, \f^{\b} \, , \hspace{10mm} \f^{\a} = \ve^{\a \b} \, \f_{\b} \, .
\end{align}
The $\g$-matrices are chosen to be real, and are expressed in terms of the Pauli matrices $\s$ as follows:
\begin{subequations}
	\begin{align}
		(\g_{0})_{\a}{}^{\b} = - \text{i} \s_{2} = 
		\begingroup
		\setlength\arraycolsep{4pt}
		\begin{pmatrix}
			\, 0 & -1 \, \\
			\, 1 & 0 \,
		\end{pmatrix}
		\endgroup 
		\, , & \hspace{8mm}
		(\g_{1})_{\a}{}^{\b} = \s_{3} = 
		\begingroup
		\setlength\arraycolsep{4pt}
		\begin{pmatrix}
			\, 1 & 0 \, \\
			\, 0 & -1 \,
		\end{pmatrix}
		\endgroup 
		\, , \\[3mm]
		(\g_{2})_{\a}{}^{\b} = - \s_{1} &= 
		\begingroup
		\setlength\arraycolsep{4pt}
		\begin{pmatrix}
			\, 0 & -1 \, \\
			\, -1 & 0 \,
		\end{pmatrix}
		\endgroup 
		\, ,
	\end{align}
\end{subequations}
\begin{equation}
	(\g_{m})_{\a \b} = \ve_{\b \d} (\g_{m})_{\a}{}^{\d} \, , \hspace{10mm} (\g_{m})^{\a \b} = \ve^{\a \d} (\g_{m})_{\d}{}^{\b} \, .
\end{equation}
The $\g$-matrices are traceless and symmetric
\begin{equation}
	(\g_{m})^{\a}{}_{\a} = 0 \, , \hspace{10mm} (\g_{m})_{\a \b} = (\g_{m})_{\b \a} \, ,
\end{equation} 
and also satisfy the Clifford algebra
\begin{equation}
	\g_{m} \g_{n} + \g_{n} \g_{m} = 2 \eta_{mn} \, .
\end{equation}
For products of $\g$-matrices we make use of the identities
\begin{subequations}
	\begin{align}
		(\g_{m})_{\a}{}^{\r} (\g_{n})_{\r}{}^{\b} &= \eta_{mn} \d_{\a}{}^{\b} + \e_{mnp} (\g^{p})_{\a}{}^{\b} \, , \\[2mm]
		(\g_{m})_{\a}{}^{\r} (\g_{n})_{\r}{}^{\s} (\g_{p})_{\s}{}^{\b} &= \eta_{mn} (\g_{p})_{\a}{}^{\b} - \eta_{mp} (\g_{n})_{\a}{}^{\b} + \eta_{np} (\g_{m})_{\a}{}^{\b} + \e_{mnp} \d_{\a}{}^{\b} \, ,
	\end{align}
\end{subequations}
where we have introduced the 3D Levi-Civita tensor $\e$, with $\e^{012} = - \e_{012} = 1$. We also have the orthogonality and completeness relations for the $\g$-matrices
\begin{equation}
	(\g^{m})_{\a \b} (\g_{m})^{\r \s} = - \d_{\a}{}^{\r} \d_{\b}{}^{\s}  - \d_{\a}{}^{\s}  \d_{\b}{}^{\r} \, , \hspace{5mm} (\g_{m})_{\a \b} (\g_{n})^{\a \b} = -2 \eta_{mn} \, .
\end{equation}
Finally, the $\g$-matrices are used to swap from vector to spinor indices. For example, given some three-vector $x_{m}$, it may equivalently be expressed in terms of a symmetric second-rank spinor $x_{\a \b}$ as follows:
\begin{align}
	x^{\a \b} = (\g^{m})^{\a \b} x_{m}  \, , \hspace{5mm} x_{m} = - \frac{1}{2} (\g_{m})^{\a \b} x_{\a \b} \, , \\[2mm]
	\det (x_{\a \b}) = \frac{1}{2} x^{\a \b} x_{\a \b} = - x^{m} x_{m} = -x^{2} \, .
\end{align}
The same conventions are also adopted for the spacetime partial derivatives $\partial_{m}$
\begin{align}
	\partial_{\a \b} = (\g^{m})^{\a \b} \partial_{m}  \, , \hspace{5mm} \partial_{m} = - \frac{1}{2} (\g_{m})^{\a \b} \partial_{\a \b} \, , \\[2mm]
	\partial_{m} x^{n} = \d_{m}^{n} \, , \hspace{5mm} \partial_{\a \b} x^{\r \s} = - \d_{\a}{}^{\r} \d_{\b}{}^{\s}  - \d_{\a}{}^{\s}  \d_{\b}{}^{\r} \, ,
\end{align}
\begin{equation}
	\x^{m} \partial_{m} = - \frac{1}{2} \x^{\a \b} \partial_{\a \b} \, .
\end{equation}
%

\newpage

	
%



\section{More examples of higher-spin correlators}\label{AppB}
\noindent In this appendix we provide further examples of three-point functions of higher-spin currents using our formalism.

\textbf{Correlation function} $\langle J^{}_{5} J'_{5} J''_{5} \rangle$\textbf{:}
\begin{align}\label{5-5-5}
	\includegraphics[width=0.9\textwidth, valign=c]{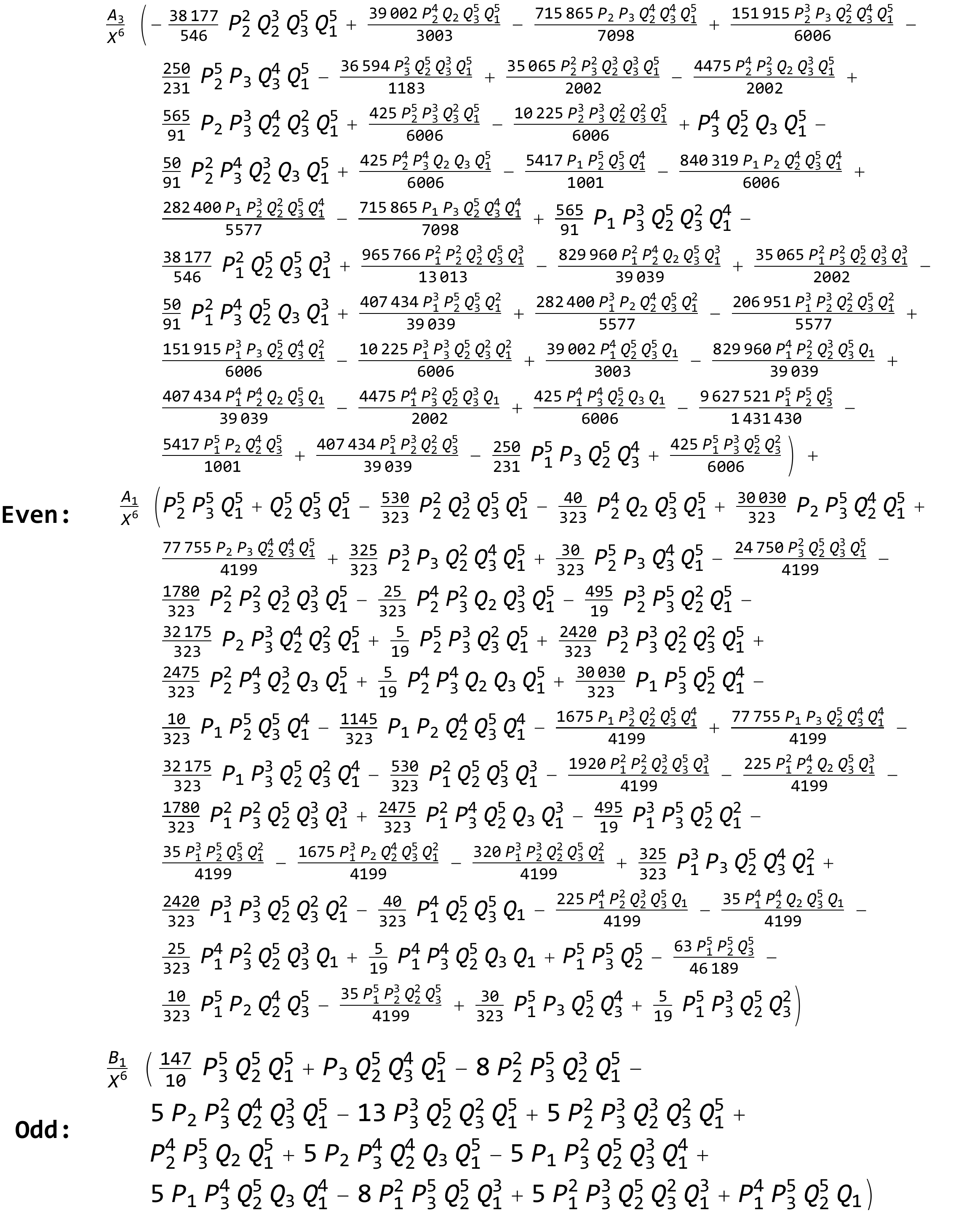}
\end{align}
\textbf{Correlation function} $\langle J^{}_{6} J'_{6} J''_{6} \rangle$\textbf{:}
\begin{align}\label{6-6-6}
	\includegraphics[width=0.9\textwidth, valign=c]{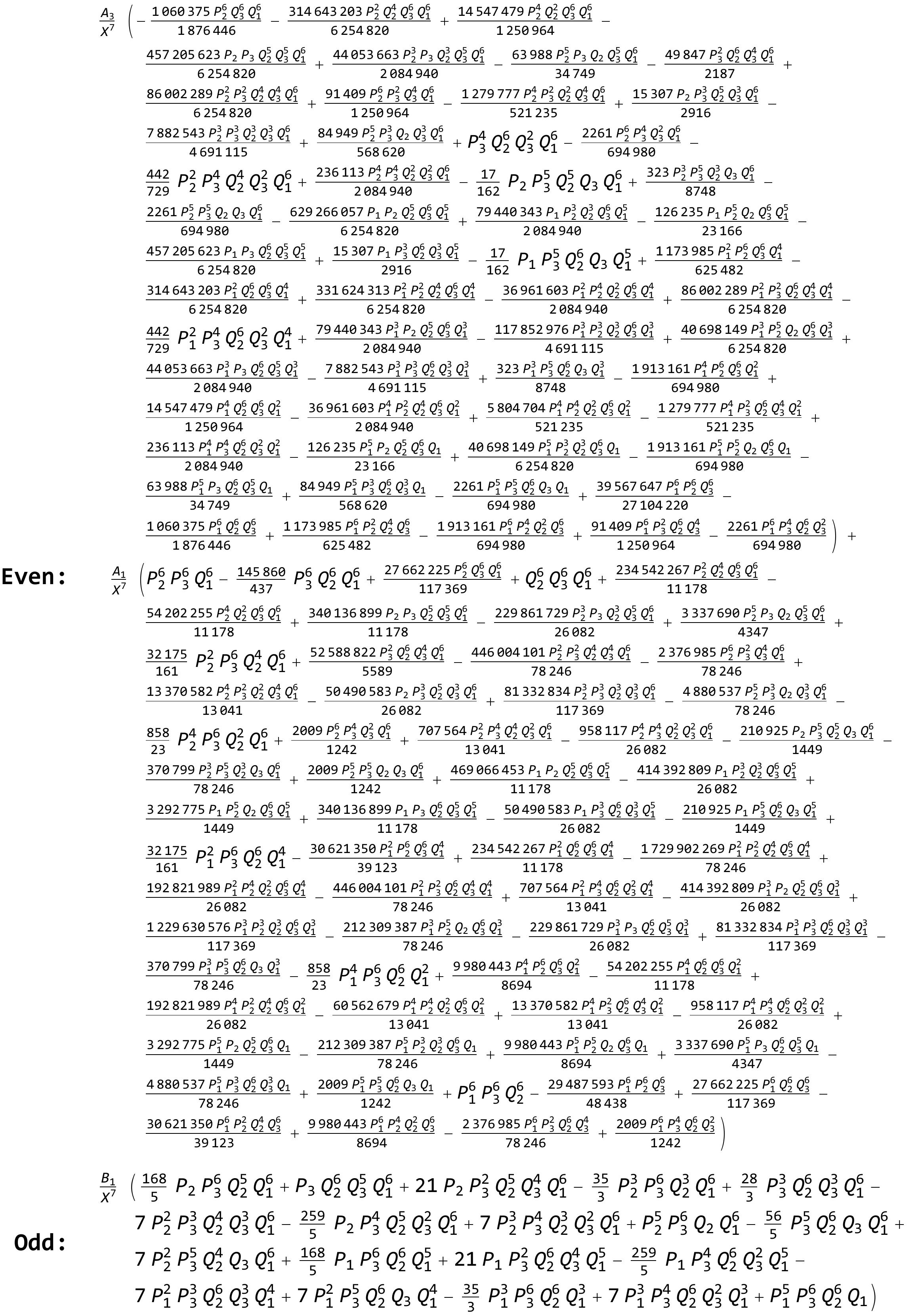}
\end{align}
\textbf{Correlation function} $\langle J^{}_{7/2} J'_{7/2} J''_{6} \rangle$\textbf{:}
\begin{align}\label{7/2-7/2-6}
	\includegraphics[width=0.9\textwidth, valign=c]{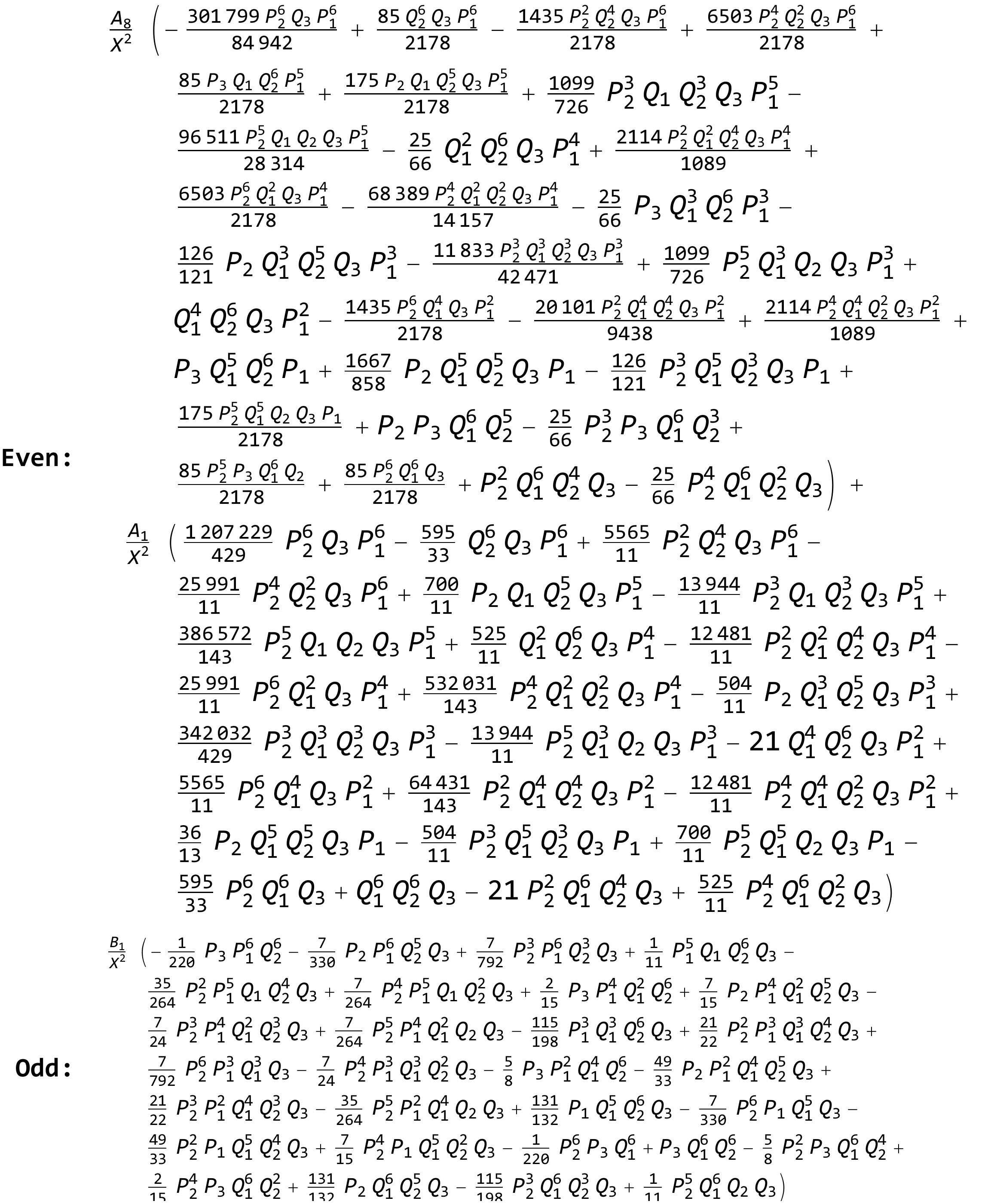}
\end{align}
\textbf{Correlation function} $\langle J^{}_{9/2} J'_{9/2} J''_{6} \rangle$\textbf{:}
\begin{align}\label{9/2-9/2-12}
	\includegraphics[width=0.9\textwidth, valign=c]{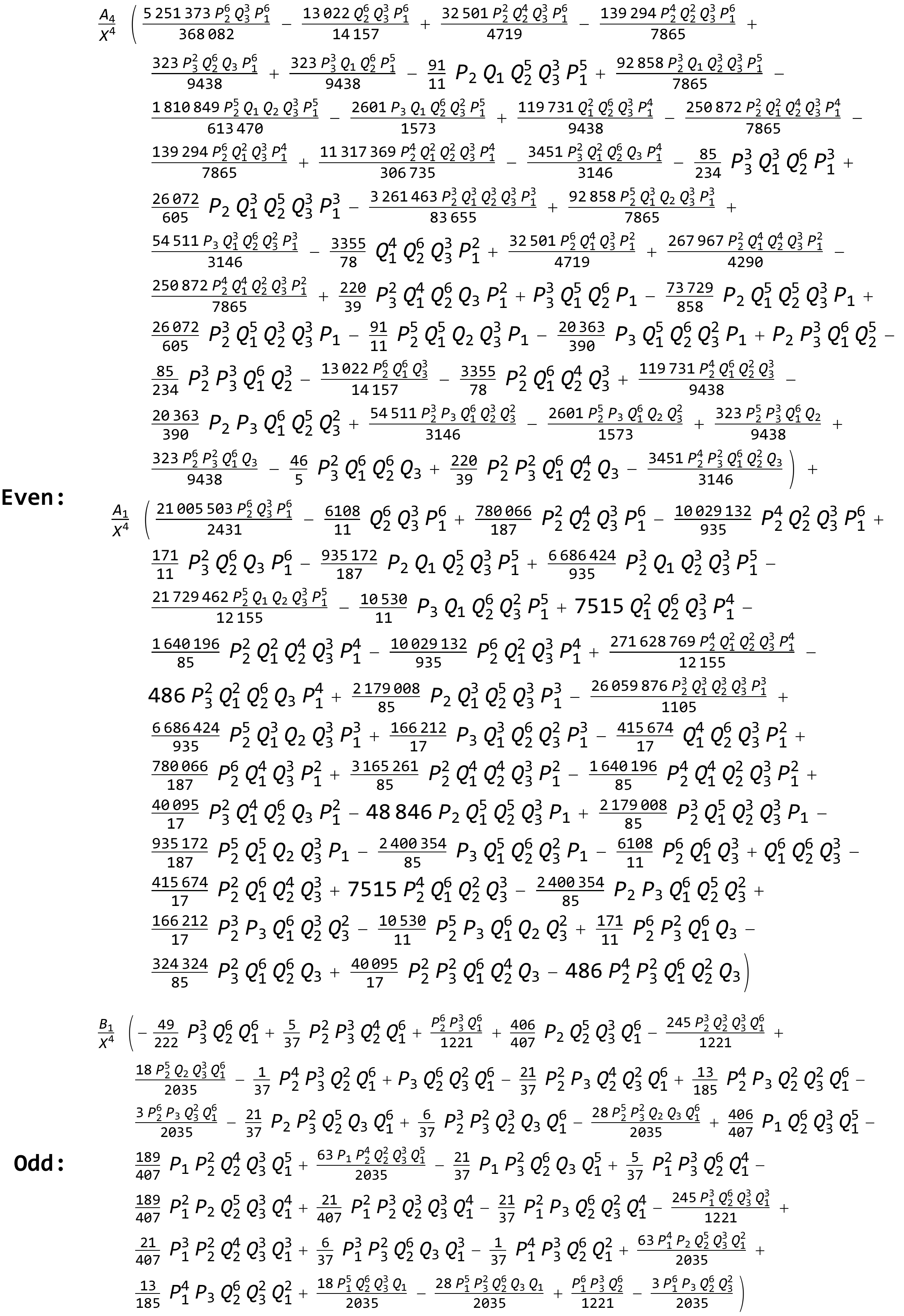}
\end{align}


%


\printbibliography[heading=bibintoc,title={References}]



\end{document}